\documentclass[prb,twocolumn,groupedaddress,showpacs,aps,floatfix,fleqn]{revtex4}
\usepackage{amsmath}
\usepackage{amstext}
\usepackage{epsfig}
\usepackage{graphicx}
\usepackage{dcolumn}
\usepackage{rotating}
\usepackage{draftcopy}
\usepackage{longtable}
\usepackage{caption2}
\usepackage{float}
\usepackage{flafter}

\begin{document}
\bibliographystyle{apsrev}

\title{Electronic structures and electron spin decoherence in $(001)$-grown layered zincblende semiconductors}


\author{Wayne H. Lau}
\altaffiliation{To whom correspondence should be addressed. Current address: California Nanosystems Institute, The University of California, Santa Barbara, California, 93106. E-mail: wlau@mailaps.org.}
\author{J. T. Olesberg}
\author{ Michael E. Flatt\'e}

\affiliation{Optical Science and Technology Center and Department of Physics and Astronomy, The University of Iowa, Iowa City, Iowa 52242}

\date{\today}

\begin{abstract}
Electronic structure calculations for layered zincblende semiconductors are described within a restricted basis formalism which naturally and non-perturbatively accomodates both crystalline inversion asymmetry and cubic anisotropy. These calculations are applied to calculate the electron spin decoherence times $T_1$ and $T_2$ due to precessional decoherence in quantum wells. Distinctly different dependences of spin coherence times on mobility, quantization energy, and temperature are found from perturbative calculations. Quantitative agreement between these calculations and experiments is found for GaAs/AlGaAs, InGaAs/InP, and GaSb/AlSb $(001)$-grown quantum wells. The electron spin coherence times for CdZnSe/ZnSe II-VI quantum wells are calculated, and calculations of InGaAs/GaAs quantum wells appropiate for comparison with spin-LED structures are also presented.

\end{abstract}

\maketitle



%
%

\section{INTRODUCTION}
\label{sec:intro}
In recent years there has been considerable interest in exploiting not only charge but also spin in solid-state electronics, which has led to new ultrafast optical studies of electron spin dynamics in bulk and quantum-well semiconductors and possible applications to ultrafast spin-dependent switching, quantum computing, or other "spintronic" devices \cite{awschalom02, wolf01, sham97}. The magnitude and persistence of such effects are governed partly by the spin coherence times $T_1$ and $T_2$, describing the decay of longitudinal and transverse spin order, respectively. Ultrafast optical measurements have been performed of both $T_1$ and $T_2$, although in different geometries \cite{kikkawa97, kikkawa98, kikkawa99, kohl91, barad92, heberle94, wagner93, britton98, malinowski00, hyland99, hall99, terauchi99, tackeuchi97, tackeuchi99, nishikawa96, boggess00}.

We recently described an approach to calculate these times for layered zincblende semiconductor structures based on fourteen-band non-perturbative restricted basis set calculations~\cite{awschalom02, boggess00, lau01, lau02}. Success was reported in resolving the discrepencies between theoretical and experimental results for several material systems, both in trends and in absolute magnitude. Here we provide full details of the calculational method, including both the general restricted basis calculations,
which can be applied to a variety of optoelectronic properties, and the specific calculations of spin coherence times.

The mechanism of electron spin decoherence we consider occurs via the spin precession of carriers with finite crystal momentum {\bf k} in the effective momentum-dependent crystal magnetic field of an inversion-asymmetric material. This effective ${\bf k}$-dependent crystal magnetic field is a direct result of the spin splitting of the conduction band at finite momentum caused by lack of a spatial inversion symmetry in the zincblende lattice. Thus a proper calculation of this spin splitting requires a proper electronic structure.

In this work, we employ a non-perturbative heterostructure theory based on a generalized ${\bf K{\cdot}p}$ envelope function theory~\cite{flatte96, johnson90, johnson88, johnson87} solved in a fourteen-band restricted basis set. The electronic structure of quantum wells is obtained by expressing the electronic states as spatially dependent linear combinations of the fourteen states in the basis. The full Hamiltonian is projected onto this restricted basis set, which produces a set of fourteen coupled differential equations for the spatially dependent coefficient of the basis states (generalized envelope functions). These equations are then solved in Fourier space in a similar method to that of Winkler and R$\ddot{\rm o}$ssler \cite{winkler93, olesberg99}. The expressions for electron spin coherence times $T_1$ and $T_2$ are obtained by solving the momentum-dependent spin density matrix equation, and the calculations of electron spin coherence times are carried out within the framework of the fourteen-band theory. 

Our paper is organized in the following way: in Sec.~\ref{sec:esd} a detailed description of the theory of electron spin decoherence in bulk and quantum-well zincblende semiconductors is given. This includes a description of the D'yakonov-Perel'(DP) formulae for bulk~\cite{dyakonov71, dyakonov72, meier84} and the D'yakonov-Kachorovskii (DK) formulae for quantum wells~\cite{dyakonov86} to facilate the comparison between our results and those of DP and DK. We also describe the relationship between the longitudinal spin relaxation time $T_1$ and the transverse spin relaxation time $T_2$, and  the connection between the longitudinal spin relaxation time $T_1$ and the spin-flip time $\tau_s$. The calculation of electronic structure and the momentum-dependent effective magnetic field based on a fourteen-band non-perturbative ${\bf K \cdot p}$ theory is described in Sec.~\ref{sec:fbt}. Calculated electron spin coherence times for a variety of material system, both bulk and quantum-well semiconductors, are presented in Sec.~\ref{sec:rd}, and comparison with experiments is made where possible. 
\section{ELECTRON SPIN DYNAMICS}
\label{sec:esd}
Spin relaxation of conduction electrons in inversion asymmetric bulk semiconductors due to rotation of spins during scattering of electrons was originally studied by D'yakonov and Perel'~\cite{dyakonov71, dyakonov72} using a momentum-dependent spin density matrix approach based on a simple electronic band structure approximation (see also Ref.~\onlinecite{meier84}). This work was later extended by D'yakonov and Kachorovskii~\cite{dyakonov86} to study electron spin relaxation in inversion asymmetric quantum-well semiconductors (see also Refs.~\onlinecite{averkiev99}, \onlinecite{averkiev02}, and \onlinecite{bournel00}). In this work, we generalize the analysis of electron spin relaxation for both bulk and quantum-well semiconductors to include multiband effects.
\subsection{DENSITY MATRIX FORMALISM}
\label{sec:dmf}
To study electron spin decoherence in bulk and layered zincblende semiconductors, we consider a population of $N_\text{e}$ electrons in the system. Each electron is described by the total Hamiltonian:
\begin{equation}
\hat{H}^\text{T} = \hat{H}^\text{0}  +\hat{H}^\text{SO} + \hat{H}^\text{SC},
\label{totalHam}
\end{equation}
where $\hat{H}^\text{0}$ is the periodic crystal Hamiltonian without spin-orbit interaction, $\hat{H}^\text{SO}$ is the Hamiltonian for the spin-orbit interaction, and $\hat{H}^\text{SC}$ is the spin-independent Hamiltonian describing any scattering interaction. The crystal Hamiltonian is $\hat{H} = \hat{H}^\text{0} + \hat{H}^\text{SO}$ with $\hat{H}^\text{0}$ and $\hat{H}^\text{SO}$ written as
\begin{subequations}
\label{hamCryT}
\begin{equation}
\hat{H}^\text{0} = \frac{\hat{\bf p}^{2}}{2m_\text{e}} + V(\hat{\bf r}),
\label{ham0}
\end{equation}
\begin{equation}
\hat{H}^\text{SO}  = \frac{\hbar}{4m_\text{e}^{2}c^{2}} [{\nabla} V(\hat{\bf r})  {\times}  \hat{\bf p}] \cdot \hat{\mbox{\boldmath$\sigma$}},
\label{hamSO}
\end{equation}
\end{subequations}
where $m_\text{e}$ is the free electron mass, $c$ is the velocity of light, $\hbar$ is the Planck's constant divided by $2\pi$, $\hat{\bf p}$ is the electron momentum operator, $V(\hat{\bf r})$ is the periodic crystal potential, and $\hat{\mbox{\boldmath$\sigma$}}$ is the Pauli spin operator. We consider the case where the spin-independent scattering potential is weak ($\hat{H}^\text{SO}$ $\ll$ $\hat{H}^\text{SC}$ $\ll$ $\hat{H}^\text{0}$) so that both $\hat{H}^\text{SO}$ and $\hat{H}^\text{SC}$ can be treated as small perturbations. To describe the electron spin dynamics in bulk and layered zincblende semiconductors, we start from the von Neumann equation~\cite{rammer98, kohn57, krieger87}
\begin{equation}
i \hbar \frac{\partial}{ \partial t} \hat{\rho}(t) = \Big[ \hat{H}^\text{T}, \hat{\rho}(t) \Big],
\label{vne} 
\end{equation}
where $\hat{\rho}(t)$ is the time-dependent density operator for the electron population. 

To derive the spin-dependent distribution function for the electrons, we shall express the spin density matrix $\underline{\underline{\rho}}(t)$ in the $\arrowvert n, \sigma, {\bf k} \rangle$ representation in which the Hamiltonian $\underline{\underline{H}}^\text{0} $ is diagonal:
\begin{equation}
\hat{H}^\text{0}  \arrowvert n, \sigma, {\bf k} \rangle 
= E^0_{n \sigma} ({\bf k}) \arrowvert n, \sigma, {\bf k} \rangle,
\label{seh0}
\end{equation}
where $n$ and $\sigma$ are the band and spin indices respectively. The associated eigenfunction is given by
\begin{equation}
\langle {\bf r} \arrowvert n, \sigma, {\bf k} \rangle
=  \exp(i {\bf k \cdot r}) \langle {\bf r} \arrowvert u_{n \sigma {\bf k} }\rangle,
\label{efh0}
\end{equation}
where $\arrowvert u_{n \sigma {\bf k} } \rangle$ is expressed in terms of the product of the Bloch states $\arrowvert  u^0_\text{n} \rangle$ at ${\bf k}$ $=$ $0$ and the spin-1/2 states $\arrowvert  U_\sigma \rangle$:
\begin{equation}
\arrowvert u_{n \sigma {\bf k} } \rangle
=  \sum_{n' \sigma'} a_{n \sigma n' \sigma'} ({\bf k}) 
\arrowvert u^0_{n'} \rangle  \otimes \arrowvert U_{\sigma'}  \rangle.
\label{bfh0}
\end{equation}
In the fourteen-band restricted basis set, $u^0_n$ $\in$ $\{S, X^\text{v}, Y^\text{v}, Z^\text{v}, X^\text{c}, Y^\text{c}, Z^\text{c}\}$ and $U_\sigma$ $\in$ $\{\uparrow, \downarrow\}$ (see Sec.~\ref{sec:fbt} for detailed dissussion). The eigenstates of the electrons $\arrowvert \nu, t \rangle$ under the action of the total Hamiltonian $H^\text{T}$ can be expanded in this complete set of time-independent states $\arrowvert n, \sigma, {\bf k} \rangle$:
\begin{equation}
\arrowvert \nu, t \rangle 
=  \sum_{n \sigma {\bf k} } c_{\nu n \sigma} ({\bf k}, t) \arrowvert n, \sigma, {\bf k} \rangle,
\label{esht}
\end{equation}
and the matrix elements of the time-dependent spin density operator $\hat{\rho}(t)$ are then given by
\begin{eqnarray}
\rho_{n' \sigma' {\bf k}' n \sigma {\bf k}} (t) 
&\equiv& \langle n', \sigma', {\bf k}'  
\arrowvert  \hat{\rho}(t) \arrowvert n, \sigma, {\bf k} \rangle \nonumber \\
&=&  \frac{1}{N_\text{e}} \sum_{\nu = 1} ^{N_\text{e}} 
c^\ast_{\nu n' \sigma'} ({\bf k}', t) c_{\nu n \sigma} ({\bf k}, t).
\label{dme}
\end{eqnarray}

The spin density matrix equation for Eq.~(\ref{vne}) in this representation is
\begin{equation}
i \hbar \frac{\partial}{ \partial t} \rho_{n' \sigma' {\bf k}' n \sigma {\bf k}}(t)
 =  \langle n', \sigma', {\bf k}' \arrowvert \Big[ \hat{H}^\text{T}, \hat{\rho}(t) \Big] 
\arrowvert n, \sigma, {\bf k} \rangle.
\label{dmeh0} 
\end{equation} 
Substituting Eq.~(\ref{totalHam}) into the right-hand side of (\ref{dmeh0}) and expressing each operator in terms of individual matrix elements we have 
\begin{eqnarray}
&& i \hbar \frac{\partial}{ \partial t} \rho_{n' \sigma' {\bf k}' n \sigma {\bf k}}(t) \nonumber \\
&& =  \sum_{n'' \sigma'' {\bf k}''} \Big(E^\text{0}_{n' \sigma'} ({\bf k}') 
\delta_{n' n''} \delta_{\sigma' \sigma''} \delta_{{\bf k}' {\bf k}''}
\rho_{n'' \sigma'' {\bf k}'' n \sigma {\bf k}}(t) \nonumber \\
&& - \rho_{n' \sigma' {\bf k}' n'' \sigma'' {\bf k}''}(t)
E^\text{0}_{n \sigma} ({\bf k}) \delta_{n'' n}
 \delta_{\sigma'' \sigma} \delta_{{\bf k}'' {\bf k}} \Big)  \nonumber \\
&& + \sum_{n'' \sigma'' {\bf k}''}  
\Big(H^\text{SO}_{n' \sigma' {\bf k}' n'' \sigma'' {\bf k}''} 
 \rho_{n'' \sigma'' {\bf k}'' n \sigma {\bf k}}(t) \nonumber \\ 
&& - \rho_{n' \sigma' {\bf k}' n'' \sigma'' {\bf k}''}(t) 
H^\text{SO}_{n'' \sigma'' {\bf k}'' n \sigma {\bf k}} \Big)  \nonumber \\
&& +  \sum_{n'' \sigma'' {\bf k}''}  
\Big(H^\text{SC}_{n' \sigma' {\bf k}' n'' \sigma'' {\bf k}''}  
\rho_{n'' \sigma'' {\bf k}'' n \sigma {\bf k}}(t) \nonumber \\
&& - \rho_{n' \sigma' {\bf k}' n'' \sigma'' {\bf k}''}(t) 
H^\text{SC}_{n'' \sigma'' {\bf k}'' n \sigma {\bf k}} \Big).
\label{dmeh02} 
\end{eqnarray}
For compactness, we may rewrite Eq.~(\ref{dmeh02}) in matrix form:
\begin{eqnarray} 
&& i \hbar \frac{\partial}{ \partial t} \underline{\underline{\rho}}\text{\@}
_{n' {\bf k}' n {\bf k}}(t) 
= \Big(\underline{\underline{E}}\text{\@}^\text{0}_{n'} ({\bf k}') 
- \underline{\underline{E}}\text{\@}^\text{0}_{n} ({\bf k})  \Big) 
\underline{\underline{\rho}}\text{\@}_{n' {\bf k}' n {\bf k}}(t) \nonumber \\
&& +  \sum_{n'' {\bf k}''}  
\Big(\underline{\underline{H}}\text{\@}^\text{SO}_{n' {\bf k}' n'' {\bf k}''} \underline{\underline{\rho}}\text{\@}_{n'' {\bf k}'' n {\bf k}}(t) 
- \underline{\underline{\rho}}\text{\@}_{n' {\bf k}' n'' {\bf k}''}(t) 
\underline{\underline{H}}\text{\@}^\text{SO}_{n'' {\bf k}'' n {\bf k}} \Big)  \nonumber \\
&& + \sum_{n'' {\bf k}''}  
\Big(\underline{\underline{H}}\text{\@}^\text{SC}_{n' {\bf k}' n'' {\bf k}''}  \underline{\underline{\rho}}\text{\@}_{n'' {\bf k}'' n {\bf k}}(t) 
- \underline{\underline{\rho}}\text{\@}_{n' {\bf k}' n'' {\bf k}''}(t) 
\underline{\underline{H}}\text{\@}^\text{SC}_{n'' {\bf k}'' n {\bf k}} \Big),  \nonumber \\
\label{dmeh0matrix} 
\end{eqnarray}
where $\left[\underline{\underline{E}}\text{\@}^\text{0}_{n} ({\bf k})\right]_{\sigma'' \sigma}$ $\equiv$ 
$E^\text{0}_{n \sigma''} ({\bf k}) \delta_{\sigma'' \sigma}$ are the matrix elements of $\underline{\underline{E}}\text{\@}^\text{0}_{n} ({\bf k})$, and $\underline{\underline{\rho}}\text{\@}
_{n' {\bf k}' n {\bf k}}(t)$ is a $2 \times 2$ matrix with respect to the indices $\sigma'$ and $\sigma$.
It is convenient to separate the diagonal terms with respect to the indices $\{n, {\bf k}\}$ and $\{n', {\bf k}'\}$ from the off-diagonal terms in the summations of Eq.~(\ref{dmeh0matrix}) and rewrite it as
\begin{eqnarray}
&& i \hbar \frac{\partial}{ \partial t} 
\underline{\underline{\rho}}\text{\@}_{n' {\bf k}' n {\bf k}}(t)
= \Big(\underline{\underline{\varepsilon}}\text{\@}_{n'} ({\bf k}') 
- \underline{\underline{\varepsilon}}\text{\@}_{n} ({\bf k})  \Big) 
\underline{\underline{\rho}}\text{\@}_{n' {\bf k}' n {\bf k}}(t) \nonumber \\
&& +  \Big(\underline{\underline{H}}\text{\@}^\text{SO}_{n' {\bf k}' n' {\bf k}'} \underline{\underline{\rho}}\text{\@}_{n' {\bf k}' n {\bf k}}(t) 
- \underline{\underline{\rho}}\text{\@}_{n' {\bf k}' n {\bf k}}(t) 
\underline{\underline{H}}\text{\@}^\text{SO}_{n {\bf k} n {\bf k}} \Big) \nonumber \\
&& +  \Big(\underline{\underline{H}}\text{\@}^\text{SO-SC}_{n' {\bf k}' n {\bf k}} \underline{\underline{\rho}}\text{\@}_{n {\bf k} n {\bf k}}(t) 
- \underline{\underline{\rho}}\text{\@}_{n' {\bf k}' n' {\bf k}'}(t) 
\underline{\underline{H}}\text{\@}^\text{SO-SC}_{n' {\bf k}' n {\bf k}} \Big) \nonumber \\
&& +  \sideset{}{'} \sum_{n'' {\bf k}''}  
\Big(\underline{\underline{H}}\text{\@}^\text{SO-SC}_{n' {\bf k}' n'' {\bf k}''}  \underline{\underline{\rho}}\text{\@}_{n'' {\bf k}'' n {\bf k}}(t) 
- \underline{\underline{\rho}}\text{\@}_{n' {\bf k}' n'' {\bf k}''}(t) 
\underline{\underline{H}}\text{\@}^\text{SO-SC}_{n'' {\bf k}'' n {\bf k}} \Big), \nonumber \\
\label{dmeh0matrix2} 
\end{eqnarray}
where $\underline{\underline{\varepsilon}}\text{\@}_{n} ({\bf k})$ $\equiv$ $\underline{\underline{E}}\text{\@}^\text{0}_{n} ({\bf k})$ 
$+$ $\underline{\underline{H}}\text{\@}^\text{SC}_{n {\bf k} n {\bf k}}$, $\underline{\underline{H}}\text{\@}^\text{SO-SC}_{n' {\bf k}' n {\bf k}}$ $\equiv$ $\underline{\underline{H}}\text{\@}^\text{SO}_{n' {\bf k}' n {\bf k}}$ $+$ $\underline{\underline{H}}\text{\@}^\text{SC}_{n' {\bf k}' n {\bf k}}$, and the prime on the summation denotes the sum is over $\{n'', {\bf k''}\}$ $\neq$ $\{n, {\bf k}\}$, $\{n', {\bf k'}\}$. Note that $\underline{\underline{H}}\text{\@}^\text{SC}_{n {\bf k} n {\bf k}}$ commutes with $\underline{\underline{\rho}}\text{\@}_{n' {\bf k}' n {\bf k}}(t)$ because $\underline{\underline{H}}\text{\@}^\text{SC}_{n {\bf k} n {\bf k}}$ is spin independent and hence it is also diagonal with respect to the indices $\sigma'$ and $\sigma$. In addition, the matrix $\underline{\underline{\varepsilon}}\text{\@}_{n} ({\bf k})$ has identical diagonal elements in the $\arrowvert n, \sigma, {\bf k} \rangle$ representation (i.e., the spin-up and spin-down eigenstates are degenerate), therefore the first term on the right-hand side of Eq.~(\ref{dmeh0matrix2}) can be rewritten as $(\varepsilon\text{\@}_{n' \sigma'} ({\bf k}')$ 
$-$ $\varepsilon\text{\@}_{n \sigma} ({\bf k}) )$ $\underline{\underline{\rho}}\text{\@}_{n' {\bf k}' n {\bf k}}(t)$.

Because the scattering potential $\hat{H}^\text{SC}$ is weak, the diagonal elements of the density matrix are larger than the off-diagonal elements by a factor of the scattering strength (see Ref.~\onlinecite{kohn57} for detailed discussion). Further, the spin-orbit scattering $\hat{H}^\text{SO}$ is usually much weaker than the momentum scattering $\hat{H}^\text{SC}$ in zincblende semiconductors; therefore, we may neglect the second and the fourth terms on the right-hand side of Eq.~(\ref{dmeh0matrix2}) and obtain the solution for the off-diagonal elements for the spin matrix $\underline{\underline{\rho}}\text{\@}_{n' {\bf k}' n {\bf k}}(t)$:
\begin{eqnarray}
&& \left[ \underline{\underline{\rho}}\text{\@}_{n' {\bf k}' n {\bf k}}(t) \right]
_{\{n', {\bf k}'\} \neq \{n, {\bf k}\}} \nonumber \\
&& =  \frac{-i}{\hbar} \int^{t}_{t_\text{o}} dt'
\Big(\underline{\underline{H}}\text{\@}^\text{SO-SC}_{n' {\bf k}' n {\bf k}} \underline{\underline{\rho}}\text{\@}_{n {\bf k} n {\bf k}}(t') 
- \underline{\underline{\rho}}\text{\@}_{n' {\bf k}' n' {\bf k}'}(t') 
\underline{\underline{H}}\text{\@}^\text{SO-SC}_{n' {\bf k}' n {\bf k}} \Big) \nonumber \\
&& \times \exp \left[ \frac{- i}{\hbar}
\Big( \varepsilon\text{\@}_{n' \sigma'} ({\bf k}') 
- \varepsilon\text{\@}_{n \sigma} ({\bf k}) \Big) \Big( t - t' \Big) \right],
\label{dmeOffSol} 
\end{eqnarray}
where we have assumed that the initial condition at $t$ $=$ $t_\text{o}$ is given by
\begin{equation}
\left[ \underline{\underline{\rho}}\text{\@}_{n' {\bf k}' n {\bf k}}(t_\text{o} = - \infty) \right]
_{\{n', {\bf k}'\} \neq \{n, {\bf k}\}}  = 0.
\label{init}
\end{equation}
To study the electron spin decoherence we need to evaluate the spin-dependent distribution function for the electrons, which is related to the diagonal elements of the spin density matrix Eq.~(\ref{dmeh0matrix2}) with respect to indices $n$ and ${\bf k}$. Inserting Eq.~(\ref{dmeOffSol}) into the right-hand side of Eq.~(\ref{dmeh0matrix2}) for the diagonal elements we obtain
\begin{eqnarray}
&& \frac{\partial}{ \partial t} \underline{\underline{\rho}}\text{\@}_{n {\bf k} n {\bf k}}(t)  
=  \frac{- i}{\hbar} \Big[\underline{\underline{H}}\text{\@}^\text{SO}_{n {\bf k} n {\bf k}}, 
\underline{\underline{\rho}}\text{\@}_{n {\bf k} n {\bf k}}(t) \Big] 
+ \frac{-1}{\hbar^2} \sideset{}{'} \sum_{n' {\bf k}'}  \int^{t}_{t_\text{o}} dt'  \nonumber \\
&& \times \Bigg\{ \underline{\underline{H}}\text{\@}^\text{SO-SC}_{n {\bf k} n' {\bf k}'} 
\Big(\underline{\underline{H}}\text{\@}^\text{SO-SC}_{n' {\bf k}' n {\bf k}} \underline{\underline{\rho}}\text{\@}_{n {\bf k} n {\bf k}}(t') 
- \underline{\underline{\rho}}\text{\@}_{n' {\bf k}' n' {\bf k}'}(t') 
\underline{\underline{H}}\text{\@}^\text{SO-SC}_{n' {\bf k}' n {\bf k}} \Big) \nonumber \\
&& \times  \exp \left[ \frac{- i}{\hbar}
\Big(\varepsilon\text{\@}_{n' \sigma'} ({\bf k}') 
-\varepsilon\text{\@}_{n \sigma} ({\bf k}) \Big) \Big( t - t' \Big) \right]  \nonumber \\
&& + \Big(\underline{\underline{H}}\text{\@}^\text{SO-SC}_{n {\bf k} n' {\bf k}'} \underline{\underline{\rho}}\text{\@}_{n' {\bf k}' n' {\bf k}'}(t') 
- \underline{\underline{\rho}}\text{\@}_{n {\bf k} n {\bf k}}(t') 
\underline{\underline{H}}\text{\@}^\text{SO-SC}_{n {\bf k} n' {\bf k}'} \Big) 
\underline{\underline{H}}\text{\@}^\text{SO-SC}_{n' {\bf k}' n {\bf k}} \nonumber \\
&& \times  \exp \left[ \frac{i}{\hbar}
\Big(\varepsilon\text{\@}_{n' \sigma'} ({\bf k}') 
- \varepsilon\text{\@}_{n \sigma} ({\bf k}) \Big) \Big( t - t' \Big) \right] \Bigg\}.
\label{distribution} 
\end{eqnarray}
Taking the limit ${\hbar \rightarrow 0}$, we obtain the kinetic equation for the diagonal spin density matrix
\begin{eqnarray}
&& \frac{\partial}{ \partial t} \underline{\underline{\rho}}\text{\@}_{n {\bf k}}(t) 
=  \frac{- i}{\hbar} \Big[ \underline{\underline{H}}\text{\@}^\text{SO}_{n {\bf k}}, 
\underline{\underline{\rho}}\text{\@}_{n {\bf k}}(t) \Big] \nonumber \\
&& +  \frac{2 \pi}{\hbar}  \sideset{}{'} \sum_{n' {\bf k}'}  
 \Big[ \underline{\underline{H}}\text{\@}^\text{SO-SC}_{n {\bf k} n' {\bf k}'} 
\underline{\underline{\rho}}\text{\@}_{n' {\bf k}'}(t)
\underline{\underline{H}}\text{\@}^\text{SO-SC}_{n' {\bf k}' n {\bf k}} \nonumber \\
&& -  \frac{1}{2} \left(\underline{\underline{H}}\text{\@}^\text{SO-SC}_{n {\bf k} n' {\bf k}'}
\underline{\underline{H}}\text{\@}^\text{SO-SC}_{n' {\bf k}' n {\bf k}}
\underline{\underline{\rho}}\text{\@}_{n {\bf k}}(t)
+  \underline{\underline{\rho}}\text{\@}_{n {\bf k}}(t)
\underline{\underline{H}}\text{\@}^\text{SO-SC}_{n {\bf k} n' {\bf k}'}
\underline{\underline{H}}\text{\@}^\text{SO-SC}_{n' {\bf k}' n {\bf k}} \right) \Big] \nonumber \\
&& \times \delta \Big(\varepsilon\text{\@}_{n' \sigma'} ({\bf k}') 
- \varepsilon\text{\@}_{n \sigma} ({\bf k}) \Big),
\label{kinetic} 
\end{eqnarray}
where $\underline{\underline{\rho}}\text{\@}_{n {\bf k}}(t)$ $\equiv$ $\underline{\underline{\rho}}\text{\@}_{n {\bf k} n {\bf k}}(t)$. In deriving Eq.~(\ref{kinetic}) we have made use of an identity for the delta function
\begin{equation}
\theta (- t) \lim_{\hbar \rightarrow 0} \frac{1}{\hbar} 
\exp \left[ \frac{-i}{\hbar} (E + i \epsilon)t \right] 
= \pi \delta(t) \delta(E) + i {\cal P} \left[ \frac{1}{E} \right],
\label{deltaf} 
\end{equation}
where $\theta(t)$ and $\delta(t)$ are the usual step and delta funcitons respectively, and ${\cal P} [1/E]$ denotes the principal value of $1/E$.

Multiplying both sides of Eq.~(\ref{kinetic}) by the Pauli spin matrices $\underline{\underline{\mbox{\boldmath$\sigma$}}}$ and taking the trace with respect to the spin indices $\sigma$ $\in$ $\{\uparrow, \downarrow\}$, we obtain the momentum-dependent spin-polarization vector
\begin{eqnarray}
&& \frac{\partial}{ \partial t} {\bf s}\text{\@}_{n {\bf k}}(t) 
= \frac{- i}{\hbar} \text{Tr}\bigg[ \Big[ \underline{\underline{H}}\text{\@}^\text{SO}_{n {\bf k}}, 
\underline{\underline{\rho}}\text{\@}_{n {\bf k}}(t) \Big]
\hat{ \underline{\underline{\mbox{\boldmath$\sigma$}}}}  \bigg] \nonumber \\
&& + \frac{2 \pi}{\hbar}  \sideset{}{'} \sum_{n' {\bf k}'}  
\text{Tr}\bigg[ \Big[ \underline{\underline{H}}\text{\@}^\text{SO-SC}_{n {\bf k} n' {\bf k}'} 
\underline{\underline{\rho}}\text{\@}_{n' {\bf k}'}(t)
\underline{\underline{H}}\text{\@}^\text{SO-SC}_{n' {\bf k}' n {\bf k}}  \nonumber \\
&& - \frac{1}{2} \left(\underline{\underline{H}}\text{\@}^\text{SO-SC}_{n {\bf k} n' {\bf k}'}
\underline{\underline{H}}\text{\@}^\text{SO-SC}_{n' {\bf k}' n {\bf k}}
\underline{\underline{\rho}}\text{\@}_{n {\bf k}}(t) 
+ \underline{\underline{\rho}}\text{\@}_{n {\bf k}}(t)
\underline{\underline{H}}\text{\@}^\text{SO-SC}_{n {\bf k} n' {\bf k}'}
\underline{\underline{H}}\text{\@}^\text{SO-SC}_{n' {\bf k}' n {\bf k}} \right) \Big]  
\underline{\underline{\mbox{\boldmath$\sigma$}}}  \bigg] \nonumber \\
&& \times \delta \Big(\varepsilon\text{\@}_{n' \sigma'} ({\bf k}') 
- \varepsilon\text{\@}_{n \sigma} ({\bf k}) \Big),
\label{spve} 
\end{eqnarray}
where Tr stands for trace and the momentum-dependent spin-polarization vector is defined as $ {\bf s}\text{\@}_{n {\bf k}}(t)$ $\equiv$  $\text{Tr}[\underline{\underline{\rho}}\text{\@}_{n {\bf k}}(t) \underline{\underline{\mbox{\boldmath$\sigma$}}}]$. This general kinetic equation is now further approximated depending on the relevant situation.
\subsection{SPIN-INDEPENDENT SCATTERING}
\label{sec:sis}
Electron spin relaxation in zincblende-type semiconductors near room temperature is dominated by the  D'yakonov-Perel' (DP) precessional mechanism~\cite{dyakonov71, dyakonov72, meier84}, which is a direct result of the spin splitting of the conduction band that occurs at zero external magnetic field due to spin-orbit interaction in inversion asymmetric crystals. This can be viewed as having an effective momentum-dependent magnetic field ${\bf H}_{n}({\bf k})$ acting on the electron spins, which causes Zeeman splitting between the spin states $E_{n \uparrow} ({\bf k})$ and $E_{n \downarrow} ({\bf k})$. Consequently, non-equilibrium electron spins at finite momentum ${\bf k}$ will precess about these effective momentum-dependent magnetic fields at the Larmor frequencies proportional to the spin splittings. The presence of scattering causes random fluctuation of spin precession frequencies and directions leading to decay of the non-equilibrium electron spin population.

To study the electron spin dynamics in zincblende-type semiconductors, we shall consider only the dominant scattering mechanism (spin-independent scattering) in Eq.~(\ref{kinetic}) pertaining to the DP spin relaxation mechanism. For spin-independent scattering (i.e., the off-diagonal elements of $\underline{\underline{H}}\text{\@}^\text{SO}_{n' {\bf k}' n {\bf k}}$ vanish), equation~(\ref{spve}) reduces to
\begin{eqnarray}
&& \frac{\partial}{ \partial t} {\bf s}\text{\@}_{n {\bf k}}(t)
= \frac{- i}{\hbar} \text{Tr}\bigg[ \Big[\underline{\underline{H}}\text{\@}^\text{SO}_{n {\bf k}}, 
\underline{\underline{\rho}}\text{\@}_{n {\bf k}}(t) \Big] 
\underline{\underline{\mbox{\boldmath$\sigma$}}} \bigg] \nonumber \\
&& - \sideset{}{} \sum_{n' {\bf k}'} \text{Tr}\Big[  
\underline{\underline{W}}\text{\@}^{(\delta)}_{n n'} ({\bf k}, {\bf k}')
\left(\underline{\underline{\rho}}\text{\@}_{n {\bf k}}(t) 
-\underline{\underline{\rho}}\text{\@}_{n' {\bf k}'}(t) \right) 
\underline{\underline{\mbox{\boldmath$\sigma$}}}\Big],
\label{kineticSI} 
\end{eqnarray}
where $\underline{\underline{W}}\text{\@}^{(\delta)}_{n n'} ({\bf k}, {\bf k}')$ is the transition probability of electron scattering and is given by
\begin{equation}
\underline{\underline{W}}\text{\@}^{(\delta)}_{n n'} ({\bf k}, {\bf k}')
\equiv
\underline{\underline{W}}\text{\@}_{n n'} ({\bf k}, {\bf k}')
\delta \Big(\varepsilon\text{\@}_{n' \sigma'} ({\bf k}') 
- \varepsilon\text{\@}_{n \sigma} ({\bf k}) \Big),
\label{estp} 
\end{equation}
and $\underline{\underline{W}}\text{\@}^{(\delta)}_{n n'} ({\bf k}, {\bf k}')$ $\equiv$ $\underline{\underline{H}}\text{\@}^\text{SC}_{n {\bf k} n' {\bf k}'}
\underline{\underline{H}}\text{\@}^\text{SC}_{n' {\bf k}' n {\bf k}}$.
\subsection{KRAMERS DEGENERACY}
\label{sec:kd}
Time reversal invariance of the crystal Hamiltonian $\hat{H}$ requires that a spin state at a given ${\bf k}$ must have the same energy (Kramers degeneracy) as that of the opposite spin at $-{\bf k}$: $E_{n \uparrow} ({\bf k})$ $=$ $E_{n \downarrow} (-{\bf k})$. For spatially inversion-symmetric crystals (diamond crystal structure) such as Si and Ge, the states of the opposite spin at ${\bf k}$ are degenerate $E_{n \uparrow} ({\bf k})$ $=$ $E_{n \downarrow} ({\bf k})$. On the other hand, for spatially inversion asymmetric crystals (zincblende crystal structure) such as GaAs and InAs, $E_{n \uparrow} ({\bf k})$ $\ne$ $E_{n \downarrow} ({\bf k})$. Consequently, there is a nonzero spin splitting [$\Delta E_{n} ({\bf k})$ $=$ $E_{n \uparrow} ({\bf k})$ $-$ $E_{n \downarrow} ({\bf k})$] at finite momentum ${\bf k}$ due to spin-orbit interaction. The momentum-dependent effective spin-splitting matrix Hamiltonian due to any spin-orbit coupling can be written as
\begin{equation}
\underline{\underline{H}}\text{\@}^\text{SO}_{n {\bf k}}
= \frac{\hbar}{2} {\bf \Omega} (n, {\bf k}) \cdot \underline{\underline{\mbox{\boldmath$\sigma$}}}, 
\label{bssh} 
\end{equation}
where the momentum-dependent vector ${\bf \Omega} (n, {\bf k})$ is interpreted as the precession vector whose length equals to the spin precession frequency (Larmor frequency) and direction defines the spin precession axis. One consequence of Kramers degeneracy is that the precession vector ${\bf \Omega} (n, {\bf k})$ is an odd function of ${\bf k}$:
\begin{equation}
{\bf \Omega} (n, {\bf k}) = -{\bf \Omega} (n, -{\bf k}).
\label{bpv} 
\end{equation}
It is obvious from Eqs.~(\ref{bssh}) and (\ref{bpv}) that the spin-orbit matrix Hamiltonian $\underline{\underline{H}}\text{\@}^\text{SO}_{n {\bf k}}$ is also an odd function of ${\bf k}$. 
\subsection{ELECTRON SPIN DENSITY}
\label{sec:esde}
In zincblende-type semiconductors, the spin coherence time for thermalized electrons is typically much longer than the momentum relaxation time; therefore, it usually satisfies the condition
\begin{equation}
\tau(n, {\bf k}) \Omega(n, {\bf k}) \ll 1,
\label{condsp} 
\end{equation}
where $\tau (n, {\bf k})$ is the momentum scattering time (see Sec.~\ref{sec:sdbzc}). This allows us to solve Eq.~(\ref{kineticSI}) by making successive approximations to the spin density matrix $\underline{\underline{\rho}}\text{\@}_{n {\bf k}}(t)$ in the small parameter $\lambda (n, {\bf k})$ $\equiv$ $\tau (n, {\bf k}) \Omega(n, {\bf k})$. Expanding the spin density matrix $\underline{\underline{\rho}}\text{\@}_{n {\bf k}}(t)$ in powers of  $\lambda (n, {\bf k})$, the power series of the density matrix may be written in the usual form as
\begin{equation}
\underline{\underline{\rho}}\text{\@}_{n {\bf k}}(t) 
= \underline{\underline{\rho}}\text{\@}^{(0)}_{n {\bf k}}(t) 
+ \underline{\underline{\rho}}\text{\@}^{(1)}_{n {\bf k}}(t)
+ \underline{\underline{\rho}}\text{\@}^{(2)}_{n {\bf k}}(t)
+ \cdots
+ \underline{\underline{\rho}}\text{\@}^{(m)}_{n {\bf k}}(t)
+ \cdots. 
\label{seriesDM} 
\end{equation}
Note that $\lambda (n, {\bf k})$ $\propto$ $H^\text{SO}_{n {\bf k}}$ [see Eq.~(\ref{bssh})]. It  was shown by D'yakonov and Perel'~\cite{dyakonov71, dyakonov72} (see also Refs.~\onlinecite{averkiev99} and \onlinecite{pikus84}) that it is sufficient to expand the spin density matrix to first order in $\lambda (n, {\bf k})$ (i.e., $m$ $=$ $1$). After substituting Eq.~(\ref{seriesDM}) in (\ref{kineticSI}), we have 
\begin{eqnarray}
&& \frac{\partial}{ \partial t} {\bf s}\text{\@}_{n {\bf k}}(t) \nonumber \\
&& =  \frac{- i}{\hbar} \text{Tr} \bigg[ \Big[\underline{\underline{H}}\text{\@}^\text{SO}_{n {\bf k}}, 
\underline{\underline{\rho}}\text{\@}^{(0)}_{n {\bf k}}(t) \Big]
\underline{\underline{\mbox{\boldmath$\sigma$}}} \bigg]
+ \frac{- i}{\hbar} \text{Tr} \bigg[ \Big[\underline{\underline{H}}\text{\@}^\text{SO}_{n {\bf k}}, 
\underline{\underline{\rho}}\text{\@}^{(1)}_{n {\bf k}}(t) \Big] 
\underline{\underline{\mbox{\boldmath$\sigma$}}} \bigg] \nonumber \\
&& - \sideset{}{} \sum_{n'} \int \frac{d{\bf k}'}{(2 \pi)^3}
\text{Tr} \bigg[ \underline{\underline{W}}\text{\@}^{(\delta)}_{n n'} ({\bf k}, {\bf k}')
\left(\underline{\underline{\rho}}\text{\@}^{(0)}_{n {\bf k}}(t) 
-\underline{\underline{\rho}}\text{\@}^{(0)}_{n' {\bf k}'}(t) \right)
\underline{\underline{\mbox{\boldmath$\sigma$}}} \bigg] \nonumber \\
&& - \sideset{}{} \sum_{n'} \int \frac{d{\bf k}'}{(2 \pi)^3}
\text{Tr} \bigg[ \underline{\underline{W}}\text{\@}^{(\delta)}_{n n'} ({\bf k}, {\bf k}')
\left(\underline{\underline{\rho}}\text{\@}^{(1)}_{n {\bf k}}(t) 
-\underline{\underline{\rho}}\text{\@}^{(1)}_{n' {\bf k}'}(t) \right) 
\underline{\underline{\mbox{\boldmath$\sigma$}}} \bigg], \nonumber \\
\label{kineticSIpower}
\end{eqnarray}
where the zeroth order term $\underline{\underline{\rho}}\text{\@}^{(0)}_{n {\bf k}}(t)$ depends only on energy $\varepsilon\text{\@}_{n \sigma} ({\bf k})$, while the first order term $\underline{\underline{\rho}}\text{\@}^{(1)}_{n {\bf k}}(t)$ $\propto$ $\underline{\underline{H}}\text{\@}^\text{SO}_{n {\bf k}}$. In Eq.~(\ref{kineticSIpower}) we have converted the summation over ${\bf k}'$ into an integral  $\sum_{\bf k}$ $\rightarrow$ $(2 \pi)^{-3}\int d {\bf k}$. In the first order approximation, the spin density matrix can be written as 
\begin{subequations}
\label{dmSI}
\begin{equation}
\underline{\underline{\rho}}\text{\@}_{n {\bf k}}(t)
 = \langle \underline{\underline{\rho}}\text{\@}_{n {\bf k}}(t) \rangle_{\bf \hat{k}} 
+ \underline{\underline{\rho}}\text{\@}^{(1)}_{n {\bf k}}(t),
\label{dmSIa}
\end{equation}
\begin{equation}
\langle \underline{\underline{\rho}}\text{\@}^{(1)}_{n {\bf k}}(t) \rangle_{\bf \hat{k}}  = 0,
\label{dmSIb}
\end{equation}
\end{subequations}
where $\langle \underline{\underline{\rho}}\text{\@}_{n {\bf k}}(t) \rangle_{\bf \hat{k}}$ denotes the averaging of $\underline{\underline{\rho}}\text{\@}_{n {\bf k}}(t)$ over all directions of the crystal momentum ${\bf k}$. Equation~(\ref{dmSIb}) is zero because $\underline{\underline{\rho}}\text{\@}^{(1)}_{n {\bf k}}(t)$ is  an odd function of ${\bf k}$ [see Eqs.~(\ref{bssh}) and (\ref{bpv})]. Note that since $\underline{\underline{\rho}}\text{\@}^{(0)}_{n {\bf k}}(t)$ depends only on energy, it is independent of the direction of ${\bf k}$ (i.e., $\underline{\underline{\rho}}\text{\@}^{(0)}_{n {\bf k}}(t)$ $=$ $\langle \underline{\underline{\rho}}\text{\@}^{(0)}_{n {\bf k}}(t) \rangle_{\bf \hat{k}}$ $=$ $\langle \underline{\underline{\rho}}\text{\@}_{n {\bf k}}(t) \rangle_{\bf \hat{k}}$). Integrating Eq.~(\ref{kineticSIpower}) over the momentum variable ${\bf k}$ and summing over $n$, we obtain the kinetic equation for the electron spin density ${\bf S}(t)$:
\begin{equation}
\frac{\partial}{ \partial t} {\bf S}(t)
= \frac{-i}{\hbar}  \text{Tr} \bigg[  \sideset{}{} \sum_{n} \int \frac{d{\bf k}}{(2 \pi)^3}
\Big[\underline{\underline{H}}\text{\@}^\text{SO}_{n {\bf k}}, 
\underline{\underline{\rho}}\text{\@}^{(1)}_{n {\bf k}}(t) \Big] 
\underline{\underline{\mbox{\boldmath$\sigma$}}} \bigg],
\label{kineticSD}
\end{equation}
where the electron spin density is defined as
\begin{eqnarray}
{\bf S}(t) 
& \equiv &
\sideset{}{} \sum_{n} \int \frac{d{\bf k}}{(2 \pi)^3} {\bf s}\text{\@}_{n {\bf k}}(t) \nonumber \\
& = & \text{Tr} \bigg[  \sideset{}{} \sum_{n} \int \frac{d{\bf k}}{(2 \pi)^3}
\left( \underline{\underline{\rho}}\text{\@}^{(0)}_{n {\bf k}}(t) 
+ \underline{\underline{\rho}}\text{\@}^{(1)}_{n {\bf k}}(t) \right)
\underline{\underline{\mbox{\boldmath$\sigma$}}} \bigg] \nonumber \\
& = &  \text{Tr} \bigg[  \sideset{}{} \sum_{n} \int \frac{d{\bf k}}{(2 \pi)^3}
\underline{\underline{\rho}}\text{\@}^{(0)}_{n {\bf k}}(t)
\underline{\underline{\mbox{\boldmath$\sigma$}}} \bigg].
\label{defSD}
\end{eqnarray}
Equation~(\ref{kineticSD}) is obtained by noting (i) the averaging of the first term on the right-hand side of Eq.~(\ref{kineticSIpower}) over momentum ${\bf k}$ vanishes because the spin-splitting matrix Hamiltonian $\underline{\underline{H}}\text{\@}^\text{SO}_{n {\bf k}}$ is an odd function of ${\bf k}$ [see Eqs.~(\ref{bssh}) and (\ref{bpv})], and (ii) the last two terms on the right-hand of (\ref{kineticSIpower}) also disappear, for they describe the scattering in and out of a momentum state, hence the average of the total rate of change of occupation of a momentum state due to scattering is identically zero.

To solve the kinetic equation for the electron spin density ${\bf S}(t)$, we first need to evaluate the density matrix $\underline{\underline{\rho}}\text{\@}^{(1)}_{n {\bf k}}(t)$ in Eq.~(\ref{kineticSD}). In the first order approximation in $\lambda(n, {\bf k})$, the density matrix $\underline{\underline{\rho}}\text{\@}^{(1)}_{n {\bf k}}(t)$ is determined by the following equation:
\begin{eqnarray}
&& \frac{\partial}{ \partial t} \underline{\underline{\rho}}\text{\@}^{(1)}_{n {\bf k}}(t)
=  \frac{- i}{\hbar} \Big[\underline{\underline{H}}\text{\@}^\text{SO}_{n {\bf k}}, 
\underline{\underline{\rho}}\text{\@}^{(0)}_{n {\bf k}}(t) \Big] \nonumber \\
&&- \sideset{}{} \sum_{n'}  \int \frac{d{\bf k}'}{(2 \pi)^3}
\underline{\underline{W}}\text{\@}^{(\delta)}_{n n'} ({\bf k}, {\bf k}')
\left(\underline{\underline{\rho}}\text{\@}^{(1)}_{n {\bf k}}(t) 
-\underline{\underline{\rho}}\text{\@}^{(1)}_{n' {\bf k}'}(t) \right), \nonumber \\
\label{rho1a}
\end{eqnarray}
which is obtained from Eq.~(\ref{kineticSIpower}). From Eqs.~(\ref{kineticSD})-(\ref{rho1a}) it can be seen that the relaxation rate of the spin density matrix $\underline{\underline{\rho}}\text{\@}^{(0)}_{n {\bf k}}(t)$ is governed by the spin-splitting matrix Hamiltonian $\underline{\underline{H}}\text{\@}^\text{SO}_{n {\bf k}}$, while the relaxation rate of $\underline{\underline{\rho}}\text{\@}^{(1)}_{n {\bf k}}(t)$ is dominated by the momentum scattering rate $\underline{\underline{W}}\text{\@}^{(\delta)}_{n n'} ({\bf k}, {\bf k}')$ under the condition (\ref{condsp}) [i.e., $\tau(n,{\bf k})$ $\ll$ $\Omega^{-1}(n,{\bf k})$]. Therefore, on the time scale of electron spin relaxation, which is of order $\Omega^{-1}(n,{\bf k})$, Eq.~(\ref{rho1a}) reduces to \cite{averkiev99}
\begin{eqnarray}
&& \frac{- i}{\hbar} \Big[\underline{\underline{H}}\text{\@}^\text{SO}_{n {\bf k}}, 
\underline{\underline{\rho}}\text{\@}^{(0)}_{n {\bf k}}(t) \Big] \nonumber \\
&& =  \sideset{}{} \sum_{n'} \int \frac{d{\bf k}'}{(2 \pi)^3}
\underline{\underline{W}}\text{\@}^{(\delta)}_{n n'} ({\bf k}, {\bf k}')
\left(\underline{\underline{\rho}}\text{\@}^{(1)}_{n {\bf k}}(t) 
-\underline{\underline{\rho}}\text{\@}^{(1)}_{n' {\bf k}'}(t) \right). \nonumber \\
\label{rho1b}
\end{eqnarray}
\subsection{QUASI-ELASTIC SCATTERING}
\label{sec:qelsm}
To make the problem more tractable and to express the solution to the spin density matrix equations [Eqs.~(\ref{kineticSD}) and (\ref{rho1b})] in a physically revealing way, the collision integral on the right-hand side of Eq.~(\ref{rho1b}) is further approximated by assuming the momentum scattering is quasi elastic~\cite{gantmakher87}. In addition we assume that the energy spectrum is isotropic (i.e., a spherical Fermi surface, $\varepsilon_{n \sigma} (\bf k)$ $=$ $\varepsilon_{n \sigma} (|\bf k|)$), and this implies that $\left| {\bf k'} \right| = \left| {\bf k} \right|$ as demanded by energy conservation. Under these assumptions, the momentum scattering rate $\underline{\underline{W}}\text{\@}_{n n'} ({\bf k}, {\bf k}')$ appearing in Eq.~(\ref{estp}) depends only on the magnitude of ${\bf k}$ and the scattering angle $\xi_{{\bf k} {\bf k}'}$ between ${\bf k}$ and ${\bf k}'$. Thus the momentum scattering rate can be written as 
\begin{equation}
\underline{\underline{W}}\text{\@}_{n n'} ({\bf k}, {\bf k}') 
= \underline{\underline{W}}\text{\@}_{n n'} ({\bf \hat{k} \cdot \hat{k}}', k),
\label{mspm} 
\end{equation}
where $\cos \xi_{{\bf k} {\bf k}'}$ $\equiv$ ${\bf \hat{k} \cdot \hat{k}}'$. All spin coherence times will be calculated in this paper assuming quasi-elastic scattering. Using Eq.~(\ref{mspm}) in (\ref{rho1b}), the resulting simplified expression is given by
\begin{eqnarray}
&& \frac{- i}{\hbar} \Big[\underline{\underline{H}}\text{\@}^\text{SO}_{n {\bf k}}, 
\underline{\underline{\rho}}\text{\@}^{(0)}_{n {\bf k}}(t) \Big] \nonumber \\
&& = \sideset{}{} \sum_{n'}   \int \frac{d {\bf k}'}{(2 \pi)^3}
\underline{\underline{W}}\text{\@}^{(\delta)}_{n n'} ({\bf \hat{k} \cdot \hat{k}}', k)
\left(\underline{\underline{\rho}}\text{\@}^{(1)}_{n {\bf k}}(t) 
-\underline{\underline{\rho}}\text{\@}^{(1)}_{n' {\bf k}'}(t) \right). \nonumber \\
\label{rho1c}
\end{eqnarray}
The coupled equations (\ref{kineticSD}) and (\ref{rho1c}) are quite general and they can be applied to both three-dimensional (bulk semiconductors) and two-dimensional (quantum wells and superlattices) systems. We will treat these two cases separately in the subsequent sections.
\subsection{SPIN DECOHERENCE IN BULK ZINCBLENDE CRYSTALS}
\label{sec:sdbzc}
In the case of bulk semiconductors, the coupled Eqs.~(\ref{kineticSD}) and (\ref{rho1c}) are explicitly written as
\begin{eqnarray}
\frac{\partial}{ \partial t} {\bf S}(t) 
&=& \text{Tr} \bigg[  \sideset{}{} \sum_{n} \int \frac{d k k^2}{2 \pi^2} 
\frac{\partial}{ \partial t}
\langle \underline{\underline{\rho}}\text{\@}_{n {\bf k}}(t) \rangle_{\bf \hat{k}}
\underline{\underline{\mbox{\boldmath$\sigma$}}} \bigg] \nonumber \\
& =& \frac{-i}{\hbar}  \text{Tr} \bigg[  \sideset{}{} \sum_{n} \int \frac{d k k^2}{2 \pi^2}
\int \frac{d O_{\hat{\bf k}}}{4 \pi}
\Big[\underline{\underline{H}}\text{\@}^\text{SO}_{n {\bf k}}, 
\underline{\underline{\rho}}\text{\@}^{(1)}_{n {\bf k}}(t) \Big] 
\underline{\underline{\mbox{\boldmath$\sigma$}}} \bigg], \nonumber \\
\label{kineticSD3D}
\end{eqnarray}
\begin{eqnarray}
&& \frac{- i}{\hbar} \Big[\underline{\underline{H}}\text{\@}^\text{SO}_{n {\bf k}}, 
\underline{\underline{\rho}}\text{\@}^{(0)}_{n {\bf k}}(t) \Big] \nonumber \\
&& = \sideset{}{} \sum_{n'} 
\int \frac{d k' k'^2}{2 \pi^2} 
\int \frac{d O_{\bf \hat{k}}}{4 \pi} 
\underline{\underline{W}}\text{\@}^{(\delta)}_{n n'} ({\bf \hat{k} \cdot \hat{k}}', k) \nonumber \\
&& \times \left(\underline{\underline{\rho}}\text{\@}^{(1)}_{n {\bf k}}(t) 
-\underline{\underline{\rho}}\text{\@}^{(1)}_{n' {\bf k}'}(t) \right),
\label{rho13D}
\end{eqnarray}
where $d O_{\hat{\bf k}}$ is the differential solid angle in ${\bf k}$-space and $\langle \underline{\underline{\rho}}\text{\@}_{n {\bf k}}(t) \rangle_{\bf \hat{k}}$ $\equiv$ $(4 \pi)^{-1}$ $\int d O_{\hat{\bf k}} \underline{\underline{\rho}}\text{\@}_{n {\bf k}}(t)$. In order to solve the coupled equations~(\ref{kineticSD3D}) and (\ref{rho13D}), we shall expand
$\underline{\underline{H}}\text{\@}^\text{SO}_{n {\bf k}}$, $\underline{\underline{W}}\text{\@}^{(\delta)}_{n n'} ({\bf \hat{k} \cdot \hat{k}}', k)$, and $\underline{\underline{\rho}}\text{\@}^{(1)}_{n {\bf k}}(t)$ in a series of spherical harmonics \cite{meier84}
\begin{equation}
\underline{\underline{H}}\text{\@}^\text{SO}_{n {\bf k}}
= \sum_{lm} \widetilde{\underline{\underline{H}}}\text{\@}^\text{SO}_{lm}(n,k) Y_{lm}(\theta, \phi),
\label{besh} 
\end{equation}
\begin{equation}
\underline{\underline{\rho}}\text{\@}^{(1)}_{n {\bf k}}(t)
= \sum_{lm} \widetilde{\underline{\underline{\rho}}}\text{\@}^{(1)}_{lm}(n,k, t) Y_{lm}(\theta, \phi),
\label{besd} 
\end{equation}
\begin{eqnarray}
&& \underline{\underline{W}}\text{\@}^{(\delta)}_{n n'} ({\bf \hat{k} \cdot \hat{k}}', k) 
 = \sum_{l} \widetilde{\underline{\underline{W}}}\text{\@}^{(\delta)}_{n n' l}(k)
P_{l}(\cos \xi_{{\bf k} {\bf k}'})  \nonumber \\
&& =  \sum_{lm} \widetilde{\underline{\underline{W}}}\text{\@}^{(\delta)}_{n n' l}(k) 
\frac{4 \pi}{2l + 1} Y^{\ast}_{lm}(\theta', \phi')Y_{lm}(\theta, \phi),
\label{bemsp} 
\end{eqnarray}
where $P_{l}(\xi_{{\bf k} {\bf k}'})$ is the Legendre polynomial and the second equality in Eq.~(\ref{bemsp}) is obtained by using the addition theorem for spherical harmonics \cite{arfken95}.

Substituting Eqs.~(\ref{besh})-(\ref{bemsp}) into Eq.~(\ref{rho13D}), Multiplying both sides of the resulting equation by $Y^{\ast}_{l', m'}(\theta, \phi)$, and integrating over the solid angle $d O_{\hat{\bf k}}$, we have
\begin{equation}
\widetilde{\underline{\underline{\rho}}}\text{\@}^{(1)}_{lm}(n,k, t)
= \frac{-i}{\hbar} \widetilde{\underline{\underline{\tau}}}\text{\@}_l(n,k) 
\Big[ \widetilde{\underline{\underline{H}}}\text{\@}^\text{SO}_{lm}(n,k), 
\langle \underline{\underline{\rho}}\text{\@}_{n {\bf k}}(t) \rangle_{\bf \hat{k}} \Big],
\label{bsdp} 
\end{equation}
where the momentum scattering time matrix $\widetilde{\underline{\underline{\tau}}}\text{\@}_l(n,k)$ is given by
\begin{eqnarray} 
\widetilde{\underline{\underline{\tau}}}\text{\@}^{(-1)}_l(n,k) 
&=& \underline{\underline{{\cal D}}}_{n}(k) \sum_{n'} \int \frac{d O_{\bf \hat{k}'}}{4 \pi} 
\underline{\underline{W}}\text{\@}_{n n'} ({\bf \hat{k} \cdot \hat{k}}', k) \nonumber \\
&\times& \Big[1 - P_l({\bf \hat{k}' \cdot \bf \hat{k}})  \Big]. 
\label{btn} 
\end{eqnarray}
In obtaining Eq.~(\ref{btn}), the delta function $\delta \left(\varepsilon\text{\@}_{n' \sigma'} ({\bf k}') - \varepsilon\text{\@}_{n \sigma} ({\bf k}) \right)$ contained in the definition of the momentum scattering rate $\underline{\underline{W}}\text{\@}^{(\delta)}_{n n'}({\bf k}, {\bf k}')$ in Eq.~(\ref{rho13D}) is eliminated by the integration $d k'$. Here $\underline{\underline{{\cal D}}}_{n}(k)$ $=$ ${\cal D}^{(3D)}_{n \sigma}(k) \underline{\underline{I}}$, where ${\cal D}^{(3D)}_{n \sigma}(k)$ is the three-dimensional density of states of the $n$th band with spin index $\sigma$, and $\underline{\underline{I}}$ is the $2 \times 2$ identity matrix. Note that the density of states ${\cal D}^{(3D)}_{n \sigma}(k)$ is the same for both up spin and down spin due to Kramers degeneracy: ${\cal D}^{(3D)}_{n \uparrow}(k)$ $=$ ${\cal D}^{(3D)}_{n \downarrow}(k)$. As the scattering potential is spin independent, the momentum scattering time matrix is diagonal with respect to the spin indices and can be rewritten as $\widetilde{\underline{\underline{\tau}}}\text{\@}_l(n,k)$ $=$ $\widetilde{\tau}_l(n,k) \underline{\underline{I}}$, where $\widetilde{\tau}_l(n,k)$ is the momentum relaxation time. The $l = 1$ component of the momentum relaxation time $\widetilde{\tau}_1(n,k)$ corresponds to the transport time $\tau_\text{tr}(n,k)$ appearing in the mobility expression [see Eq.~(\ref{bm}) in Sec~\ref{sec:mrtm}]. 

Inserting Eq.~(\ref{bsdp}) into Eq.~(\ref{kineticSD3D}) and performing the integration over the solid angle $dO_{\bf \hat{k}}$ and the sum over all states $n$, the resulting expression is
\begin{eqnarray}
&&\frac{\partial}{\partial t} 
{\bf S}(t)
=  \frac{-1}{(2 \pi)^{3} \hbar^2}  
\text{Tr} \Bigg[ \sum_{l m n} \int dk k^2 (-1)^m \widetilde{\tau}_{l}(n,k)  \nonumber\\
&&\times \bigg[\widetilde{\underline{\underline{H}}}\text{\@}^\text{SO}_{lm}(n,k),
\Big[ \widetilde{\underline{\underline{H}}}\text{\@}^\text{SO}_{l-m}(n,k), 
\langle \underline{\underline{\rho}}\text{\@}_{n {\bf k}}(t) \rangle_{\bf \hat{k}} \Big] \bigg]
\underline{\underline{\mbox{\boldmath$\sigma$}}} \Bigg],
\label{kineticSI3D} 
\end{eqnarray}
where $\widetilde{\underline{\underline{H}}}\text{\@}^\text{SO}_{lm}(n,k)$ is written as
\begin{eqnarray}
\widetilde{\underline{\underline{H}}}\text{\@}^\text{SO}_{lm}(n,k)
& = & \int d O_{\hat{\bf k}} 
\underline{\underline{H}}\text{\@}^\text{SO}_{n {\bf k}} Y_{lm}(\theta, \phi) \nonumber \\
& = & \frac{\hbar}{2} \int d O_{\hat{\bf k}} 
\Big[{\bf \Omega} (n, {\bf k}) \cdot \underline{\underline{\mbox{\boldmath$\sigma$}}} \Big] 
Y_{lm}(\theta, \phi).
\label{bho} 
\end{eqnarray}
To evaluate the trace on the right-hand side of Eq.~(\ref{kineticSI3D}), we shall write the spin density matrix $\langle \underline{\underline{\rho}}\text{\@}_{n {\bf k}}(t) \rangle_{\bf \hat{k}}$ in a general form as  
\begin{equation}
\langle \underline{\underline{\rho}}\text{\@}_{n {\bf k}}(t) \rangle_{\bf \hat{k}}
= \frac{1}{2} \Big[ \text{g}(n, k) \underline{\underline{I}} + h(n, k) {\bf S}(t) \cdot 
\underline{\underline{\mbox{\boldmath$\sigma$}}}  \Big ],
\label{bdmi} 
\end{equation}
where $\text{g}(n, k)$ and $h(n, k)$ are electron distribution functions which are determined by the initial condition at $t = t_\text{o}$ [Eq.~(\ref{init})] and carrier statistics. In general these distribution functions can be written as~\cite{dyakonov72, averkiev99}
\begin{subequations}
\label{bdf} 
\begin{equation}
\text{g}(n, k) = \frac{f^{(+)}_{0}(n, k) + f^{(-)}_{0}(n, k)}
{\displaystyle {\sum_{n} \int \frac{dk k^2}{2 \pi^2} }
\left(f^{(+)}_{0}(n, k) + f^{(-)}_{0}(n, k)\right)},
\label{bdf1} 
\end{equation}
\begin{equation}
 h(n, k) = \frac{f^{(+)}_{0}(n, k) - f^{(-)}_{0}(n, k)}
{\displaystyle {\sum_{n} \int \frac{dk k^2}{2 \pi^2} }
\left(f^{(+)}_{0}(n, k) - f^{(-)}_{0}(n, k)\right)},
\label{bdf2} 
\end{equation}
\end{subequations}
where $f^{(\pm)}_{0}(n, k)$ are the Fermi distribution functions for spin-up and spin-down electrons with the corresponding Fermi energies $E^{(\pm)}_\text{F}$. 

After substituting Eq.~(\ref{bdmi}) into Eq.~(\ref{kineticSI3D}) and taking the trace on the right-hand side, we have the final form of the kinetic equation for the electron spin density, expressed in terms of the precession vector and momentum scattering time:
\begin{eqnarray}
\frac{\partial}{\partial t} 
{\bf S}(t)
 &=&  \frac{-1}{(2 \pi)^3}  
\sum_{l m n} \int dk k^2 (-1)^m \widetilde{\tau}_{l}(n,k)  h(n,k)  \nonumber\\
&\times& \bigg[ \left( \widetilde{\bf \Omega}_{l-m}(n,k) \cdot 
\widetilde{\bf \Omega}_{lm}(n,k) \right) {\bf S}(t) \bigg. \nonumber\\
\bigg. 
& - &\left( \widetilde{\bf \Omega}_{l-m}(n,k) \cdot {\bf S}(t) \right)
\widetilde{\bf \Omega}_{lm} (n,k) \bigg],
\label{kineticSI3Dfinal} 
\end{eqnarray}
where 
\begin{equation}
\widetilde{\bf \Omega}_{lm} (n,k)
= \int d O_{\hat{\bf k}} 
{\bf \Omega} (n, {\bf k}) Y_{lm}(\theta, \phi),
\label{fcbpv} 
\end{equation}
and 
\begin{equation}
{\bf \Omega} (n, {\bf k}) 
=\sum_{lm} \widetilde{\bf \Omega}_{lm} (n,k) Y_{lm}(\theta, \phi).
\label{fsebpv} 
\end{equation}
In spherical coordinates, Eq.~(\ref{bpv}) can be written as ${\bf \Omega} (n, k, \theta, \phi)$ $=$ $-{\bf \Omega} (n, k, \pi - \theta, \pi + \phi)$. Therefore, from Eqs.~(\ref{bpv}) and (\ref{fsebpv}), we have
\begin{eqnarray}
& &\sum_{lm} \widetilde{\bf \Omega}_{lm} (n,k) Y_{lm}(\theta, \phi) \nonumber\\
&=& \sum_{lm} - \widetilde{\bf \Omega}_{lm} (n,k) Y_{lm}(\pi - \theta, \pi + \phi) \nonumber\\
&=& \sum_{lm} \widetilde{\bf \Omega}_{lm} (n,k) Y_{lm}(\theta, \phi) (-1)^{l+1},
\label{fsebpvi} 
\end{eqnarray}
where the last equality is obtained using $Y_{lm}(\pi - \theta, \pi + \phi)$ $=$ $Y_{lm}(\theta, \phi) (-1)^{l}$. Equation~(\ref{fsebpvi}) yields 
\begin{equation}
\widetilde{\bf \Omega}_{lm} (n,k)$ $=$ $\widetilde{\bf \Omega}_{lm} (n,k) (-1)^{l+1}.
\label{bpvc} 
\end{equation}
In order to satisfy this condition, $\widetilde{\bf \Omega}_{lm} (n,k)$ must vanish for even $l$. 

Rewriting Eq.~(\ref{kineticSI3Dfinal}) in tensorial form, we have
\begin{equation}
\frac{\partial}{\partial t} S_{i}(t)
= - \sum_{j} \Gamma_{ij} S_{j}(t),
\label{bspt} 
\end{equation}
where $(i, j)$ $\in$ $\{x, y, z\}$ and the components of the spin relaxation rate tensor $\underline{\underline{\Gamma}}$ are given by
\begin{eqnarray}
\Gamma_{ij} 
&=& \frac{1}{(2 \pi)^3} \sum_{l m n} \int dk k^2 \widetilde{\tau}_l (n,k) h(n ,k) \nonumber \\
&\times&\bigg[ \sum_{j'}  \widetilde{\Omega}^{\ast}_{j' l m}(n, k) 
\widetilde{\Omega}_{j' l m}(n,k) \delta_{ij} \bigg. \nonumber\\
\bigg.
&-& \widetilde{\Omega}_{ilm}(n,k) \widetilde{\Omega}^{\ast}_{jlm}(n,k) \bigg].
\label{bsrt1}
\end{eqnarray}
In deriving the last expression, we used $\widetilde{\bf \Omega}_{l-m}(n,k)$ $=$ $(-1)^{m} \widetilde{\bf \Omega}^{\ast}_{lm}(n,k)$, noting that ${\bf \Omega}(n,{\bf k})$ is real. The longitudinal $T_1$ and transverse $T_2$ spin relaxation times are now directly obtained from Eq.~(\ref{bsrt1}) [see Sec.~\ref{sec:t1t2} for detailed discussion]. In our spin coherence time calculations for bulk semiconductors $h(n,k)$ $=$ $f_{0}(n,k)$ $[1- f_{0}(n,k)]$ $(n^{(3D)})^{-1}$, where $f_{0}(n,{\bf k})$ is the Fermi occupation function and $n^{(3D)}$ is the three-dimensional electron density.
\subsection{PERTURBATIVE EXPANSION FOR BULK ZINCBLENDE CRYSTALS}
\label{sec:pebzc}
In general the electron spin splitting must be calculated numerically; however it was argued by D'yakonov and Perel' \cite{dyakonov71, dyakonov72, meier84} that, to a good approximation, the electron spin splitting for bulk zincblende crystals can be evaluated using a perturbative approach. In a third order perturbation approximation (the lowest order for non-vanishing spin splitting), the spin splitting is proportional to $k^3$ and the components of the precession vector are \cite{dyakonov71, dyakonov72, meier84}
\begin{subequations}
\label{bssf}
\begin{equation}
\Omega_{x}(1,{\bf k})
= \frac{\hbar^2 \alpha_c }{(2m^{3}_{c}E_{\text{g}})^{\frac{1}{2}}} k_{x} \Big( k^{2}_{y} - k^2_{z} \Big),
\label{bssfx}
\end{equation}
\begin{equation}
\Omega_{y}(1,{\bf k})
= \frac{\hbar^2 \alpha_c }{(2m^{3}_{c}E_{\text{g}})^{\frac{1}{2}}} k_{y} \Big( k^{2}_{z} - k^2_{x} \Big),
\label{bssfy}
\end{equation}
\begin{equation}
\Omega_{z}(1,{\bf k})
= \frac{\hbar^2 \alpha_c }{(2m^{3}_{c}E_{\text{g}})^{\frac{1}{2}}} k_{z} \Big( k^{2}_{x} - k^2_{y} \Big)
\label{bssfz},
\end{equation}
\end{subequations}
where $m_c$ is the electron effective mass, $E_{\text{g}}$ is the band-gap energy, $\alpha_c = \gamma_{c} \hbar^{-3} E^{1/2}_{\text{g}} (2m_c)^{3/2}$ is a dimensionless constant related to the spin splitting of the conduction band, and $\gamma_{c}$ is the conduction-band spin-splitting coefficient (see, for example, Ref.~\onlinecite{cardona88} for calculations of $\gamma_{c}$ for zincblende semiconductors). This corresponds to taking only $l = 3$ in the series expansion of Eq.~(\ref{fsebpv}), and the non-zero expansion coefficients are
\begin{subequations}
\label{omegaylm}
\begin{equation}
\widetilde{\Omega}_{y 3 \pm 1}(1,k) 
= i \widetilde{\Omega}_{x 3 \pm 1}(1,k)
= \frac{i \pi^{\frac{1}{2}} \hbar^2 \alpha_c k^3}{(42 m^{3}_{c}E_{\text{g}})^{\frac{1}{2}}},
\label{omegaylm1} 
\end{equation}
\begin{equation}
\widetilde{\Omega}_{y 3 \pm 3}(1,k) 
= i \widetilde{\Omega}_{x 3 \pm 3}(1,k)
= \frac{i \pi^{\frac{1}{2}} \hbar^2 \alpha_c k^3}{(70 m^{3}_{c}E_{\text{g}})^{\frac{1}{2}}},
\label{omegaylm2} 
\end{equation}
\begin{equation}
\widetilde{\Omega}_{z 3 \pm 2}(1,k) 
= \frac{2 \pi^{\frac{1}{2}} \hbar^2 \alpha_c k^3}{(105 m^{3}_{c}E_{\text{g}})^{\frac{1}{2}}}.
\label{omegaylm3} 
\end{equation}
\end{subequations}
In this case, the kinetic equation [Eq.~(\ref{bsrt1})] for the spin-polarization vector can be simplified to 
\begin{eqnarray}
\frac{\partial}{\partial t} {S}_{i}(t)
& = & \frac{-1}{2 \pi^{2}} 
\int dk k^2 \widetilde{\tau}_3 (1,k)  h(1, k) \nonumber \\
& \times & \sum_{j} \bigg[ 
\langle \Omega_{j}(1,{\bf k}) \Omega_{j}(1,{\bf k}) \rangle_{\bf \hat{k}} {S}_{i}(t) \nonumber\\
&-& \langle \Omega_{j}(1,{\bf k}) \Omega_{i}(1,{\bf k}) \rangle_{\bf \hat{k}} {S}_{j}(t) 
\bigg],
\label{bspv2} 
\end{eqnarray}
and the components of the spin relaxation rate tensor $\Gamma_{ij}$ in Eq.~(\ref{bspt}) are reduced to
\begin{equation}
\Gamma_{ij} = 
\begin{cases}
{\displaystyle \frac{1}{2 \pi^{2}}} 
{\displaystyle \int} dk k^2 h(1, k) \widetilde{\tau}_3 (1,k) \\ 
\times \left( \sum_{j} \langle \Omega^{2}_{j}(1,{\bf k}) \rangle_{\bf \hat{k}}
- \langle \Omega^{2}_{i}(1,{\bf k}) \rangle_{\bf \hat{k}} \right) &\text{if $i = j$}, \\ \\
{\displaystyle \frac{1}{2 \pi^{2}}}
{\displaystyle \int} dk k^2 h(1, k) \widetilde{\tau}_3 (1,k)  \\
\times \langle \Omega_{i}(1,{\bf k}) \Omega_{j}(1,{\bf k}) \rangle_{\bf \hat{k}} &\text{if $i \ne j$},
\end{cases}
\label{bcsrt}
\end{equation}
For a Maxwell-Boltzmann distribution for $h(1, k)$, D'yakonov and Perel'~\cite{dyakonov72, meier84} obtained the following analytical expression for $T_1$:
\begin{equation}
\frac{1}{T_1} = \frac{\Xi \alpha_c^{2} \tau_\text{p}( k_\text{B} T)^3}{\hbar^2 E_g},
\label{DPt1}
\end{equation}
with
\begin{equation}
\Xi = \frac{16}{35} \frac{1}{\eta^{(3D)}}(\nu + \frac{7}{2})(\nu + \frac{5}{2}),
\label{DPC}
\end{equation}
where $k_\text{B}$ is the Boltzmann constant, $\tau_\text{p}$ is the average transport time which is directly related to the electron mobility through $\mu = e \tau_\text{p}/m_c$, and the numerical constants $\eta^{(3D)}$ and $\nu$ depend on the momentum scattering mechanism (and are given in Table~\ref{bulkMS}). In the case of a degenerate electron density~\cite{dyakonov72, meier84}, the expression for $T_1$ above is modified by simply replacing $k_\text{B} T$ with the Fermi energy $E_\text{F}$ and  setting $\Xi = 32/(105 {\eta^{(3D)}})$. This perturbative approach provides an adequate description for electron spin decoherence in bulk zincblende crystals. The comparison between the perturbative and non-perturbative approach is made in Sec.~\ref{sec:rd}. 
\subsection{SPIN DECOHERENCE IN NANOSTRUCTURES}
\label{sec:zqws}
In this section, we consider the electron spin dynamics for both quantum wells and superlattices simultaneously. Quantum wells can be viewed as superlattices with large barrier thickness in which the wells are practically isolated; therefore, the spin relaxation in quantum wells and superlattices can be treated on the same footing. For two-dimensional systems, it is convenient to use a cylindrical coordinate system. The coupled equations (\ref{kineticSD}) and (\ref{rho1c}) for the two-dimensional systems are explicitly written as
\begin{eqnarray}
\frac{\partial}{ \partial t} {\bf S}(t)
&=& \text{Tr} \bigg[  \sideset{}{} \sum_{\cal L} 
\int \frac{d K_z}{2 \pi}
\int \frac{d K_{\|} K_{\|}}{2 \pi}
\frac{\partial}{ \partial t}
\langle \underline{\underline{\rho}}\text{\@}_{{\cal L} {\bf K}}(t) \rangle_{\bf \hat{K}_{\|}}
\underline{\underline{\mbox{\boldmath$\sigma$}}} \bigg]  \nonumber \\
&=&  \frac{-i}{\hbar}  \text{Tr} \bigg[  \sideset{}{} \sum_{\cal L} 
\int \frac{d K_z}{2 \pi}
\int \frac{d K_{\|} K_{\|}}{2 \pi} 
\int \frac{d \phi_{\bf \hat{K}_{\|}}}{2 \pi} \nonumber \\
&\times& \Big[\underline{\underline{H}}\text{\@}^\text{SO}_{{\cal L} {\bf K}}, 
\underline{\underline{\rho}}\text{\@}^{(1)}_{{\cal L} {\bf K}}(t) \Big] 
\underline{\underline{\mbox{\boldmath$\sigma$}}} \bigg], \nonumber \\
\label{kineticSD2D}
\end{eqnarray}
\begin{eqnarray}
&& \frac{- i}{\hbar} \Big[\underline{\underline{H}}\text{\@}^\text{SO}_{{\cal L} {\bf K}}, 
\underline{\underline{\rho}}\text{\@}^{(0)}_{{\cal L} {\bf K}}(t) \Big]
 = \sideset{}{} \sum_{\cal L'}   
\int \frac{d K'_z}{2 \pi}
\int \frac{d K'_{\|} K'_{\|}}{2 \pi} 
\int \frac{d \phi_{\bf \hat{K}'_{\|}}}{2 \pi} \nonumber \\
&& \times \underline{\underline{W}}\text{\@}^{(\delta)}_{\cal L L'} (\hat{{\bf K}}_{\|} \cdot \hat{{\bf K}}'_{\|}, K_{\|}, K_z)
\left(\underline{\underline{\rho}}\text{\@}^{(1)}_{{\cal L} {\bf K}}(t) 
-\underline{\underline{\rho}}\text{\@}^{(1)}_{{\cal L'} {\bf K}'}(t) \right),
\label{rho12D}
\end{eqnarray}
where ${\bf K}$ is the superlattice wavevector with $K_z$ along the growth direction ${\bf \hat{z}}$ of the superlattice, ${\bf K_{\|}}$ is the in-plane wavevector, the quantum number ${\cal L}$ is the superlattice band index that enumerates all superlattice eigenstates, and $\langle \underline{\underline{\rho}}\text{\@}_{{\cal L} {\bf K}}(t) \rangle_{\bf \hat{K}_{\|}}$ $\equiv$ $(2 \pi)^{-1}\int d \phi_{\bf \hat{K}_{\|}}$ denotes the averaging of $\underline{\underline{\rho}}\text{\@}_{{\cal L} {\bf K}}(t)$ over all directions of the in-plane wavevector ${\bf K_{\|}}$. The momentum scattering in this case is essentially two-dimensional, therefore the electron scattering transition probability $\underline{\underline{W}}\text{\@}^{(\delta)}_{\cal L L'} (\hat{{\bf K}}_{\|} \cdot \hat{{\bf K}}'_{\|}, K_{\|}, K_z)$ depends on the magnitude of the in-plane wavevector $K_{\|}$ and angle between the initial and final in-plane momenta (${\bf K}_{\|}$ and ${\bf K}_{\|}'$) for each $K_z$. Under this assumption, we are effectively solving the coupled Eqs.~(\ref{kineticSD2D}) and (\ref{rho12D}) of an ideal two-dimensional system for each value of $K_z$ and then averaging the final results over the momentum along the growth direction. This approximation is quite reasonable because the band structure along $K_z$ of the layered structures we will consider here is relatively flat (i.e., almost dispersionless)\cite{averkiev99, maialle93}.

To solve the coupled spin density matrix Eqs.~(\ref{kineticSD2D}) and (\ref{rho12D}), $\underline{\underline{H}}\text{\@}^\text{SO}_{{\cal L} {\bf K}_{\|}}$, $\underline{\underline{\rho}}\text{\@}^{(1)}_{{\cal L} {\bf K}}(t)$, and $\underline{\underline{W}}\text{\@}^{(\delta)}_{\cal L L'} (\hat{{\bf K}}_{\|} \cdot \hat{{\bf K}}'_{\|}, K_{\|}, K_z)$ are expanded in a Fourier series in the ${\bf K}_{\|}$-plane for each value of $K_z$ ~\cite{averkiev99, maialle93}
\begin{equation}
\underline{\underline{H}}\text{\@}^\text{SO}_{{\cal L} {\bf K}_{\|}}
= \sum_{l} \widetilde{\underline{\underline{H}}}\text{\@}^\text{SO}_{l}({\cal L}, K_{\|}, K_z)
\exp(i l \phi_{\bf \hat{K}_{\|}}),
\label{esh} 
\end{equation}
\begin{equation}
\underline{\underline{\rho}}\text{\@}^{(1)}_{{\cal L} {\bf K}}(t)
= \sum_{l}
\widetilde{\underline{\underline{\rho}}}\text{\@}^\text{(1)}_{l}({\cal L}, K_{\|}, K_z, t)
\exp(i l \phi_{\bf \hat{K}_{\|}}),
\label{esd} 
\end{equation}
\begin{eqnarray}
&& \underline{\underline{W}}\text{\@}^{(\delta)}_{\cal L L'} 
(\hat{{\bf K}}_{\|} \cdot \hat{{\bf K}}'_{\|}, K_{\|}, K_z) \nonumber \\
&& = \sum_{l} 
\widetilde{\underline{\underline{W}}}\text{\@}^{(\delta)}_{\cal L L' l}(K_{\|}, K_z)
\exp \Big[i l (\phi_{\bf \hat{K}_{\|}} - \phi_{\bf \hat{K}'_{\|}}) \Big].
\label{emsp} 
\end{eqnarray}

Substituting Eqs.~(\ref{esh})-(\ref{emsp}) into Eq.~(\ref{rho12D}), Multiplying both sides of the resulting equation by $\exp(- i l' \phi_{\bf \hat{K}_{\|}})$, and integrating over the $d \phi_{\bf \hat{K}_{\|}}$, we have
\begin{eqnarray}
\widetilde{\underline{\underline{\rho}}}\text{\@}^\text{(1)}_{l}({\cal L}, K_{\|}, K_z, t) 
& = & \frac{-i}{\hbar} \widetilde{\underline{\underline{\tau}}}\text{\@}_{l}({\cal L}, K_{\|}, K_z) \nonumber \\
&\times& \Big[ \widetilde{\underline{\underline{H}}}\text{\@}^\text{SO}_{l}({\cal L}, K_{\|}, K_z), 
\langle \underline{\underline{\rho}}\text{\@}_{{\cal L} {\bf K}}(t) \rangle_{\bf \hat{K}_{\|}} \Big],\nonumber \\
\label{sdp} 
\end{eqnarray}
where the momentum scattering time matrix $\widetilde{\underline{\underline{\tau}}}\text{\@}_{l}({\cal L}, K_{\|}, K_z)$ is given by
\begin{eqnarray}
&& \widetilde{\underline{\underline{\tau}}}\text{\@}^{-1}_{l}({\cal L}, K_{\|}, K_z)  \nonumber \\
&& = \underline{\underline{{\cal D}}}_{\cal L} \sum_{n'} 
\int^{1}_{-1} \frac{d \Xi_{{\bf K}_{\|} {\bf K}'_{\|}}}{2 \pi} 
\underline{\underline{W}}\text{\@}_{\cal L L} (\hat{{\bf K}}_{\|}  
\cdot \hat{{\bf K}}'_{\|}, K_{\|}, K_z) \nonumber \\
&& \times \Big[1 - \cos (l \Xi_{{\bf K}_{\|} {\bf K}'_{\|}})  \Big], 
\label{tn} 
\end{eqnarray}
where $\cos \Xi_{{\bf K}_{\|} {\bf K}'_{\|}} \equiv \hat{{\bf K}}_{\|} \cdot \hat{{\bf K}}'_{\|}$. Here $\underline{\underline{{\cal D}}}_{\cal L}$ $=$ ${\cal D}^{(2D)}_{L\sigma}\underline{\underline{I}}$ is the two-dimensional density of states of the $L$th band with spin index $\sigma$. Note that the constant density of states ${\cal D}^{(2D)}_{L\sigma}$ is independent of the superlattice wavevector and is the same for both up spin and down spin due to Kramers degeneracy: ${\cal D}^{(2D)}_{L \uparrow}$ $=$ ${\cal D}^{(2D)}_{L \downarrow}$. As in the three-dimensional case the momentum scattering time matrix is diagonal with respect to the spin indices and hence it can be written as $\widetilde{\underline{\underline{\tau}}}\text{\@}_{l}({\cal L}, K_{\|}, K_z)$ $=$ $\widetilde{\tau}\text{\@}_{l}({\cal L}, K_{\|}, K_z) \underline{\underline{I}}$. For superlattices the transport time $\tau_\text{tr}({\cal L}, K_{\|}, K_z)$ from the two-dimensional mobility is given by $\widetilde{\underline{\underline{\tau}}}\text{\@}_{l=1}({\cal L}, K_{\|}, K_z)$ as in Eq.~(\ref{tn}) (see Sec.~\ref{sec:mrtm}). Notice the different angular dependence of the transport time in bulk crystals and superlattices. 

Substituting Eq.~(\ref{sdp}) into Eq.~(\ref{kineticSD2D}) and performing the integration over $d \phi_{\bf \hat{k}}$, we obtain
\begin{eqnarray}
&& \frac{\partial}{ \partial t} {\bf S}(t)
= \frac{-1}{(2 \pi \hbar)^2} \text{Tr} \Bigg[ \sum_{l {\cal L}} 
\int d K_z \int d K_{\|} K_{\|} 
\widetilde{\underline{\underline{\tau}}}\text{\@}_{l}({\cal L}, K_{\|}, K_z) \nonumber\\
&& \times \bigg[\widetilde{\underline{\underline{H}}}\text{\@}^\text{SO}_{l}({\cal L}, K_{\|}, K_z),
\Big[ \widetilde{\underline{\underline{H}}}\text{\@}^\text{SO}_{-l}({\cal L}, K_{\|}, K_z), 
\langle \underline{\underline{\rho}}\text{\@}_{{\cal L} {\bf K}}(t) \rangle_{\hat{{\bf K}}_{\|}} \Big] \bigg]
\underline{\underline{\mbox{\boldmath$\sigma$}}} \Bigg],  \nonumber \\
\label{kineticSI2D}
\end{eqnarray}
where $\widetilde{\underline{\underline{H}}}\text{\@}^\text{SO}_{l}({\cal L}, K_{\|}, K_z)$ is obtained from the relation
\begin{eqnarray}
&& \widetilde{\underline{\underline{H}}}\text{\@}^\text{SO}_{l}({\cal L}, K_{\|}, K_z) 
= \int \frac{d \phi_{\bf \hat{K}_{\|}}}{2 \pi}
\underline{\underline{H}}\text{\@}^\text{SO}_{{\cal L} {\bf K}_{\|}}
\exp(- i l \phi_{\bf \hat{K}_{\|}}) \nonumber \\
&& = \frac{\hbar}{2}  \int \frac{d \phi_{\bf \hat{K}_{\|}}}{2 \pi}
\left[{\bf \Omega} ({\cal L}, {\bf K}) \cdot \underline{\underline{\mbox{\boldmath$\sigma$}}} \right] 
\exp(- i l \phi_{\bf \hat{K}_{\|}}).
\label{ho} 
\end{eqnarray}
To evaluate the trace on the right-hand side of Eq.~(\ref{kineticSI2D}), we write the spin density matrix $\langle \underline{\underline{\rho}}\text{\@}_{{\cal L} {\bf K}}(t) \rangle_{\hat{{\bf K}}_{\|}}$ in the form of Eq.~(\ref{bdmi})
\begin{equation}
\langle \underline{\underline{\rho}}\text{\@}_{{\cal L} {\bf K}}(t) \rangle_{\hat{{\bf K}}_{\|}}
= \frac{1}{2} \Big[ g({\cal L}, K_{\|}, K_z) \underline{\underline{I}} + h({\cal L}, K_{\|}, K_z) {\bf S}(t) \cdot 
\underline{\underline{\mbox{\boldmath$\sigma$}}}  \Big ],
\label{dmi} 
\end{equation}
where $g({\cal L}, K_{\|}, K_z)$ and $h({\cal L}, K_{\|}, K_z)$ are the electron distribution functions which are written similar to Eqs.~(\ref{bdf}). Substituting Eq.~(\ref{dmi}) into Eq.~(\ref{kineticSI2D}) and taking the trace with respect to the spin indices, we obtain the kinetic equation for the electron spin density for superlattices
\begin{eqnarray}
&& \frac{\partial}{\partial t} {\bf S}(t)  \nonumber \\
&& = \frac{-1}{(2 \pi)^2} \sum_{l {\cal L}} 
\int d K_z \int d K_{\|} K_{\|} 
\widetilde{\underline{\underline{\tau}}}\text{\@}_{l}({\cal L}, K_{\|}, K_z) h({\cal L}, K_{\|}, K_z) \nonumber\\
&& \times \bigg[ \left( \widetilde{\bf \Omega}_{-l}({\cal L}, K_{\|}, K_z) \cdot 
\widetilde{\bf \Omega} _{l}({\cal L}, K_{\|}, K_z) \right) {\bf S}(t) \bigg. \nonumber\\
\bigg. && - \left( \widetilde{\bf \Omega}_{-l}({\cal L}, K_{\|}, K_z) \cdot {\bf S}(t) \right) 
 \widetilde{\bf \Omega}_{l}({\cal L}, K_{\|}, K_z) \bigg].
\label{kineticSI2Dfinal} 
\end{eqnarray}
Rewriting Eq.~(\ref{kineticSI2Dfinal}) in a matrix form equivalent to Eq.~(\ref{bspt}), we have
\begin{equation}
\frac{\partial}{\partial t} {\bf S}(t)
= - \underline{\underline{\Gamma}} {\bf S}(t),
\label{spt} 
\end{equation}
where the components of the spin relaxation rate $\underline{\underline{\Gamma}}$ are given by
\begin{eqnarray}
&&\Gamma_{ij} 
= \frac{1}{(2 \pi)^2} \sum_{l {\cal L}} \int d K_z \int d K_{\|} K_{\|} 
\widetilde{\underline{\underline{\tau}}}\text{\@}_{l}({\cal L}, K_{\|}, K_z) \nonumber\\
&&\times   h({\cal L}, K_{\|}, K_z) \bigg[ \sum_{j'} \widetilde{\Omega}^\ast_{j' l}({\cal L}, K_{\|}, K_z)
\widetilde{\Omega} _{j' l}({\cal L}, K_{\|}, K_z) \delta_{i j}   \nonumber \\
&& -\widetilde{\Omega}^\ast_{j l}({\cal L}, K_{\|}, K_z)
\widetilde{\Omega} _{i l}({\cal L}, K_{\|}, K_z) \bigg].
\label{srt}
\end{eqnarray}
The last expression is obtained using $\widetilde{\Omega} _{i -l}({\cal L}, K_{\|}, K_z)$ $=$ $\widetilde{\Omega}^{\ast}_{i l}({\cal L}, K_{\|}, K_z)$. In our spin coherence time calculations for quantum wells and superlattices $h({\cal L}, K_{\|}, K_z)$ $=$ $f_{0}({\cal L}, K_{\|}, K_z)$ $[1- f_{0}({\cal L}, K_{\|}, K_z)]$ $(n^{(3D)})^{-1}$, where $f_{0}({\cal L}, K_{\|}, K_z)$ is the Fermi occupation function and $n^{(3D)}$ is the three-dimensional electron density.
\subsection{PERTURBATIVE APPROACH FOR QUANTUM WELLS}
\label{sec:paqw}
The extension of the perturbation approximation in bulk crystals to two-dimensional systems is originally due to 
D'yakonov and Kachorovskii \cite{dyakonov86}. In a third order approximation, the precession vector for structures grown along $(001)$ direction is\cite{ivchenko97}
\begin{eqnarray}
&& {\bf \Omega}(1, {\bf K}_{\|}) \nonumber \\ 
&& = \Big[ \widetilde{\Omega}_{x1}(1, K_{\|}) \cos(\phi_{\bf \hat{K}_{\|}}) 
+ \widetilde{\Omega}_{x3}(1, K_{\|}) \cos(3 \phi_{\bf \hat{K}_{\|}}), \nonumber \\ 
&& \quad  \widetilde{\Omega}_{y1}(1, K_{\|}) \sin(\phi_{\bf \hat{K}_{\|}}) 
+ \widetilde{\Omega}_{y3}(1, K_{\|}) \sin(3 \phi_{\bf \hat{K}_{\|}}), 0 \Big], \nonumber \\
\label{spv}
\end{eqnarray}
with the angular components $\widetilde{\Omega}_{jl}(1, K_{\|})$ given by 
\begin{subequations}
\label{spvs}
\begin{eqnarray}
&& \widetilde{\Omega}_{x1}(1, K_{\|})
= - \widetilde{\Omega}_{y1}(1, K_{\|}) \nonumber \\ 
&& = \frac{- \hbar^2 \alpha_c}
{8 (2 m^{3}_{e}E_{\text{g}})^{\frac{1}{2}}}
\Big(\frac{8 m_c E_{\text{1e}}}{\hbar^2} - K^{2}_{\|} \Big) K_{\|}, 
\label{spvsa}
\end{eqnarray}
\begin{eqnarray}
 \widetilde{\Omega}_{x3}(1, K_{\|})
= \widetilde{\Omega}_{y3}(1, K_{\|}) 
= \frac{- \hbar^2 \alpha_c }
{8 (2 m^{3}_{c}E_{\text{g}})^{\frac{1}{2}}} K^{3}_{\|},
\label{spvsb}
\end{eqnarray}
\end{subequations}
where $E_{\text{1e}}$ is the electron confinement energy of the first quantum-well state. In this approach, the crystal effective magnetic field along the growth direction vanishes because the z-component of the precession vector is identically zero. Using Eqs.~(\ref{spvs}) in Eq.~(\ref{srt}), the components of the spin relaxation rate tensor are
\begin{eqnarray}
&& \Gamma_{xx}
= \Gamma_{yy}
= \frac{1}{2} \Gamma_{zz} \nonumber \\ 
&& = \frac{1}{4 \pi} \sum_{l=1}^{3} \int d K_{\|} K_{\|} h(1, K_{\|})
\widetilde{\tau}_{l}(1, K_{\|}) 
\widetilde{\Omega}^{2}_{xl}(1, K_{\|}). \nonumber \\
\label{spvspa}
\end{eqnarray}
In Ref. \onlinecite{dyakonov86}, the $K^{3}_{\|}$ term in Eqs.~(\ref{spvs}) was considered to be small compared to the linear $K_{\|}$ term, and consequently it was dropped, as well as the $l = 3$ term. Under this assumption, D'yakonov and Kachorovskii~\cite{dyakonov86} obtained an analytical expression for the $T_1$ of a non-degenerate electron gas after averaging Eq.~(\ref{spvspa}) over the Maxwell-Boltzmann distribution
\begin{equation}
\frac{1}{T_1} = \frac{2 \alpha_c^{2} E_{\text{1e}}^2 \tau_{p} k_\text{B} T}{\hbar^2 E_g},
\label{DKt1}
\end{equation}
and 
\begin{equation}
\tau_{p} \equiv 
\frac{{\displaystyle \int d K_{\|} K_{\|}^{3} h(1, K_{\|}) \widetilde{\tau}_{1}(1, K_{\|})}} 
{{\displaystyle \int d K_{\|} K_{\|}^{3} h(1, K_{\|})}},
\label{DKTauP}
\end{equation}
where $\tau_\text{p}$ (see, for example, Ref.~\onlinecite{ferry97}) is the average transport time, related to the electron mobility via $\mu = e \tau_\text{p}/m_c$ [see Eq.~(\ref{bm}) in Sec.~\ref{sec:mrtm}]. In the case of a degenerate electron gas, Eq.~(\ref{DKt1}) is modified by replacing $k_\text{B} T$ with the Fermi energy $E_\text{F}$. The validity of these approximations is examined in Sec.~\ref{sec:rd}.

\subsection{TRANSPORT TIME AND MOBILITY}
\label{sec:mrtm}
To evaluate the momentum relaxation time $\widetilde{\tau}_{l}(n, k)$ for bulk semiconductors [see Eqs.~(\ref{btn})], we first extract the transport time $\tau_\text{tr}(n, k)$ [i.e., $\widetilde{\tau}_{1}(n, k)$] from the experimental measured mobility by solving the Boltzmann equation using a fourteen-band non-parabolic electronic structure for the bulk semiconductor (see Sec.~\ref{sec:fbt}). Once the transport time is determined, the momentum relaxation time $\widetilde{\tau}_{l}(n, k)$ for $l >1$ is expressed in terms of the transport time $\widetilde{\tau}_{1}(n, k)$. The Boltzmann equation is written in the standard form (see, for example, Ref.~\onlinecite{rammer98})
\begin{eqnarray}
&& \frac{\partial f({\bf r}, n, {\bf k}, t)}{\partial t}
+ {\bf v}(n, {\bf k}) \cdot \Big[ {\nabla}_{\bf r} f({\bf r}, n, {\bf k}, t) \Big] \nonumber \\ 
&& + e \hbar^{-1} {\bf F}({\bf r}, t) \cdot \Big[ {\nabla}_{\bf k} f({\bf r}, n, {\bf k}, t) \Big] 
= I_{{\bf r}, n, {\bf k}, t} \left[ f \right],
\label{boltzmann}
\end{eqnarray}
where $e$ is the elementary charge, ${\bf v}(n,{\bf k})$ is the electron velocity in the $n$th band, 
$ {\bf F}({\bf r}, t)$ is the electric field, and the collision integral on the right-hand side of Eq.~(\ref{boltzmann}) is given by
\begin{eqnarray}
 I_{{\bf r}, n, {\bf k}, t} \left[ f \right]
&=& - \sum_{n'} \int  \frac{d {\bf k'}}{(2\pi)^3}  W_{n n'}({\bf k'}, {\bf k}) \nonumber \\ 
&\times& \Big[  f({\bf r}, n, {\bf k}, t) -  f({\bf r}, n', {\bf k'}, t)  \Big].
\label{collision}
\end{eqnarray}

In the relaxation time approximation, the collision integral takes the form
\begin{equation}
I_{{\bf r}, n, {\bf k}, t} \left[ f \right]
= - \frac{ f({\bf r}, n, {\bf k}, t) -  f_{0}({\bf r}, n, {\bf k}, t)}{\tau_\text{tr} (n,{\bf k})},
\label{collisionR}
\end{equation}
where $f_{0}({\bf r}, n, {\bf k}, t)$ is the thermal equilibrium distribution function and 
$\tau_\text{tr} (n,{\bf k})$
is the transport time. In steady state, for a spatially homogeneous and time-independent electric field, the solution of the Boltzmann equation (\ref{boltzmann}) linearized with respect to the electric field $\bf F$ is written as
\begin{equation}
f(n, {\bf k}) = f_{0}(n, {\bf k}) - e \hbar^{-1} \tau_\text{tr}(n,{\bf k}) {\bf F} 
\cdot \Big[ {\nabla}_{\bf k} f(n,{\bf k}) \Big].
\label{linearboltzmann}
\end{equation}

The electron mobility tensor $\mbox{\boldmath$\mu$}$ is defined as
\begin{equation}
{\bf v}_\text{drift} = \mbox{\boldmath$\mu$} \cdot {\bf F},
\label{mobilitytensor}
\end{equation}
where ${\bf v}_\text{drift}$ is the average drift velocity and it is obtained by averaging the momentum-dependent electron velocity over the non-equilibrium distribution function
\begin{equation}
{\bf v}_\text{drift} = \frac{{\displaystyle \sum_{n} \int  d {\bf k}  f(n,{\bf k}) {\bf v}(n,{\bf k})}}{\displaystyle {\sum_{n} \int  d {\bf k}  f(n,{\bf k})}}.
\label{driftvelocity}
\end{equation} 
Thus the electron mobility tensor for bulk semiconductors can be obtained from Eqs.~(\ref{linearboltzmann}), (\ref{mobilitytensor}), and (\ref{driftvelocity}), and the final expression is written as \cite{rammer98, ridley99}
\begin{eqnarray}
\mbox{\boldmath$\mu$}^\text{Bulk} 
&=& \frac{e}{n^{(3D)} \hbar^2} \sum_n \int  \frac{d {\bf k}}{(2\pi)^3} \tau_\text{tr}(n,{\bf k})
\left[- \frac{\partial f_{0}(n,{\bf k})} {\partial E_{n \sigma} ({\bf k})} \right] \nonumber \\ 
&\times&\Big[{\nabla}_{\bf k} E_{n \sigma} ({\bf k})  \Big] \Big[{\nabla}_{\bf k} E_{n \sigma} ({\bf k}) \Big],
\label{bm}
\end{eqnarray}
where $n^{(3D)}$ is the three-dimensional carrier density and $E_{n \sigma} ({\bf k})$ is the electron energy in the $n$th band. Equation~(\ref{bm}) connects the transport time $\tau_\text{tr}(n,{\bf k})$ to the experimental measured mobility $\mbox{\boldmath$\mu$}^\text{Bulk}$. For zincblende semiconductors, the mobility tensor [Eq.(~\ref{bm})] is diagonal with identical elements owing to the cubic symmetry of the zincblende materials.

The electron mobility tensor $\mu^\text{SL}$ for superlattices is obtained in a similar fashion, solving the linearized Boltzmann equation within the framework of a generalized ${\bf K{\cdot}p}$ envelope function approach (see Sec.~\ref{sec:fbt}). In this case, the electron mobility tensor for superlattices takes the same form as that for the bulk case, Eq.~(\ref{bm}) (see Sec.~\ref{sec:spv} for a brief discussion on symmetry). For zincblende type superlattices, the cubic symmetry of the constituent semiconductors is preserved in the in-plane direction, but it is broken along the growth direction. As a result, the diagonal element of the mobility tensor along the growth direction takes a different value than those along the in-plane directions. 
\subsection{ORBITAL SCATTERING TIME}
\label{sec:mrt}
To calculate the orbital relaxation time [see Eq.~(\ref{btn})], the functional form of the orbital scattering rate $W_{n n'}({\bf \hat{k} \cdot \bf \hat{k}'}, k)$ is taken from standard expressions for ionized impurity (II), neutral impurity (NI), or optical phonon (OP) scattering \cite{dyakonov72, meier84}. The scattering time $\widetilde{\tau}_l(n,k)$ for $l > 1$ is expressed in terms of the transport time $\tau_\text{tr}(n,k)$ $=$ $\widetilde{\tau}_1(n,k)$, and the ratios for ionized impurity, neutral impurity, and optical phonon scattering are summarized in Table~\ref{bulkMS}. The transport time $\tau_\text{tr}(n,k)$ is obtained by inverting the Boltzmann transport equation (\ref{bm}) using a fourteen-band nonparabolic electronic structure for the semiconductor (see Sec.~\ref{sec:fbt}). In Eq.~(\ref{bm}), the transport time is expressed as a function of energy and it has the form of $\tau_\text{tr}(n, k)$ $\propto$ $E^{\nu}_{n \sigma}({\bf k})$. The values of $\nu$ for various scattering mechanisms are listed in Table~\ref{bulkMS}. 

The approach to evaluate the momentum relaxation time $\widetilde{\tau}_l({\cal L}, K_{\|}, K_z)$ for two-dimensional structures is very similar to that of the bulk case. The functional form of the scattering rate $W_{\cal L L'}({\bf \hat{K}_{\|} \cdot \bf \hat{K}'_{\|}}, K_{\|}, K_z)$ is again taken from standard expression for ionized impurity, neutral impurity (such as arises from superlattice interface roughness), or optical phonon scattering. The $\widetilde{\tau}_l({\cal L}, K_{\|}, K_z)$'s differ for different mechanisms and the relationship between the momentum relaxation time and the transport time for various scattering mechanisms is tabulated in Table~\ref{mqwMS}. The connection between the momentum relaxation time and the transport time in two-dimensional systems is very simple, and the ratio of $\widetilde{\tau}_l({\cal L}, K_{\|}, K_z)$ and $\tau_\text{tr}({\cal L}, K_{\|}, K_z)$ can be expressed as a function of the angular index $l$ of the field component. The effective time for field reversal $\widetilde{\tau}_l ({\cal L}, K_{\|}, K_z)$ depends on the angular indices $l$ of the field component $\widetilde{\Omega}_{jl}({\cal L}, K_{\|}, K_z)$. For example, an $l=1$ component $\widetilde{\Omega}_{j1}({\cal L}, K_{\|}, K_z)$ requires a $180^{\rm o}$ change in the angle of ${\bf K}$ to change sign, whereas an $l=3$ component $\widetilde{\Omega}_{j3}({\cal L}, K_{\|}, K_z)$ only requires a $60^{\rm o}$ change, so typically $\widetilde{\tau}_3 ({\cal L}, K_{\|}, K_z)$ $<$ $\widetilde{\tau}_1 ({\cal L}, K_{\|}, K_z)$. The momentum-dependent transport time has the form of $\tau_\text{tr}({\cal L}, K_{\|}, K_z)$ $\propto$ $E^{\nu}_{L S}({\bf K})$, and the values of $\nu$ for ionized impurity, neutral impurity, and optical phonon scattering are given in Table~\ref{mqwMS}.

\subsection{GENERAL FEATURES OF THE LONGITUDINAL $T_{1}$ AND TRANSVERSE $T_{2}$ SPIN RELAXATION TIMES}
\label{sec:t1t2}
This section considers both bulk and quantum-well systems. In the motional narrowing regime the electronic spin system is subject to an effective {\it time-dependent,} randomly oriented magnetic field ${\bf H}(n, {\bf k})$ which changes direction with a time $\tau(n, {\bf k})$ that is much shorter than the precession time  of either the constant applied field ${\bf H_o}$ or the random field. The coherence times in bulk zincblende semiconductors depend on the transverse $H_\perp (n, {\bf k})$ and longitudinal $H_\parallel (n, {\bf k})$ components of the random field, according to \cite{yafet63}
\begin{subequations}
\label{t1t2}
\begin{eqnarray} 
T_1^{-1}  
= \Upsilon \sum_{n} \int d {\bf k} H_\perp^2(n, {\bf k}) \tau(n, {\bf k})  h(n, {\bf k}),
\label{t1}
\end{eqnarray}
\begin{eqnarray} 
T_2^{-1}  
&=&  \Upsilon  \sum_{n} \int d {\bf k} 
\left[ \frac{1}{2}H_\perp^2(n, {\bf k}) + H_\parallel^2(n, {\bf k}) \right]  \nonumber \\ 
&\times& \tau(n, {\bf k}) h(n, {\bf k}),
\label{t2}
\end{eqnarray}
\end{subequations} 
where the electron distribution function $h(n, {\bf k})$ is defined similar to Eq.~(\ref{bdf2}), $\Upsilon \equiv (\text{g} \mu_{\text{B}} \hbar^{-1})^2$, $\mu_{\text{B}}$ is the Bohr magneton and $\text{g}$ is the electron g-factor. In a crystal with inversion asymmetry and spin-orbit coupling there is a spin splitting described by the Hamiltonian given in Eq.~(\ref{bssh}). The momentum-dependent effective magnetic field due to inversion asymmetry is ${\bf H}(n, {\bf k})$ $=$ $\Upsilon^{-1/2}{\bf \Omega}(n, {\bf k})$.
As the electron is scattered from ${\bf k}$ to ${\bf k'}$ via ordinary orbital (not spin-dependent) elastic scattering, the effective magnetic field ${\bf H}(n, {\bf k})$ changes direction with time. If the crystal is cubic, then $H_x^2(n, {\bf k}) = H_y^2 (n, {\bf k}) = H_z^2 (n, {\bf k})$, so $H_\perp^2 (n, {\bf k}) = 2H_\parallel^2 (n, {\bf k})$ and $T_2=T_1$ (see Ref.~\onlinecite{yafet63}). 

The relationship between $T_1$ and $T_2$ differs, however, for systems of lower symmetry, such as quantum wells or superlattices. For a $(001)$-grown quantum well, one finds in perturbative expressions~\cite{dyakonov72, ivchenko97} that the fluctuating field along the growth direction vanishes. In this case~\cite{lau01}
\begin{subequations}
\label{t1at2a}
\begin{equation}
T_1^{-1}(\alpha) = T_{1}^{-1}(\alpha=0) \Big[ \frac{1}{2}(1 + \cos^2 \alpha) \Big],
\label{t1a}
\end{equation}
\begin{equation}
T_2^{-1}(\alpha) = T_{1}^{-1}(\alpha=0) \Big[ \frac{1}{4}(2+ \sin^2 \alpha) \Big],
\label{t2a}
\end{equation}
\end{subequations}
where $\alpha$ is the direction between ${\bf H_o}$ and the growth direction. Thus $T_2$ ranges from $2T_1/3$ to $2T_1$ depending on $\alpha$.  By similar arguments, for $(110)$-grown quantum wells the effective crystal magnetic field is entirely along the growth direction, and
\begin{subequations}
\label{t1bt2b}
\begin{equation}
T_1^{-1}(\alpha) = T_{1}^{-1}(\alpha=0) \sin^2 \alpha ,
\label{t1b}
\end{equation}
\begin{equation}
T_2^{-1}(\alpha) = T_{1}^{-1}(\alpha=0) \Big[ \frac{1}{2}(1+ \cos^2 \alpha) \Big],
\label{t2b}
\end{equation}
\end{subequations}
Thus although $T_1^{-1}(\alpha=0)$ vanishes, the same is not the case for $T_2^{-1}(\alpha)$ for any $\alpha$. From Eqs.~(\ref{spt}), (\ref{srt}), (\ref{t1at2a}), and (\ref{t1bt2b}), one can immediately identify
\begin{subequations}
\label{sct12}
\begin{equation}
T_{1}^{-1} (0) = \Gamma_{zz}
\label{sct1} 
\end{equation}
\begin{equation}
T_{2}^{-1} (0) = \frac{1}{2}
\Big( \Gamma_{xx} + \Gamma_{yy}  \Big).
\label{sct2} 
\end{equation}
\end{subequations}
\subsection{CONNECTION BETWEEN $T_{1}$ AND $\tau_{s}$ TIMES}
\label{sec:ts}
In this section, we briefly discuss the relationship between the longitudinal spin relaxation time $T_1$ and the spin-flip time $\tau_s$, which is usually refered to as the ``spin relaxation time" in optical experiments \cite{britton98, hyland99, hall99, terauchi99, tackeuchi97, tackeuchi99}. In spin-dependent pump-probe experiments, a non-equilibrium spin population is created by circularly polarized light, and the spin orientation of the excited electrons depends on the helicity of the incident light. Once the unbalanced electron spin population is excited in the conduction band, it relaxes toward equilibrium both via carrier recombination and via flipping spins. It is usually assumed that the unbalanced spin population is described by the rate equations \cite{meier84, hyland99}
\begin{subequations}
\label{spr}
\begin{equation}
\frac{dN_+(t)}{dt} 
= - \frac{N_+(t)}{\tau_r} - \frac{N_+(t)}{\tau_s} + \frac{N_-(t)}{\tau_s},
\label{spr1} 
\end{equation}
\begin{equation}
\frac{dN_-(t)}{dt} 
= - \frac{N_-(t)}{\tau_r} + \frac{N_+(t)}{\tau_s} - \frac{N_-(t)}{\tau_s},
\label{spr2} 
\end{equation}
\end{subequations}
where $N_{+(-)}(t)$ is the number of circularly polarized electrons in spin-up (down) state, $\tau_r$ is the electron-hole recombination time, and $\tau_s$ is the spin-flip time \cite{meier84, yafet63}. Taking the difference between Eqs.~(\ref{spr1}) and (\ref{spr2}) gives
\begin{equation}
\frac{dD(t)}{dt} 
= - \Big( \frac{1}{\tau_r}  + \frac{2}{\tau_s} \Big)D(t),
\label{spd} 
\end{equation}
where $D(t)$ is defined as $D(t) \equiv (N_+(t) - N_-(t))$, and is directly proportional to the macroscopic magnetization of the system. The solution of Eq.~(\ref{spd}) takes a simple exponential form
\begin{equation}
D(t) = D_0 \exp{(-t/\tau_r)} \exp{(-2t/\tau_s)},
\label{spds} 
\end{equation}
where $D_0$ is the initial excited electron spin population. For an electron-hole recombination time much greater than the electron spin relaxation time ($\tau_r \gg \tau_s$), the unbalanced spin population decays exponentially toward equilibrium with a decay constant $ \tau_s / 2$. Thus the macroscopic magnetization of the system relaxes exponentially with a time constant $\tau_s / 2$, and the longitudinal spin relaxation time $T_1 = \tau_s / 2$ (see, for example, Refs.~\onlinecite{gantmakher87} and \onlinecite{yafet63}).
\section{FOURTEEN-BAND ${\bf K{\cdot}p}$ THEORY}
\label{sec:fbt}
Calculations of the longitudinal time $T_1$ and the transverse time $T_2$ require knowledge of the precession vector ${\bf \Omega} ({\cal L}, {\bf K})$, which is directly related to the energy eigenstates and eigenvectors of the system. In this section, a generalized fourteen-band non-perturbative theory is presented for the electronic band structure of zincblende semiconductor heterostructures based on the superlattice ${\bf K{\cdot}p}$ theory~\cite{flatte96, johnson90, johnson88, johnson87, olesberg99}, which is a generalization of the bulk ${\bf k{\cdot}p}$ theory for systems with periodicity on the scales larger than the bulk lattice constant. This periodicity exists only in the superlattice growth direction, which in this case is taken to be the $\hat{{\bf z}}$ direction. 
\subsection{FOURTEEN-BAND RESTRICTED BASIS}
\label{sec:fbrb}
The Hamiltonian $\hat{H}$ pertinent for the bulk constituents of heterostrutures is given by Eq.~(\ref{hamCryT}). It is solved within a fourteen-band restricted basis set of the constituents, consisting of two conduction antibonding $s$ states ($\overline{s}$, or the $\Gamma_6$ representation), six valence bonding $p$ states ($\Gamma_8\oplus \Gamma_7$), and six antibonding $p$ states ($\overline{p}$, also $\Gamma_8\oplus\Gamma_7$) (see Table~\ref{basis})~\cite{olesberg99, flatte96, cardona99}. These fourteen basis states are characterized by the total angular momentum quantum number $(J, J_z)$. The total angular momentum is ${\bf J}$ $=$ ${\bf L}$ $+$ ${\bf S}$, where ${\bf L}$ is the orbital angular momentum and ${\bf S}$ is the spin angular momentum. The orbital angular momentum takes the values of ${\bf L}$ $=$ ${\bf 0}$ for the $s$ and $\overline{s}$  states and ${\bf L}$ $=$ ${\bf 1}$ for the $p$ and $\overline{p}$  states. This is the minimum number of basis states that are required to generate spin splitting non-perturbatively (see, for example, Ref.~\onlinecite{cardona88}).

The electronic band structure of bulk GaAs at room temperature is shown in Fig.~\ref{BulkGaAsGaSbBands}(a) together with its parameters defined in the ${\bf k{\cdot}p}$ formalism. Here $E_{\text{g}}$ is the fundamental energy gap [between $\overline{s} \left(\Gamma_{6} \right)$ and $p \left(\Gamma_{8} \right)$ states], $E_{c}$ is the energy difference between $\overline{p} \left(\Gamma_{7} \right)$ and $p \left(\Gamma_{8} \right)$, $\Delta$ is the spin-orbit splitting of the $p$ states, and $\Delta_{c}$ is the spin-orbit splitting of the $\overline{p}$ states. The three interband momentum matrix elements ($P_{0}$, $P_{1}$, and $Q$) that couple the states away from the zone center in the ${\bf k{\cdot}p}$ model are
\begin{subequations}
\label{mme}
\begin{equation}
i P_{0} 
\equiv \langle S \arrowvert \hat{p}_{x} \arrowvert X^{v} \rangle 
= \langle S \arrowvert \hat{p}_{y} \arrowvert Y^{v} \rangle
=\langle S \arrowvert \hat{p}_{z} \arrowvert Z^{v} \rangle, \quad \quad \quad
\end{equation}
\begin{equation}
i P_{1} \equiv \langle S \arrowvert \hat{p}_{x} \arrowvert X^{c} \rangle = \langle S \arrowvert \hat{p}_{y} \arrowvert Y^{c} \rangle
= \langle S \arrowvert \hat{p}_{z} \arrowvert Z^{c} \rangle,  \quad \quad \quad
\label{mmeb}
\end{equation}
\begin{eqnarray}
i Q 
& \equiv & \langle Z^{v} \arrowvert \hat{p}_{x} \arrowvert Y^{c} \rangle
= \langle X^{v} \arrowvert \hat{p}_{y} \arrowvert Z^{c} \rangle  
=  \langle Y^{v} \arrowvert \hat{p}_{z} \arrowvert X^{c} \rangle \quad \quad \quad  \quad \nonumber \\
& = & \langle Y^{v} \arrowvert \hat{p}_{x} \arrowvert Z^{c} \rangle 
= \langle Z^{v} \arrowvert \hat{p}_{y} \arrowvert X^{c} \rangle
= \langle X^{v} \arrowvert \hat{p}_{z} \arrowvert Y^{c} \rangle. \quad \quad \quad \quad
\label{mmec} 
\end{eqnarray}
\end{subequations}

The energy splitting at finite crystal momentum ${\bf k}$ of GaAs shown in Fig.~\ref{BulkGaAsGaSbBands}(a) is a result of spin-orbit interaction and inversion asymmetry in zincblende semiconductors. At zone center, the eigenstates are mixtures of different spin eigenstates but  are still doubly degenerate, and so a ``pseudospin'' degeneracy can be identified. Away from zone center, the energy eigenstates of the crystal Hamiltonian ${\hat{H}}$ are certainly not spin eigenstates, and it is not even possible to choose a pseudospin basis which diagonalizes the Hamiltonian for all directions of ${\bf k}$. Tables~\ref{materials1} and \ref{materials2} present the parameters for the systems described in Sec.~\ref{sec:rd}, which are obtained from fits to experimental measurements of the conduction-band mass, the heavy- and light-hole masses, and the g-factor given in Refs.~\onlinecite{madelung96}, \onlinecite{landolt87}, and \onlinecite{blacha84}. The parameters in Tables~\ref{materials1} and \ref{materials2} obtained from these experimental observables have been used in several recent publications describing various optical and electronic properties of semiconductors\cite{olesberg2000,olesberg2001,grein2002}, and have not been altered for our spin coherence calculations in bulk and quantum-well semiconductors.
\subsection{SUPERLATTICE WAVE EQUATION}
\label{sec:slse}
The Schr$\ddot{\rm o}$dinger equation for heterostructure superlattices is written as
\begin{equation}
\hat{H}^\text{SL}\langle {\bf r} \arrowvert {\cal L}, {\cal S}, {\bf K} \rangle 
= E_{\cal L S} ({\bf K}) \langle {\bf r} \arrowvert {\cal L}, {\cal S}, {\bf K} \rangle,
\label{sch}
\end{equation}
and the superlattice Hamiltonian $H^{\text{SL}}$ is written as
\begin{subequations}
\label{hamSL}
\begin{equation}
\hat{H}^{\text{SL}} = \sum^{N_\text{layer}}_{i = 1} \hat{H}_{i} \theta_i ({\bf r}), 
\label{hamSLA}
\end{equation}
\begin{equation}
\theta_i ({\bf r}) =
\begin{cases}
1& \text{if ${\bf r} \in$ $i$th layer}, \\
0& \text{if ${\bf r} \not \in$ $i$th layer},
\end{cases}
\label{hamSLB}
\end{equation}
\end{subequations}
where $\hat{H}_{i}$ is the crystal Hamiltonian of the $i$th layer [Eq.~(\ref{hamCryT})], $N_\text{layer}$ is the number of layers in the superlattice unit cell, $\arrowvert {\cal L}, S, {\bf K} \rangle$ is the superlattice eigenstate for a carrier with wavevector ${\bf K}$ and spin ${\cal S}$ $\in$ $\{\uparrow, \downarrow\}$ in the ${\cal L}$th band, and $E_{\cal L S} ({\bf K})$ is the corresponding superlattice eigenenergy. To obtain  $\arrowvert {\cal L}, {\cal S}, {\bf K} \rangle$ and $E_{\cal L S} ({\bf K})$, we first solve the Schr$\ddot{\rm o}$dinger equation [Eq.~(\ref{sch})] at the zone center in which the superlattice wavevector ${\bf K}$ is identically zero. 
\subsection{ZONE-CENTER SUPERLATTICE STATES}
\label{sec:zcs}
A zone-center superlattice wavefunction is expressed in terms of bulk electronic structure parameters \cite{olesberg99, flatte96}  as 
\begin{equation}
\langle {\bf r} \arrowvert N, S, {\bf 0} \rangle 
= \sum_{n \sigma} F_{N S n \sigma} ({\bf r}) \langle {\bf r} \arrowvert n, \sigma, {\bf 0} \rangle, 
\label{zcs}
\end{equation}
where $\arrowvert n, \sigma, {\bf 0} \rangle$, defined in Table~\ref{basis}, are the corresponding zone-center Bloch states of the constituent bulk semiconductors, the expansion coefficients $F_{N S n \sigma} ({\bf r})$ are slowly-varying envelope functions on the scale of the constituent bulk semiconductor lattice constant, the quantum number $N$ is the superlattice zone-center band index, and $S$ is the spin quantum number that enumerates both the spin-up and spin-down states. The ordering of the basis states given in Table~\ref{basis} is such that the matrix Hamiltonian is block diagonal at ${\bf k}={\bf 0}$. Here, for convenience we adopt the convention that lower and upper cases label bulk and superlattice states respectively. Equation~(\ref{zcs}) implies that the zone-center superlattice states $\langle {\bf r} \arrowvert N, S, {\bf 0} \rangle$ are not spin eigenstates due to the mixing of different angular momenta. Therefore, the zone-center states $\langle {\bf r} \arrowvert N, \uparrow, {\bf 0} \rangle$ and $\langle {\bf r} \arrowvert N, \downarrow, {\bf 0} \rangle$ are generally referred to as pseudospin-up and pseudospin-down states, respectively. By substituting Eq.~(\ref{zcs}) into (\ref{sch}), Multiplying from the left-hand side of the resulting equation by the corresponding wavefunction $\langle n, \sigma, {\bf 0} \arrowvert {\bf r} \rangle$ of the zone-center Bloch state and integrating over a unit cell of the constituent bulk semiconductor, we obtain the zone-center Schr$\ddot{\rm o}$dinger equation as \cite{olesberg99}
\begin{equation}
\sum_{n' \sigma'} \hat{H}_{n \sigma n' \sigma'}^\text{ZC}({\bf r}) F_{N S n' \sigma'} ({\bf r}) 
=  E_{N S} ({\bf 0}) F_{N S n \sigma} ({\bf r}) 
\label{zcse}
\end{equation}
with 
\begin{eqnarray}
\hat{H}_{n \sigma n' \sigma'}^\text{ZC}({\bf r}) 
&\equiv& \frac{\hat{\bf p}^2}{2m_{e}} \delta_{n n'} \delta_{\sigma \sigma'} \nonumber \\
&+& E^\text{Bulk}_{n \sigma n' \sigma'} ({\bf r})
+ \frac{\hbar}{m_{e}} \hat{\bf p} \cdot {\bf P}^\text{Bulk}_{n \sigma n' \sigma'} ({\bf r}),
\label{zcme}
\end{eqnarray} 

\begin{equation}
{\bf P}^\text{Bulk}_{n \sigma n' \sigma'} ({\bf r}) \equiv 
\langle n, \sigma, {\bf 0} \arrowvert \hat{{\bf p}} \arrowvert n', \sigma', {\bf 0} \rangle,
\label{bmme}
\end{equation}
and
\begin{equation}
E^\text{Bulk}_{n \sigma n' \sigma'} ({\bf r}) \equiv 
\langle n, \sigma, {\bf 0} \arrowvert \hat{H} \arrowvert n', \sigma', {\bf 0} \rangle.
\label{beme}
\end{equation}
Note that within the total angular momentum basis states (i.e., zone-center eigenstates given in Table \ref{basis}), the bulk energy term $\underline{\underline{E}}\text{\@}^\text{Bulk}({\bf r})$ in $\underline{\underline{\hat{H}}}\text{\@}^\text{ZC}({\bf r})$ is a diagonal matrix whose diagonal terms are the eigenenergies of the constituent bulk states at zone center and they are
\begin{equation}
E^\text{Bulk}_{n \sigma n' \sigma'}({\bf r}) 
= E^\text{Bulk}_{n \sigma}({\bf r}) \delta_{n n'} \delta_{\sigma \sigma'},
\end{equation}
where $E^\text{Bulk}_{n \uparrow}({\bf r})$ $=$ $E^\text{Bulk}_{n \downarrow}({\bf r})$. 

The presence of strain in superlattices due to lattice mismatch modifies the electronic band structure and can have dramatic effects on the electron spin coherence time. Consequently, it is important to take into account the strain induced change in the electronic band structure of superlattices. The implementation of the lattice-mismatched-induced strain within the framework of generalized ${\bf K{\cdot}p}$ formalism is accomplished by minimizing the strain energy in the superlattice under the condition that the interfaces of the system are pseudomorphic~\cite{vandewalle89}. Consider a superlattice with periodicity $D_{\text{SL}}$ along the $z$-axis. Projecting the full Hamiltonian~(\ref{hamCryT}) onto this fourteen-band restricted basis set with strain effects due to lattice mismatch taken into account, the effective strained ${\bf k} \cdot {\bf p}$ matrix Hamiltonian of the bulk constituents for $(001)$-growth superlattices takes the form of~\cite{olesberg99}
\begin{equation}
\underline{\underline{\hat{H}}}\text{\@}^{\text{Bulk}}= 
\left[
\begin{array}{cc}
\underline{\underline{\hat{H}}}\text{\@}_{\uparrow}
&\underline{\underline{\hat{H}}}\text{\@}_{+} \\

\underline{\underline{\hat{H}}}\text{\@}_{-} 
& \underline{\underline{\hat{H}}}\text{\@}_{\downarrow}
\end{array}
\right].
\label{zcmh}
\end{equation}
The sub-matrices $\underline{\underline{\hat{H}}}\text{\@}_{\uparrow (\downarrow)}$ and $\underline{\underline{\hat{H}}}\text{\@}_{+} $ are given respectively by
\begin{widetext}
\begin{subequations}
\label{allequations}
\begin{equation}
\underline{\underline{\hat{H}}}\text{\@}_{\uparrow (\downarrow)} = 
\left[
\begin{array}{ccccccc}
\hat{\cal E}^\text{Bulk}_{c \uparrow (\downarrow)}
& 0 
& i \sqrt{\frac{2}{3}}P_0 \hat{v}_z 
& i \sqrt{\frac{1}{3}}P_0 \hat{v}_z 
& 0 
& i \sqrt{\frac{2}{3}}P_1 \hat{v}_z 
& i \sqrt{\frac{1}{3}}P_1 \hat{v}_z \\

0 
& \hat{\cal E}^\text{Bulk}_{h \downarrow (\uparrow)}
& 0 
& 0 
& 0 
& \sqrt{\frac{1}{3}} Q \hat{v}_z 
& - \sqrt{\frac{2}{3}} Q \hat{v}_z \\

-i \sqrt{\frac{2}{3}}P_0 \hat{v}_z 
& 0 
& \hat{\cal E}^\text{Bulk}_{l \uparrow (\downarrow)} 
& 0 
& - \sqrt{\frac{1}{3}} Q \hat{v}_z 
& 0 
& 0 \\

-i \sqrt{\frac{1}{3}}P_0 \hat{v}_z 
& 0 
& 0 
& \hat{\cal E}^\text{Bulk}_{s \uparrow (\downarrow)}
& \sqrt{\frac{2}{3}} Q \hat{v}_z 
& 0 
& 0 \\

0 
& 0 
& - \sqrt{\frac{1}{3}} Q \hat{v}_z 
& \sqrt{\frac{2}{3}} Q \hat{v}_z 
& \hat{\cal E}^\text{Bulk}_{\overline{h} \downarrow (\uparrow)} 
& 0 
& 0 \\

- i \sqrt{\frac{2}{3}}P_1 \hat{v}_z 
& \sqrt{\frac{1}{3}} Q \hat{v}_z 
& 0 
& 0 
& 0 
& \hat{\cal E}^\text{Bulk}_{\overline{l} \uparrow (\downarrow)}
& 0 \\

-i \sqrt{\frac{1}{3}}P_1 \hat{v}_z 
& - \sqrt{\frac{2}{3}} Q \hat{v}_z 
& 0 
& 0 
& 0 
& 0 
& \hat{\cal E}^\text{Bulk}_{\overline{s} \uparrow (\downarrow)} 
\end{array}
\right],
\label{zcsmz}
\end{equation}
\begin{equation}
\underline{\underline{\hat{H}}}\text{\@}_{+} = 
\left[
\begin{array}{ccccccc}
0
& i \sqrt{\frac{1}{2}}P_0 \hat{v}_+ 
& i \sqrt{\frac{1}{6}} P_0 \hat{v}_- 
& -i \sqrt{\frac{1}{3}}P_0 \hat{v}_- 
& i \sqrt{\frac{1}{2}}P_1 \hat{v}_+ 
& i \sqrt{\frac{1}{6}}P_1 \hat{v}_- 
& -i \sqrt{\frac{1}{3}}P_1 \hat{v}_- \\

-i \sqrt{\frac{1}{2}}P_0 \hat{v}_+ 
& 0 
& 0 
& 0 
& 0 
& - \sqrt{\frac{1}{3}} Q \hat{v}_- 
& - \sqrt{\frac{1}{6}} Q \hat{v}_- \\

i \sqrt{\frac{1}{6}}P_0 \hat{v}_- 
& 0 
& 0 
& 0 
& - \sqrt{\frac{1}{3}} Q \hat{v}_- 
& 0 
& - \sqrt{\frac{1}{2}} Q \hat{v}_+ \\

-i \sqrt{\frac{1}{3}}Q \hat{v}_- 
& 0 
& 0 
& 0 
& - \sqrt{\frac{1}{6}} Q \hat{v}_- 
& - \sqrt{\frac{1}{2}} Q \hat{v}_+ 
& 0 \\

-i \sqrt{\frac{1}{2}} P_1 \hat{v}_+  
& 0 
& \sqrt{\frac{1}{3}} Q \hat{v}_- 
& \sqrt{\frac{1}{6}} Q \hat{v}_- 
& 0 
& 0 
& 0 \\

i \sqrt{\frac{1}{6}} P_1 \hat{v}_- 
& \sqrt{\frac{1}{3}} Q \hat{v}_- 
& 0 
& \sqrt{\frac{1}{2}} Q \hat{v}_+
& 0 
& 0 
& 0 \\

-i \sqrt{\frac{1}{3}}P_1 \hat{v}_- 
&  \sqrt{\frac{1}{6}} Q \hat{v}_- 
& - \sqrt{\frac{1}{2}} Q \hat{v}_+ 
& 0 
& 0 
& 0 
& 0 
\end{array}
\right],
\label{zcsmpm}
\end{equation}
\end{subequations}
\end{widetext}
where $\hat{v}_z$ $\equiv$ $m^{-1}_{\text{e}}\hat{p}_{z}$, $\hat{v}_{\pm}$ $\equiv$ $m^{-1}_{\text{e}}(\hat{p}_x \pm i \hat{p}_y)$, and $\hat{\cal E}^\text{Bulk}_{n \sigma}$ $\equiv$ $E^\text{Bulk}_{n \sigma}$ $+$ $E^\text{Strain}_{n \sigma}$ $+$ $(2 m_e)^{-1} \hat{\bf p}^2$ and $E^\text{Strain}_{n \sigma}$ describes the energy shifts due to strain effects. The strained energy shifts are given by~\cite{vandewalle89} 
\begin{subequations}
\label{ses}
\begin{equation}
E^\text{Strain}_{c \uparrow (\downarrow)}
= \frac{\delta_{c}}{2},
\label{sesc} 
\end{equation}
\begin{equation}
E^\text{Strain}_{h \downarrow (\uparrow)}
= - \frac{\delta_{v}}{2},
\label{sesh} 
\end{equation}
\begin{equation}
E^\text{Strain}_{l \uparrow (\downarrow)}
= - \frac{\Delta}{2} + \frac{\delta_{v}}{4} 
+ \frac{1}{2} (\Delta^2 + \Delta \delta_{v} + \frac{9}{4} \Delta \delta_{v}^2)^{\frac{1}{2}},
\label{sesl} 
\end{equation}
\begin{equation}
E^\text{Strain}_{s \uparrow (\downarrow)}
= \frac{\Delta}{2} + \frac{\delta_{v}}{4} 
- \frac{1}{2} (\Delta^2 + \Delta \delta_{v} + \frac{9}{4} \Delta \delta_{v}^2)^{\frac{1}{2}},
\label{sess} 
\end{equation}
\end{subequations}
with
\begin{subequations}
\label{set}
\begin{equation}
\delta_{c}
= 2 V_{a} (2 \epsilon_\parallel + \epsilon_\perp),
\label{set1} 
\end{equation}
\begin{equation}
\delta_{v}
= 2 V_{b} (\epsilon_\parallel - \epsilon_\perp),
\label{set2} 
\end{equation}
\begin{equation}
\epsilon_\parallel
= \frac{a_\parallel}{a_{\text{o}}} - 1,
\label{set3} 
\end{equation}
\begin{equation}
\epsilon_\perp
= \frac{a_\perp}{a_{\text{o}}} - 1,
\label{set4} 
\end{equation}
\begin{equation}
a_\perp
= a_{\text{o}} \left[ 1 - 2 \frac{c_{12}}{c_{11}} 
\left( \frac{a_\parallel}{a_{\text{o}}} - 1 \right) \right],
\label{set5} 
\end{equation}
\end{subequations}
where $\epsilon_\parallel$ and $\epsilon_\perp$ are components of the strain tensor describing the in-plane and growth-axis strains of the superlattice, respectively, $a_{\text{o}}$, $a_\parallel$, and $a_\perp$ are the unstrained, in-plane strained, and growth-axis strained lattice constants of the constituent semiconductors, respectively, $c_{11}$ and $c_{12}$ are the elastic constants, and $V_{a}$ and $V_{b}$ are the hydrostatic and biaxial deformation potentials, respectively. The in-plane strained lattice constant $a_\parallel$ $=$ $a_{\text{sub}}$ where $a_{\text{sub}}$ is the lattice constant of the substrate. The energy shifts of the conduction bands are due to hydrostatic strain, while the energy shifts of the valence bands are due to biaxial compression [c.f., Eq.~(\ref{set1}) and Eq.~(\ref{set2})]. In the present treatment, the strain effects on the $\overline{p}$ states and the momentum matrix elements $P_{0}$, $P_{1}$, and $Q$ are neglected.

Note that $\underline{\underline{\hat{H}}}\text{\@}_{\uparrow}$ $=$ $\underline{\underline{\hat{H}}}\text{\@}_{\downarrow}$, i.e., the opposite-spin eigenstates of the constituent bulk states at zone center are two-fold degenerate, and $\underline{\underline{\hat{H}}}\text{\@}_{-}$ $=$ $\underline{\underline{\hat{H}}}\text{\@}_{+}^{\dagger}$. From Eqs.~(\ref{zcmh}) and (\ref{allequations}) it is easy to see that the zone-center matrix Hamiltonian $\underline{\underline{\hat{H}}}\text{\@}^\text{ZC}({\bf r})$ is block diagonal; therefore, the Bloch states $\arrowvert n, \uparrow, {\bf 0} \rangle$ and $\arrowvert n, \downarrow, {\bf 0} \rangle$ are not mixed at zone center (see Table~\ref{basis}). The zone-center eigenvalue problem [Eq.~(\ref{zcse})] is solved in Fourier space in a similar method to that of Winkler and R$\ddot{\rm o}$ssler \cite{winkler93}. The spatially dependent terms in Eq.~(\ref{zcse}) can be expanded in terms of Fourier series as \cite{winkler93, olesberg99}
\begin{equation}
\hat{H}^\text{ZC}_{n \sigma n' \sigma'}(z) 
= \sum_{\mu = - \infty}^{\infty} \widetilde{\hat{H}}\text{\@}^\text{ZC}_{n \sigma n' \sigma' \mu} \exp(i \mu q z),
\label{fszcm}
\end{equation}
\begin{equation}
F_{N S n \sigma} (z) 
= \sum_{\mu = - \infty}^{\infty} \widetilde{F}_{N S n \sigma \mu} \exp(i \mu q z),
\label{fszcwf}
\end{equation}
where $q$ $\equiv$ $2 \pi / D_{\text{SL}}$.  By substituting Eqs.~(\ref{fszcm}) and (\ref{fszcwf}) into (\ref{zcse}), Multiplying from the left-hand side of the resulting equation by $\exp(- i 2 \pi \nu q z)$, and integrating over the superlattice period, we obtain the zone-center superlattice eigenvalue equation in Fourier space as 
\begin{equation}
\sum_{n' \sigma' \mu} \widetilde{\hat{H}}\text{\@}^\text{ZC}_{n \sigma n' \sigma' (\nu - \mu)} 
\widetilde{F}_{N S n' \sigma' \mu}  
= E_{N S} ({\bf 0}) \widetilde{F}_{N S n \sigma \nu},
\label{fszcse}
\end{equation}
where $\widetilde{H}\text{\@}^\text{ZC}_{n \sigma n' \sigma' \mu}$ is given by the Fourier series expansion
\begin{equation}
\widetilde{\hat{H}}\text{\@}^\text{ZC}_{n \sigma n' \sigma' \mu} 
= \int_{D_\text{SL}} dz \hat{H}^\text{ZC}_{n \sigma n' \sigma'}(z) \exp(- i \mu q z).
\label{fszch}
\end{equation}
Diagonalizing the matrix $\underline{\underline{\widetilde{\hat{H}}}}\text{\@}^\text{ZC}$ in Eq.~(\ref{fszcse}) yields the zone-center eigenvalues and the corresponding eigenvectors. Once the eigenvalues and eigenvectors are found, it is desirable to characterize the zone-center states in terms of the fourteen bulk characters designated by $c_{\uparrow}$, $h_{\downarrow}$, $l_{\uparrow}$, $so_{\uparrow}$, $\overline{h}_{\downarrow}$, $\overline{l}_{\uparrow}$,
$\overline{so}_{\uparrow}$, $c_{\downarrow}$, $h_{\uparrow}$, $l_{\downarrow}$, $so_{\downarrow}$, $\overline{h}_{\uparrow}$, $\overline{l}_{\downarrow}$, and $\overline{so}_{\downarrow}$ (see Table~\ref{basis}). Noting that 
\begin{eqnarray} 
& & \int_{V_\text{SL}} d {\bf r}
\langle N, S, {\bf 0} \arrowvert {\bf r} \rangle
\langle {\bf r} \arrowvert N, S, {\bf 0} \rangle \nonumber \\
&& =  \sum_{n \sigma}  \int_{V_\text{SL}} d {\bf r}
F^{\ast}_{N S n \sigma} ({\bf r}) F_{N S n \sigma} ({\bf r}) \nonumber \\
&& \times \langle n, \sigma, {\bf 0} \arrowvert {\bf r} \rangle
\langle {\bf r} \arrowvert n, \sigma, {\bf 0} \rangle,
\end{eqnarray} 
the fraction of each bulk character of the zone-center superlattice states is given by
\begin{equation}
\int_{V_\text{SL}} d {\bf r}
F^{\ast}_{N S n \sigma} ({\bf r}) F_{N S n \sigma} ({\bf r})
= \sum_{\mu} 
\left| 
\widetilde{F}_{N S n \sigma \mu}
\right|{^2},
\label{zcschar}
\end{equation}
where $V_{\text{SL}}$ is the volume of the superlattice unit cell.
The equality in Eq.~(\ref{zcschar}) is obtained by using Parseval's theorem for the discrete Fourier transform to replace the integral on the left-hand side by the sum over the norm of the Fourier cofficients on the right-hand side. Therefore the superlattice states at zone center can be classified according to the bulk character of their largest vector component, and the character $\chi^\text{ZC}_{N} $ of the zone-center superlattice states is given by 
\begin{equation}
\chi^\text{ZC}_{N S} 
= \max \left [ \sum_{\mu} \left| \widetilde{F}_{N S n \sigma \mu} \right|{^2} \right ],
\label{zcsbulkchar}
\end{equation}
where $\max [A_{n \sigma}]$ stands for the maximum of the set $\{A_{n \sigma}\}$. 
\subsection{SUPERLATTICE STATES AT FINITE ${\bf K}$}
\label{sec:sls}
Once the zone-center superlattice eigenstates and eigenenergies have been determined, the superlattice states at ${\bf K} \neq 0$ can be obtained by application of superlattice ${\bf K} \cdot {\bf p}$ theory~\cite{johnson90}. A superlattice state $\arrowvert {\cal L}, {\cal S}, {\bf K} \rangle$ at finite ${\bf K}$ is expressed in terms of the zone-center superlattice states
\begin{equation}
\langle {\bf r} \arrowvert {\cal L}, {\cal S}, {\bf K} \rangle 
= \exp(i {\bf K \cdot r}) \sum_{N S} C_{{\cal L S} N S} ({\bf K}) 
\langle {\bf r} \arrowvert N, S, {\bf 0} \rangle,  
\label{swf}
\end{equation}
where $C_{{\cal L S} N S} ({\bf K})$ are the expansion coefficients and $S$ $\in$ $\{\uparrow, \downarrow\}$ is the spin indices. Note that the superlattice in-plane wavevector ${\bf K_{\|}} \equiv (K_{x}, K_{y})$ is equivalent to the bulk wavevector ${\bf k_{\|}}$ $\equiv$ $(k_{x}, k_{y})$ in the in-plane direction due to in-plane translational invariance in one-dimensional superlattices, i.e., ${\bf k_{\|}}$ is a good quantum number.

By inserting Eq.~(\ref{swf}) into (\ref{sch}), Multiplying the resulting equation on the left-hand side by $\exp({\bf -K \cdot r}) \langle N, S, {\bf 0} \arrowvert {\bf r} \rangle$ and integrating over a superlattice unit cell, we have
\begin{equation}
\sum_{N' S'} 
H_{N S N' S'}^\text{SL}({\bf K}) 
C_{{\cal L S} N' S'} ({\bf K}) 
=  E_{\cal L S} ({\bf K})  C_{{\cal L S} N S} ({\bf K})
\label{sse}
\end{equation}
with 
\begin{eqnarray}
H_{N S N' S'}^\text{SL}({\bf K}) 
&\equiv &  \left(\frac{\hbar^2 K^2}{2m_{e}} 
+ E_{N S N' S'}({\bf 0}) \right) \delta_{N N'} \delta_{S S'} \nonumber \\
&+& \frac{\hbar}{m_{e}} {\bf K} 
\cdot {\bf P}_{N S N' S'} ({\bf 0}),
\label{ssme}
\end{eqnarray}
where the zone-center matrix elements of momentum and energy are respectively given by 
\begin{equation}
{\bf P}_{N S N' S'} ({\bf 0}) \equiv 
\langle N, S, {\bf 0} \arrowvert \hat{{\bf p}} 
\arrowvert N', S', {\bf 0} \rangle,
\label{smme}
\end{equation}
\begin{equation}
E_{N S N' S'} ({\bf 0}) \equiv 
\langle N, S, {\bf 0} \arrowvert 
\hat{H}\arrowvert N', S', {\bf 0} \rangle.
\label{seme}
\end{equation}
Making use of Eq.~(\ref{zcs}), the momentum matrix elements connecting the zone-center superlattice states can be expressed as
\begin{eqnarray}
&&\langle N, S, {\bf 0} 
\arrowvert \hat{{\bf p}} \arrowvert N', S', {\bf 0} \rangle \nonumber \\ 
&&=  \frac{1}{V_{\text{SL}}} \sum_{n \sigma n' \sigma'} \int_{V_\text{SL}} d {\bf r}
\Big[ \langle n, \sigma, {\bf 0} \arrowvert F^{\ast}_{N S n \sigma} ({\bf r}) \Big] 
\hat{{\bf p}} \nonumber \\ 
&& \times \Big[ F_{N S n' \sigma'} ({\bf r})\arrowvert n', \sigma', {\bf 0} \rangle \Big] \nonumber \\ 
&&=  \frac{1}{V_{\text{SL}}} \sum_{n \sigma n' \sigma'} \int_{V_\text{SL}} d {\bf r}
\Big[ F^{\ast}_{N S n \sigma} ({\bf r}) 
F_{N S n' \sigma'} ({\bf r})  \nonumber \\
&& \times  \langle n, \sigma, {\bf 0} \arrowvert \hat{{\bf p}} 
\arrowvert n', \sigma', {\bf 0} \rangle \Big]  \nonumber \\ 
&&+  \frac{1}{V_{\text{SL}}} \sum_{n \sigma n' \sigma'} \int_{V_\text{SL}} d {\bf r}
\Big[ F^{\ast}_{N S n \sigma} ({\bf r}) {\bf p} 
F_{N S n' \sigma'} ({\bf r}) \nonumber \\ 
&& \times \langle n, \sigma, {\bf 0} \arrowvert n', \sigma', {\bf 0} \rangle \Big],
\label{zcmme} 
\end{eqnarray}
The first term in Eq.~(\ref{zcmme}) is evaluated in real space as its integrand is the product of three spatially dependent functions. The evaluation of the second term is performed in Fourier space to circumvent the problem of envelope function variation on scales much smaller than the bulk lattice constant. Using the relaxation in Eq.~(\ref{fszcwf}), the second term is simplified to
\begin{eqnarray}
&& \frac{1}{V_{\text{SL}}} \sum_{n \sigma n' \sigma'} \int_{V_\text{SL}} d {\bf r}
\Big[ F^{\ast}_{N S n \sigma} ({\bf r}) {\bf p} 
F_{N S n' \sigma'} ({\bf r}) \nonumber \\
&& \times \langle n, \sigma, {\bf 0} \arrowvert n', \sigma', {\bf 0} \rangle \Big] \nonumber \\ 
&& = \hbar q \sum_{n \sigma \mu} \mu \widetilde{F}^{\ast}_{N S n \sigma \mu} 
\widetilde{F}_{N S n \sigma \mu}. 
\label{zcme2}
\end{eqnarray}
Note that the first term in Eq.~(\ref{zcmme}) is larger than the second term by a factor of $m_e/m_c$, where $m_c$ is the zone-center conduction-band mass \cite{johnson90}. 
\subsection{EVALUATION OF SPIN PRECESSION VECTOR}
\label{sec:spv}
To calculate the spin precession vector, it is convenient to introduce the pseudospin state $\arrowvert \hat{{\bf n}}; {\cal L}, {\cal S}, {\bf K} \rangle$ at finite momentum ${\bf K}$ with spin projection along the direction $\hat{\bf n}$ parallel to the growth direction $\hat{{\bf z}}$ (i.e., with quantization axis along the growth direction), 
\begin{equation}
\arrowvert \hat{{\bf n}}; {\cal L}, {\cal S}, {\bf K} \rangle 
= \sum_{N S} D_{{\cal L S} N S} ({\bf K}) \delta_{{\cal S} S}\arrowvert N, S, {\bf 0} \rangle.
\label{pss}
\end{equation}
 Any superlattice energy eigenstates $\arrowvert {\cal L}, {\cal S}, {\bf K} \rangle$ can be expressed as a linear superposition of the pseudospin-up $\arrowvert \hat{{\bf n}}; {\cal L}, \uparrow, {\bf K} \rangle$ and pseudospin-down $\arrowvert \hat{{\bf n}}; {\cal L}, \downarrow, {\bf K} \rangle$ states and vice versa
\begin{equation}
\arrowvert \hat{{\bf n}}; {\cal L}, {\cal S}, {\bf K} \rangle 
= A_{\cal S} \arrowvert {\cal L}, \uparrow, {\bf K} \rangle
+ B_{\cal S} \arrowvert {\cal L}, \downarrow, {\bf K} \rangle
\label{scs}
\end{equation}
with
\begin{subequations}
\label{ssc}
\begin{eqnarray}
A_{\uparrow} = B_{\downarrow}^{\ast} \equiv \cos \left( \frac{\Theta}{2} \right) \exp \left( \frac{-i \Phi}{2} \right),
\label{ssca}
\end{eqnarray}
\begin{eqnarray}
B_{\uparrow} = - A_{\downarrow}^{\ast} \equiv \sin \left( \frac{\Theta}{2} \right) \exp \left( \frac{i \Phi}{2} \right),
\label{sscb}
\end{eqnarray}
\end{subequations}
where $\Theta$ and $\Phi$ are respectively the polar and azimuthal angles that characterize the unit vector $\hat{{\bf H}}({\cal L}, {\bf K})$ of the momentum-dependent effective magnetic field in the $xyz$-coordinate system. The electron spin precession frequency (Larmor frequency) is proportional to the spin splitting of the superlattice eigenstates and it is given by
\begin{equation}
\Omega({\cal L}, {\bf K}) = \hbar^{-1} \Big| E_{{\cal L} \uparrow} ({\bf K}) - E_{{\cal L} \downarrow} ({\bf K}) \Big|.  
\label{psf}
\end{equation}
From Eqs.~(\ref{pss}) and (\ref{scs}), we obtain the following relations:
\begin{subequations}
\label{vsc}
\begin{equation}
\frac{1}{2} \exp(i \Phi) \sin(\Theta)
= \sum_{N} C_{{\cal L} \downarrow N \uparrow}^{\ast} ({\bf K}) C_{{\cal L} \uparrow N \uparrow} ({\bf K}),
\label{vsca}
\end{equation}
\begin{equation}
\cos^{2} \left( \frac{\Theta}{2} \right)
= \sum_{N} C_{{\cal L} \uparrow N \uparrow}^{\ast} ({\bf K}) C_{{\cal L} \uparrow N \uparrow} ({\bf K}),
\label{vscb}
\end{equation}
\begin{equation}
\sin^{2} \left( \frac{\Theta}{2} \right)
= \sum_{N} C_{{\cal L} \downarrow N \uparrow}^{\ast} ({\bf K}) C_{{\cal L} \downarrow N \uparrow} ({\bf K}).
\label{vscc}
\end{equation}
\end{subequations}
From Eq.~(\ref{vsc}), the trigonometric functions of the spherical angles $\left(\Theta, \Phi \right)$ for the $xyz$-components of ${\bf \Omega}({\cal L}, {\bf K})$ are respectively written as
\begin{subequations}
\label{pf}
\begin{equation}
\sin(\Theta) \cos(\Phi)
= \text{Re} \left[ 2 \sum_{N} C_{{\cal L} \downarrow N \uparrow}^{\ast} ({\bf K}) 
C_{{\cal L} \uparrow N \uparrow} ({\bf K}) \right],
\label{pfx}
\end{equation}
\begin{equation}
\sin(\Theta) \sin(\Phi)
= \text{Im} \left[ 2 \sum_{N} C_{{\cal L} \downarrow N \uparrow}^{\ast} ({\bf K}) 
C_{{\cal L} \uparrow N \uparrow} ({\bf K}) \right],
\label{pfy}
\end{equation}
\begin{equation}
\cos(\Theta)
= \sum_{N} \Big[ \left| C_{{\cal L} \uparrow N \uparrow} ({\bf K}) \right|^{2} 
- \left[ C_{{\cal L} \downarrow N \uparrow} ({\bf K}) \right|^{2} \Big],
\label{pfz}
\end{equation}
\end{subequations}
where $\text{Re} [A]$ and $\text{Im} [A]$, respectively, stands for the real and the imaginary parts of $A$.

Taking into account the cubic symmetry of the zincblende constituent semiconductors, the evaluation of the precession vector ${\bf \Omega}({\cal L}, {\bf K})$ for $(001)$-growth superlattices can be simplified. The symmetry elements pertaining to $(001)$-oriented structures are: One fourfold rotoinversion axis ($K_z \| [001]$), two twofold rotation axes ($K_x \| [100]$ and $K_y \| [010]$), and two reflection planes containing the $[001]$ axis and  bisecting the angles between the $[100]$ and $[010]$ axes \cite{smith90}. All eigenstates with wavevector ${\bf K}$ directed along the growth direction are twofold degenerate, and hence there are no spin splittings along this direction. Therefore, the evaluation of the in-plane precession vector ${\bf \Omega}_{\|}({\cal L}, {\bf K})$ can be simplified. One needs only evaluate $\Omega_x ({\cal L}, {\bf K})$ and $\Omega_y ({\cal L}, {\bf K})$ over one eighth of the $K_x$--$K_y$ plane (e.g., in the domain $\Phi = [0, \pi / 4]$), and the rest can be obtained by symmetry consideration. This is accomplished by breaking the in-plane precession vector into eight quadrants on the $K_x$--$K_y$ plane and by mapping each quadrant back into the first quadrant using the symmetry properties of the structure. The symmetry operations of the in-plane vector ${\bf \Omega}_{\|}({\cal L}, {\bf K})$ are summarized in Table \ref{toxy}.
\section{RESULTS AND DISCUSSION}
\label{sec:rd}
In this section the electron spin decoherence in bulk, quantum-well, and superlattice zincblende semiconductors is studied by examining the electron spin coherence times $T_1$ and $T_2$ using the fourteen-band non-perturbative theory developed in the previous sections. These theoretical calculations are compared with available experimental results. The fourteen-band ${\bf k} \cdot {\bf p}$ parameters of the material systems used in these calculations are summarized in Tables~\ref{materials1}, \ref{materials2}, and \ref{vboffset}~\cite{madelung96, landolt87}. 
\subsection{SPIN COHERENCE TIMES IN BULK ZINCBLENDE SEMICONDUCTORS}
\label{sec:bzs}
The understanding of electron spin dynamics in bulk semiconductors is essential because they are the basic building blocks of spin-dependent devices. To explore the electron spin properties of bulk zincblende semiconductors, we calculate the electron spin coherence times using Eqs.~(\ref{t1}) and (\ref{t2}) derived in Sec.~\ref{sec:t1t2}. All the calculations of electron spin coherence time for bulk semiconductors are performed assuming ionized impurity scattering unless otherwise noted.
\subsubsection{III-V SEMICONDUCTORS}
\label{sec:bulk35}
\begin{figure}[htbp]
\setcaptionwidth{8.0 cm}
\scalebox{1}{\includegraphics[width = 8.0 cm]{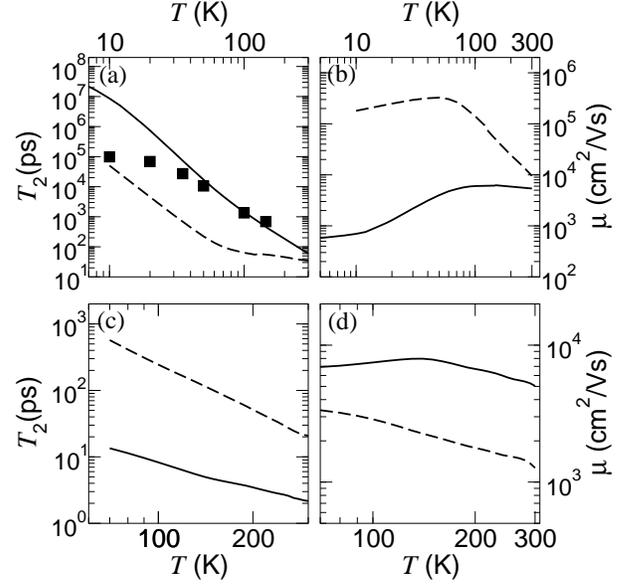}}
\caption[]{Temperature dependence of mobility and coherence time in bulk III-V semiconductors. (a) $T_2$ as a function of temperature for GaAs with $n=1.0 \times 10^{16}$ cm$^{-3}$ (solid line\cite{lau01}) and $n=1.0 \times 10^{13}$ cm$^{-3}$ (dashed line), and squares represent the results of experiments~\cite{kikkawa98}. (b) $\mu$ vs $T$ for GaAs with $n=1.0 \times 10^{16}$ cm$^{-3}$ (solid line) and $n=1.0 \times 10^{13}$ cm$^{-3}$ (dashed line)~\cite{wolfe71}. (c) $T_2$ as a function of temperature for GaSb with $n=2.0 \times 10^{18}$ cm$^{-3}$ (solid line) and $n=4.0 \times 10^{16}$ cm$^{-3}$ (dashed line). (d) $\mu$ vs $T$ for GaSb with $n=2.0 \times 10^{18}$ cm$^{-3}$ (solid line) and $n=4.0 \times 10^{16}$ cm$^{-3}$ (dashed line)~\cite{chen91}.}
\label{BulkGaAsGaSb}
\end{figure}

The electron spin coherence times $T_{2}$'s as a function of temperature $T$ for GaAs are shown in Fig.~\ref{BulkGaAsGaSb}(a), and the measured mobilities~\cite{wolfe71} are displayed in Fig.~\ref{BulkGaAsGaSb}(b). Squares represent the experimental results by Kikkawa and Awschalom~\cite{kikkawa98}. The agreement with experimental measurements for GaAs at the higher temperatures is good\cite{lau01}, whereas for low temperatures other spin relaxation mechanisms, namely the Elliot-Yafet (EY) mechanism, may dominate~\cite{meier84}. At room temperature, the calculated spin coherence time $T_{2}$ is approximately $216$ ps with an electron mobility of $5.4 \times 10^{3}$ cm$^{2}$/Vs. From Fig.~\ref{BulkGaAsGaSb}(a) and \ref{BulkGaAsGaSb}(b) the electron spin coherence times (both $T_{1}$ and $T_{2}$) are inversely proportional to the electron mobility [see Eqs.~(\ref{t1}) and (\ref{t2})], and hence the electron spin coherence times are inversely proportional to the momentum scattering times of the electrons. This implies that cleaner samples have shorter electron spin coherence times. For semiconductors with cubic symmetry, $T_{1}$ $=$ $T_{2}$ (see Sec.~\ref{sec:t1t2}).

\begin{figure}[htbp]
\setcaptionwidth{8.0 cm}
\scalebox{1}{\includegraphics[width= 8.0 cm]{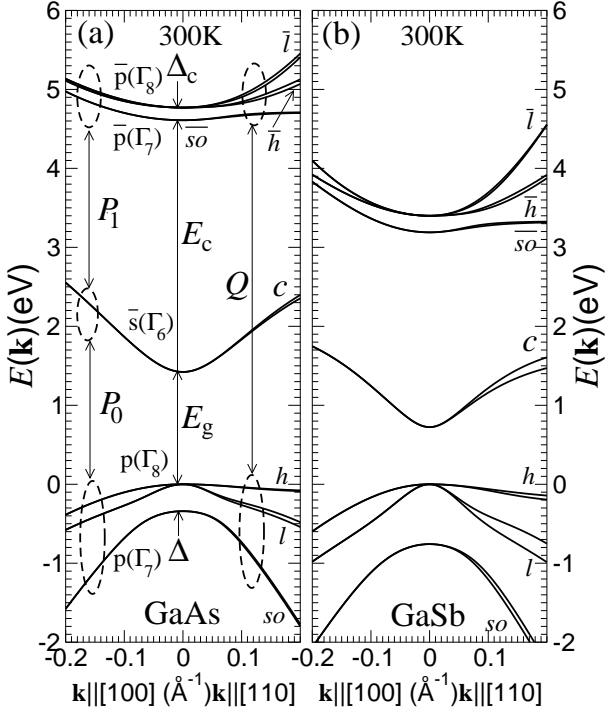}}
\caption[]{The fourteen-band electronic band structure of bulk III-V semiconductors near the $\Gamma$ point of the Brilliouin zone at 300K. (a) GaAs. (b) GaSb. The fourteen bands considered are six $p$ states, two $\overline{s}$ states, and six $\overline{p}$ states ($\overline{s}$ indicates antibonding $s$ states and $\overline{p}$ indicates antibonding $p$ states). The $\overline{s}-p$, $\overline{s}-\overline{p}$, and $p-\overline{p}$ interband interactions are characterized by the corresponding interband momentum matrix elements $P_{0}$, $P_{1}$, and $Q$, respectively. The energy parameters $E_{\text{g}}$, $E_{c}$, $\Delta$, and $\Delta_{c}$ are the energy difference of $\overline{s} \left(\Gamma_{6} \right)-p \left(\Gamma_{8} \right)$, $\overline{p} \left(\Gamma_{7} \right)-p \left(\Gamma_{8} \right)$, $p \left(\Gamma_{8} \right)-p \left(\Gamma_{7} \right)$, and $\overline{p} \left(\Gamma_{8} \right)-\overline{p} \left(\Gamma_{7} \right)$, respectively.}
\label{BulkGaAsGaSbBands}
\end{figure}

Shown in Fig.~\ref{BulkGaAsGaSb}(c) are $T_{2}$'s as a function of temperature for GaSb and the measured mobilities~\cite{chen91} are shown in Fig.~\ref{BulkGaAsGaSb}(d). The shorter spin coherence times for GaSb are due partly to the larger conduction-band spin splitting $\Delta E$, which originates from a larger ratio $R$ of the spin-orbit coupling $\Delta$ to the band gap $E_g$ for GaSb (see Ref.~\onlinecite{cardona88} for perturbative expansions of spin splittings). The band structures of GaAs and GaSb at $300$ K are shown in Figs.~\ref{BulkGaAsGaSbBands}(a)--(b), and $R_{\text{GaAs}}$~$=$~$0.24$ and $R_{\text{GaSb}}$~$=$~$1.05$. 

\begin{figure}[htbp]
\setcaptionwidth{8.0 cm}
\scalebox{1}{\includegraphics[width = 8.0 cm]{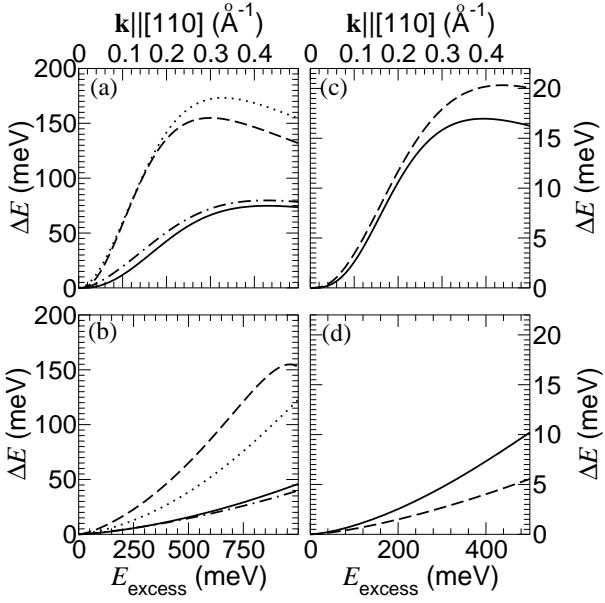}}
\caption[]{Spin splitting of the lowest conduction band for bulk III-V semiconductors at room temperature. (a) ${\Delta E}$ vs ${\bf k}_{[110]}$ and (b) ${\Delta E}$ vs $E_{\text{excess}}$ for GaAs (solid line), GaSb (dashed line), InAs (dot-dashed line), and InSb (dotted line). (c) ${\Delta E}$ vs ${\bf k}_{[110]}$ and (d) ${\Delta E}$ vs $E_{\text{excess}}$ for GaP (solid line) and InP (dashed line).}
\label{BulkDeltaE}
\end{figure}

For bulk zincblende crystals, all the electronic states at a given ${\bf k}$ are doubly degenerate along $[100]$ directions [see Figs.~\ref{BulkGaAsGaSbBands}(a)--(b) for ${\bf k}||[100]$], and the spin splitting along $[111]$ directions is usually very small (not shown) while the spin splitting in $[110]$ directions (see Figs.~\ref{BulkGaAsGaSbBands}(a)--(b) for ${\bf k}||[110]$) is the largest~\cite{cardona88}. Figure~\ref{BulkDeltaE}(a) shows the lowest conduction-band spin splitting as a function of crystal momentum (${\bf k }||[110]$) at $300$ K, and Fig.~\ref{BulkDeltaE}(b) displays the electron spin splitting as a function of excess energy for ${\bf k}$ in the same direction as in Fig.~\ref{BulkDeltaE}(a). For small momenta, $\Delta E$ $\propto$ $k^{3}$.  

\begin{figure}[htbp]
\setcaptionwidth{8.0 cm}
\scalebox{1}{\includegraphics[width = 8.0 cm]{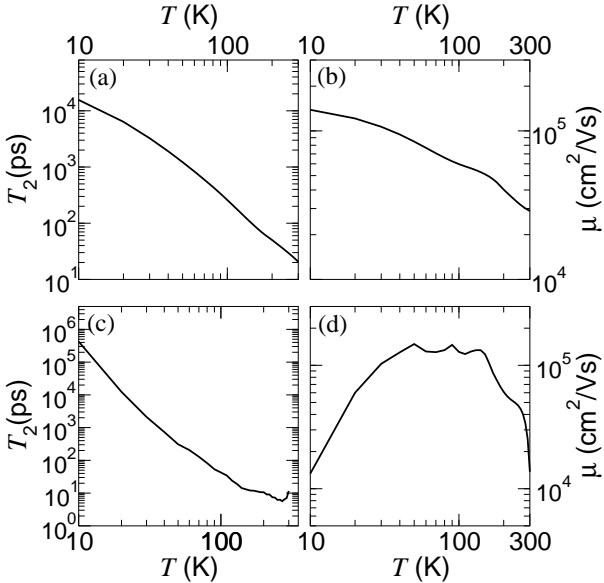}}
\caption[]{Temperature dependence of mobility and coherence time in InAs and InSb. (a) $T_2$ as a function of temperature for InAs with $n=1.7 \times 10^{16}$ cm$^{-3}$. (b) $\mu$ vs $T$ for InAs~\cite{harman56}. (c) $T_2$ as a function of temperature for InSb with $n=1.0 \times 10^{14}$ cm$^{-3}$. (d) $\mu$ vs $T$ for InSb~\cite{trifonov71}.}
\label{BulkInAsInSb}
\end{figure}

Figures~\ref{BulkInAsInSb}(a) and \ref{BulkInAsInSb}(b) respectively show the calculated electron spin coherence time and measured mobility~\cite{harman56} as a function of temperature for bulk InAs.  At $300$ K, $T_{2}$ is $20.6$ ps for $\mu$ $=$ $2.89 \times 10^{4}$ cm$^{2}$/Vs, and it is one order in magnitude shorter than the electron spin coherence time of GaAs. This is consistent with the recent measurements of $n$-type InAs at room temperature by Boggess {\it et al.}~\cite{boggess00}. The measured mobility of the sample in Ref.~\onlinecite{boggess00} is $2.6 \times 10^{4}$ cm$^{2}$/Vs and the electron density is approximately $2 \times 10^{16}$ cm$^{-3}$. The measured $T_{1}$ is $19 \pm 4$ ps which is in excellent agreement with the calculated $T_{1}$ of $21$ ps.

\begin{figure}[htbp]
\setcaptionwidth{8.0 cm}
\scalebox{1}{\includegraphics[width= 8.0 cm]{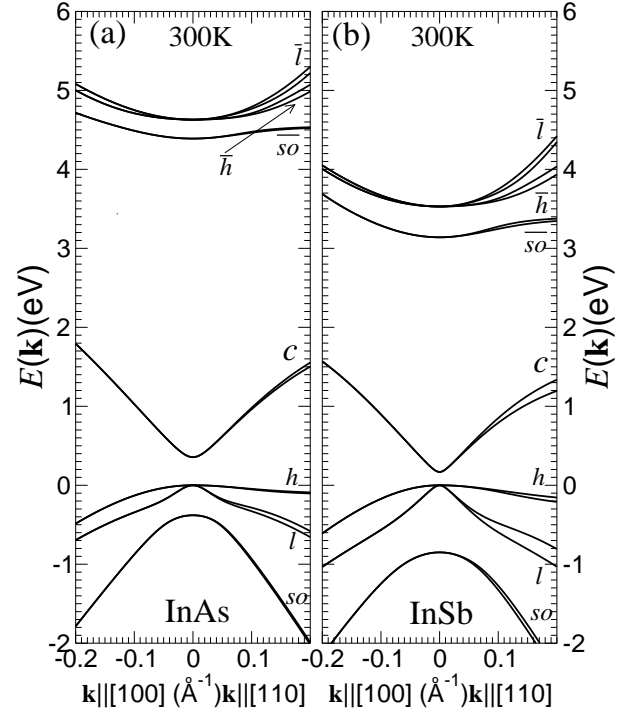}}
\caption[]{The fourteen-band electronic band structure near the $\Gamma$ point of the Brilliouin zone at 300K. (a) InAs. (b) InSb. }
\label{BulkInAsInSbBands}
\end{figure}

Shown in Figs.~\ref{BulkInAsInSb}(c) and \ref{BulkInAsInSb}(d) are the calculated electron spin coherence time and measured mobility~\cite{trifonov71} as a function of temperature for bulk InSb, respectively. The calculated $T_{2}$ is $11.6$ ps at $300$ K, and the measured mobility is $1.37 \times 10^{4}$ cm$^{2}$/Vs. The electron spin coherence time of InSb is shorter than that of InAs at $300$ K due to the larger conduction spin splitting in InSb [see Figs.~\ref{BulkDeltaE}(a) and \ref{BulkDeltaE}(b)]. Note that $T_{1}$  and $T_{2}$ scale approximately as $\mu^{-1}$. The band structures of InAs and InSb in our model are shown in Figs.~\ref{BulkInAsInSbBands}(a)--(b). The ratio of the spin-orbit coupling to the band gap $R_{\text{InAs}}$ $=$ $1.07$ for InAs and $R_{\text{InSb}}$ $=$ $5.04$ for InSb.

\begin{figure}[htbp]
\setcaptionwidth{8.0 cm}
\scalebox{1}{\includegraphics[width = 8.0 cm]{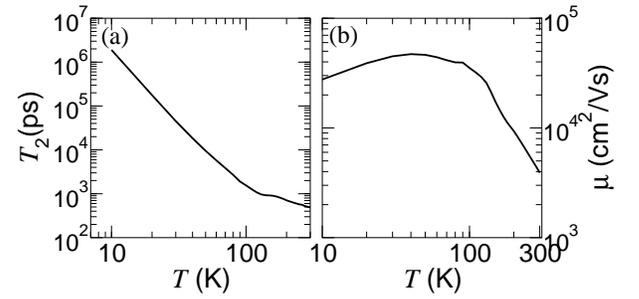}}
\caption[]{Temperature dependence of mobility and coherence time in InP. 
(a) $T_2$ as a function of temperature~\cite{rode75} for $n=2.0 \times 10^{15}$ cm$^{-3}$. (b) $\mu$ vs $T$.}
\label{BulkInPGaP}
\end{figure}

\begin{figure}[htbp]
\setcaptionwidth{8.0 cm}
\scalebox{1}{\includegraphics[width= 8.0 cm]{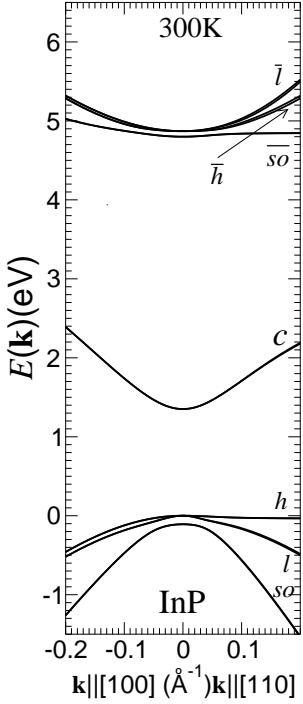}}
\caption[]{The fourteen-band electronic band structure near the $\Gamma$ point of the Brilliouin zone at 300K for InP.}
\label{BulkInPGaPBands}
\end{figure}

The calculated electron spin coherence time as a function of temperature for bulk InP is shown in Fig.~\ref{BulkInPGaP}(a) and the temperature-dependence mobility~\cite{rode75} from experiments is depicted in Fig.~\ref{BulkInPGaP}(b). The calculated electron spin coherence time $T_{2}$ is $0.49$ ns at $300$ K for $\mu$ $=$ $3.90 \times 10^{3}$ cm$^{2}$/Vs with $R_{\text{InP}}$ $=$ $0.08$ [see Fig.~\ref{BulkInPGaPBands}]. The longer electron spin coherence times in InP come from the smaller conduction spin splittings [see Figs.~\ref{BulkDeltaE} (c) and (d)].
\subsubsection{II-VI SEMICONDUCTORS}
\label{sec:bulk26}
Recently II-VI zincblende semiconductors have also gained attention due to their long electron spin coherence times and transport properties. It has been experimentally demonstrated by Malajovich {\it et al.} that spin information can be protected by transport from a region of high spin decoherence (GaAs) to a region (ZnSe) of low spin decoherence~\cite{malajovich00}. Here we calculate the electron spin coherence times of bulk II-VI semiconductors and we will look at the electron spin coherence times of II-VI quantum-well semiconductors in the next section. 

\begin{figure}[htbp]
\setcaptionwidth{8.0 cm}
\scalebox{1}{\includegraphics[width = 8.0 cm]{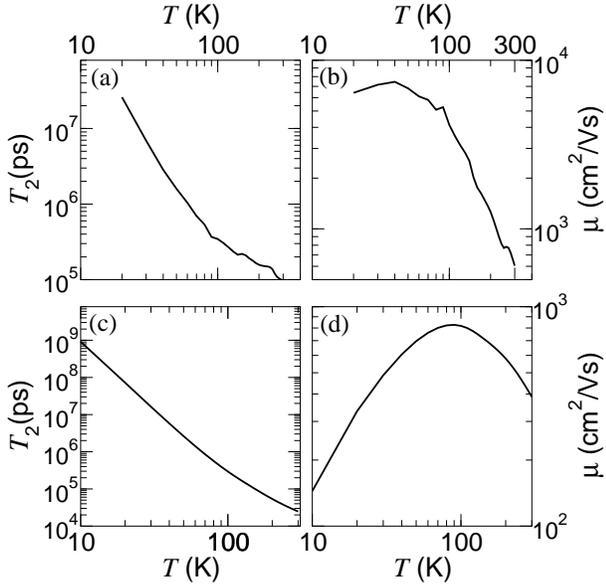}}
\caption[]{Temperature dependence of mobility and coherence time in bulk II-VI semiconductors. 
(a) $T_2$ as a function of temperature for CdSe with $n=4.0 \times 10^{14}$ cm$^{-3}$. (b) $\mu$ vs $T$ for CdSe~\cite{rode70}. (c) $T_2$ as a function of temperature for ZnSe with $n=5.0 \times 10^{15}$ cm$^{-3}$. (d) $\mu$ vs $T$ for ZnSe~\cite{ohkawa87}.}
\label{BulkCdSeZnSe}
\end{figure}

\begin{figure}[htbp]
\setcaptionwidth{8.0 cm}
\scalebox{1}{\includegraphics[width= 8.0 cm]{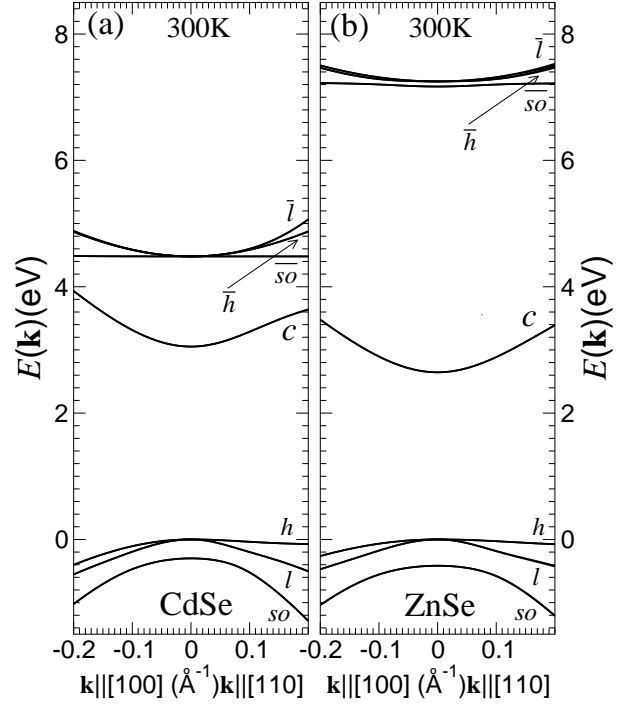}}
\caption[]{The fourteen-band electronic band structure of bulk II-VI semiconductors near the $\Gamma$ point of the Brilliouin zone at 300K. (a) CdSe. (b) ZnSe.}
\label{BulkCdSeZnSeBands}
\end{figure}

Figures~\ref{BulkCdSeZnSe}(a) and \ref{BulkCdSeZnSe}(b) show respectively the electron spin coherence time $T_{2}$ and measured mobility as a function of temperature for bulk CdSe. Due to the relatively small spin-orbit coupling (see Table~\ref{materials2}), the calculated electron spin coherence time is very long and $T_{2}$ $=$ $96$ ns at $300$ K with $\mu$ $=$ $600$ cm$^{2}$/Vs. For bulk ZnSe, the calculated electron spin coherence time $T_{2}$ $=$ $25$ ns at $300$ K with $\mu$ $=$ $388$ cm$^{2}$/Vs. The calculated electron spin coherence time $T_{2}$ and measured mobility as a function of temperature for ZnSe are respectively shown in Figs.~\ref{BulkCdSeZnSe}(c) and \ref{BulkCdSeZnSe}(d). The electronic band structures for CdSe and ZnSe are displayed in Figs.~\ref{BulkCdSeZnSeBands}(a)--(b), and the spin-orbit coupling to band-gap ratio for CdSe and ZnSe is respectively $R_{\text{CdSe}}$ $=$ $0.15$ and $R_{\text{ZnSe}}$ $=$ $0.26$. We approximate both systems in our model as zincblende lattices.

\begin{figure}[htbp]
\setcaptionwidth{8.0 cm}
\scalebox{1}{\includegraphics[width = 8.0 cm]{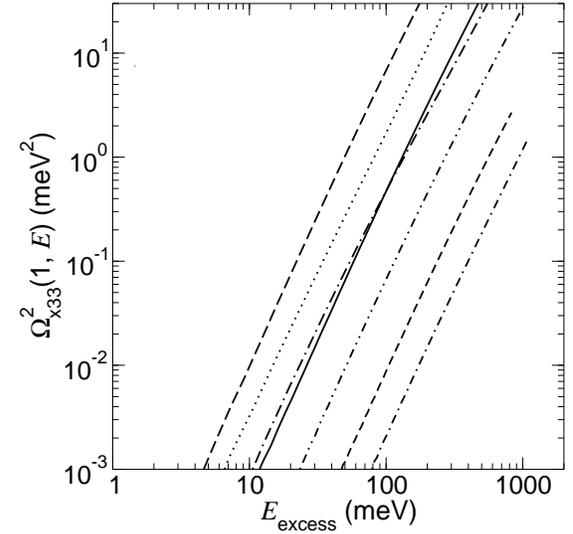}}
\caption[]{Energy dependence of $\widetilde{\Omega}_{x3}^2(1, E)$ of III-V semiconductors at room temperature for bulk GaAs (solid line), GaSb (long dashed line), InAs (long dot-dashed line), InSb (dotted line), InP (dot-dot dashed line), ZnSe (short dashed line), and CdSe (short dot-dashed line).}
\label{BulkOmegaE1}
\end{figure}

For bulk semiconductors the relevant electronic states for electron spin decoherence are the states of the two lowest conduction bands near the bulk band edge, and thus perturbative expressions~\cite{dyakonov72, meier84} for $\widetilde{\Omega}_{xl}^2(1,E)$ for these bulk semiconductors [$\widetilde{\Omega}_{x33}^2(1,E)\propto E^3$] are nearly identical to those obtained from a full fourteen-band calculation (within numerical accuracy). The energy dependence of $\widetilde{\Omega}_{x33}^2(1,E)$ for the III-V and II-VI bulk semiconductors considered above is summarized in Fig.~\ref{BulkOmegaE1}, and the cubic dependence on energy of $\widetilde{\Omega}_{x33}^2(1,E)$ [i.e., the slopes of the curves are approximately 3 and thus $\widetilde{\Omega}_{x33}(1,k)\propto k^3$] for bulk semiconductors near zone center is confirmed. The calculated spin coherence times for bulk semiconductors using the perturbative expression obtained by D'yakonov and Perel'~\cite{meier84} coincide with those calculated using the fourteen-band non-perturbative theory. The agreement between calculated and measured spin coherence times ($T_2$'s) indicates that the conduction-band electron spin splitting of bulk zincblende semiconductors is well described by the fourteen-band non-perturbative theory.
\subsection{SPIN COHERENCE TIMES IN QUANTUM WELLS AND SUPERLATTICES}
\label{sec:qws}
Electron spin dynamics in III-V and II-VI quantum wells have been extensively investigated experimentally because of their well-known electronic and optical properties. Ultrafast all-optical spin polarization switches using multiple quantum wells have been demonstrated experimentally by Nishikawa {\it et al.}~\cite{nishikawa96}, and the result suggests that these all-optical spin-dependent switches can be used for future ultrafast optical communication systems that require very high switching speeds. The switching performance of these all-optical spin-dependent switches hinges on the electron spin decoherence, and therefore, the calculations of electron spin coherence times are important. 
\subsubsection{III-V LAYERED SEMICONDUCTORS}
\label{sec:qw35}
\begin{figure}[htbp]
\setcaptionwidth{8.0 cm}
\scalebox{1}{\includegraphics[width= 8.0 cm]{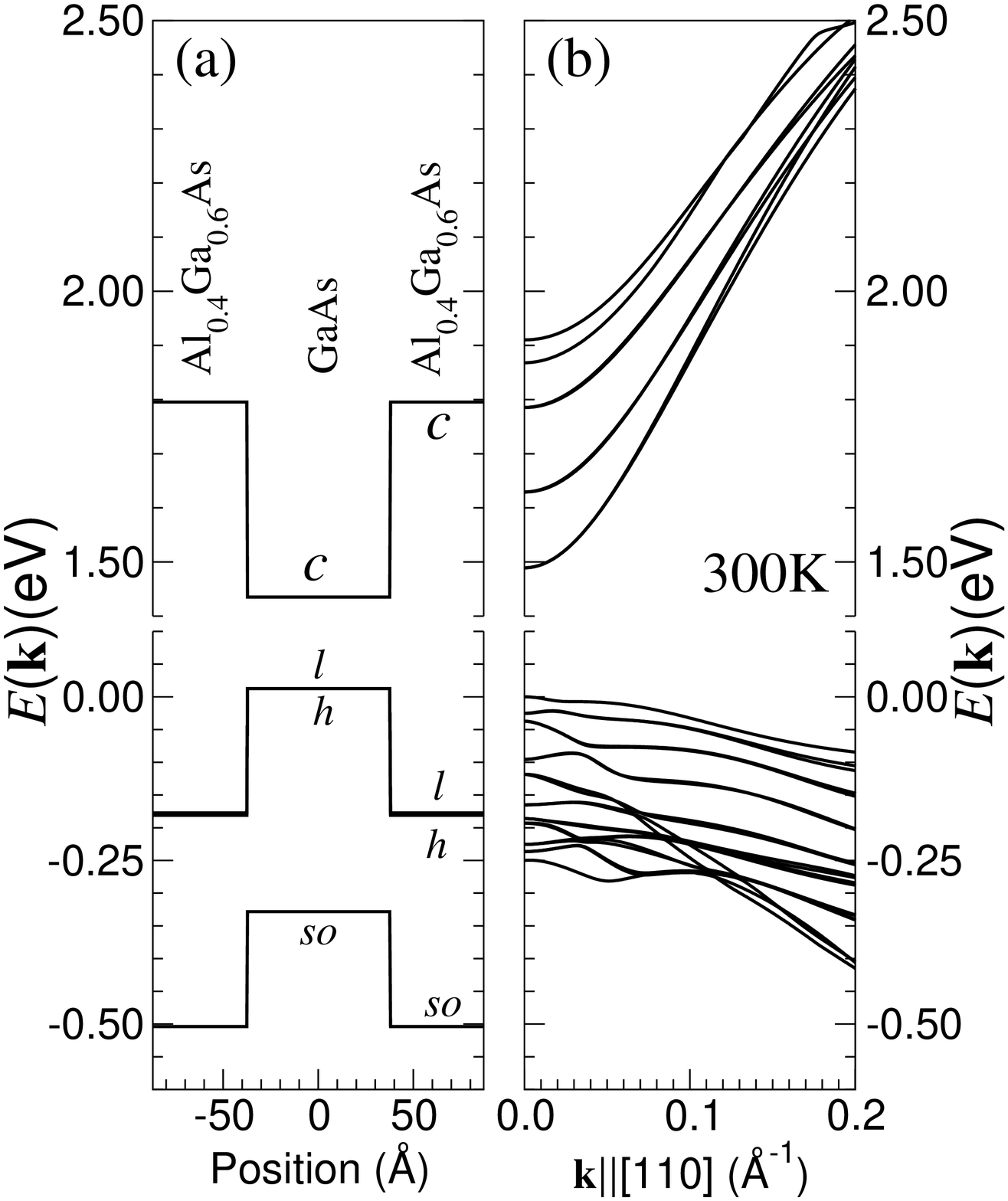}}
\caption[]{Band-edge diagram (a) and band structure (b) for a $75$-\AA\ GaAs/$100$-\AA\ Al$_{0.4}$Ga$_{0.6}$As (001)-oriented quantum well at 300 K. The labels {\it c}, {\it l}, {\it h}, and {\it so} identify the conduction, light-hole, heavy-hole, and spin-orbit split-off hole band edges, respectively, at the zone center of the bulk constituent semiconductors.}
\label{MQWGaAsAlGaAsBands}
\end{figure}

\begin{figure}[htbp]
\setcaptionwidth{8.0 cm}
\scalebox{1}{\includegraphics[width = 8.0 cm]{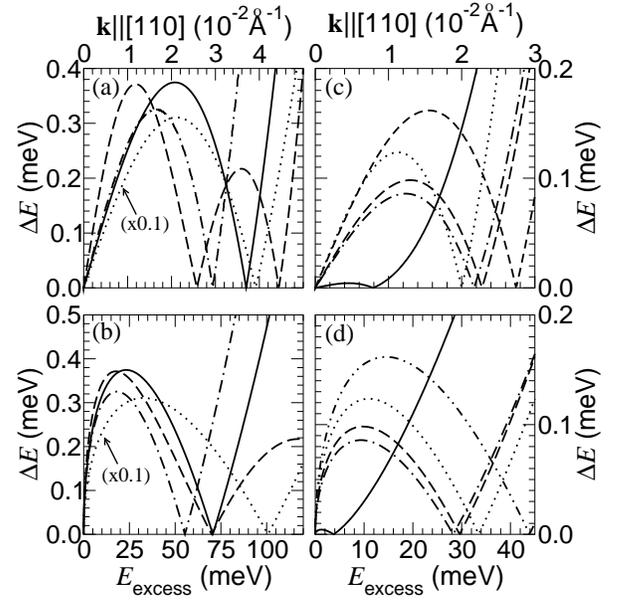}}
\caption[]{Spin splitting of the lowest conduction band for quantum wells and superlattices at room temperature. (a) ${\Delta E}$ vs ${\bf k}_{[110]}$ and (b) ${\Delta E}$ vs $E_{\text{excess}}$ for a $75$-\AA\ GaAs/$100$-\AA\ Al$_{0.4}$Ga$_{0.6}$As QW (solid line), a thin-layer $21.2$-\AA\ InAs/$36.6$-\AA\ GaSb SL (dashed line), an $70$-\AA\ In$_{0.53}$Ga$_{0.47}$As/97-\AA\ InP QW (dot-dashed line), and a $80$-\AA\ GaSb$_{0.81}$/$80$-\AA\ AlSb QW (dotted line). (c) ${\Delta E}$ vs ${\bf k}_{[110]}$ and (d) ${\Delta E}$ vs $E_{\text{excess}}$ for a $150$-\AA\ GaAs/$100$-\AA\ Al$_{0.03}$Ga$_{0.97}$As QW (solid line), a $150$-\AA\ GaAs/$100$-\AA\ Al$_{0.4}$Ga$_{0.6}$As QW (dashed line), a $100$-\AA\ In$_{0.13}$Ga$_{0.87}$As/$200$-\AA\ GaAs QW (dot-dashed), and  a $95$-\AA\ In$_{0.57}$Ga$_{0.43}$As$_{0.93}$P$_{0.07}$/$75$-\AA\ In$_{0.87}$Ga$_{0.13}$As$_{0.29}$P$_{0.71}$ QW (dotted line), and a $40$-\AA\ In$_{0.20}$Ga$_{0.80}$As/$100$-\AA\ GaAs/$40$-\AA\ In$_{0.20}$Ga$_{0.80}$As/$200$-\AA\ GaAs QW (dot-dot-dashed line).}
\label{qwDeltaE}
\end{figure}

\begin{figure}[htbp]
\setcaptionwidth{8.0 cm}
\scalebox{1}{\includegraphics[width = 8.0 cm]{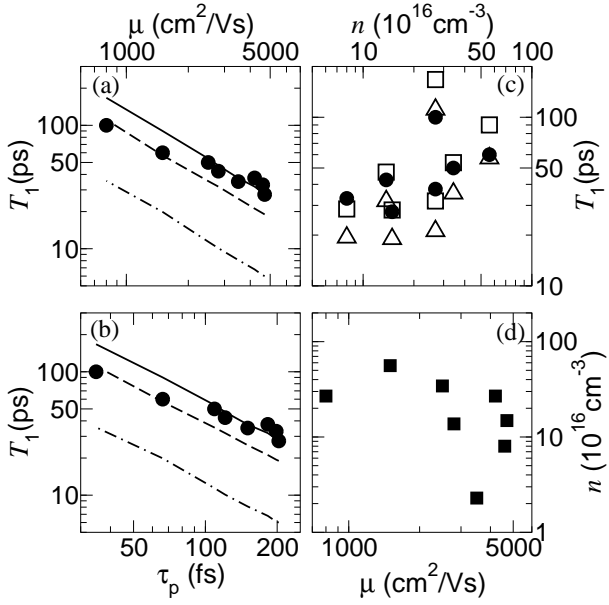}}
\caption[]{$T_1$ as a function of (a) mobility $\mu$, (b) average transport time $\tau_\text{p}$, and (c) carrier density $n$, for a $75$-\AA\ GaAs/$100$-\AA\ Al$_{0.4}$Ga$_{0.6}$As QW at room temperature. Closed circles represent the results of experiments\cite{terauchi99}. The non-perturbative theory reults with OP (solid lines and open squares) and NI scattering (dashed lines and open triangles) are shown, as well as the DK theory results (dot-dashed lines). (d) The measured carrier density and mobility at room temperature~\cite{terauchi99}.}
\label{GaAsAlGaAsT1MNP}
\end{figure}

Terauchi {\it et al.}~\cite{terauchi99} have performed measurements on GaAs/AlGaAs multiple quantum wells to study the mobility dependent electron spin coherence times. The band-edge diagram and electronic band structure of a $75$-\AA\ GaAs/$100$-\AA\ Al$_{0.4}$Ga$_{0.6}$As quantum well are shown in Figs.~\ref{MQWGaAsAlGaAsBands}(a)--(b), and the calculated band-gap energy is $1489$ meV ($0.83$ ${\mu}$m). The lowest conduction-band spin splitting as a function of crystal momentum ${\bf k}||[110]$ and excess energy for this quantum well is plotted in Figs.~\ref{qwDeltaE}(a)--(b) (along with other structures considered later). Figure~\ref{GaAsAlGaAsT1MNP}(a) shows the calculated electron spin coherence times $T_{1}$'s as a function of mobility assuming optical phonon (solid line) and neutral impurity (dashed line) scattering as the dominant process determining the mobility in the quantum well at $300$ K. In the figure, the results of the experimental measurements are represented by closed circles, and, for comparison, the theoretical results of the DK theory based on perturbative approach are also shown. Our results agree with experiment if one assumes a shift from optical phonon to neutral impurity scattering as the mobility drops --- this is the origin of the unusual experimental dependence of $T_1$ on the mobility. The calculated electron spin coherence times with optical phonon scattering are roughly two times larger than those with neutral impurity scattering, and this is directly related to the energy-dependence of the transport times for different momentum scattering mechanisms (see Table~\ref{mqwMS}).

Note that the experimental results have been adjusted from Ref.~\onlinecite{terauchi99}, for the authors defined an effective spin flip time for a single spin, $\tau_s=2T_1$, and plotted their results for $\tau_s$. The discrepency in the magnitude of $T_{1}$'s between the DK calculations and measurements is about a factor of $4$ (see Ref.~\onlinecite{lau01}). The electron spin coherence times $T_1$'s as a function of average transport time $\tau_{\text{p}}$ and electron density $n$ are displayed in Figs.~\ref{GaAsAlGaAsT1MNP}(b) and \ref{GaAsAlGaAsT1MNP}(c), respectively. Open squares and triangles respectively represent the calculated $T_{1}$'s for optical phonon and neutral impurity scattering mechanisms. The relationship between the measured electron mobility and density is depicted in Fig.~\ref{GaAsAlGaAsT1MNP}(d).

\begin{figure}[htbp]
\setcaptionwidth{8.0 cm}
\scalebox{1}{\includegraphics[width = 8.0 cm]{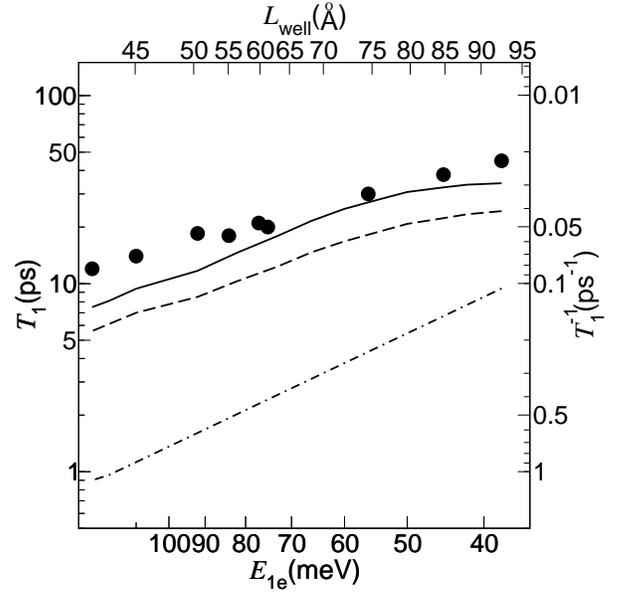}}
\caption[]{Electron spin coherence times and spin relaxation rates as a function of electron confinement energy and quantum-well width at room temperature for $(25-300)$-\AA\ GaAs/$100$-\AA\ Al$_{0.4}$Ga$_{0.6}$As quantum wells~\cite{terauchi99} with electron density $n=2.686 \times 10^{17}$ cm$^{-3}$ and electron mobility $\mu = 800$ cm$^{2}/$Vs. Closed circles represent the results of experiments. Solid and dashed lines respectively represent the results of the non-perturbative theory for OP and NI scattering. Dot-dashed lines represent results of the DK theory.}
\label{GaAsAlGaAsT1R1EL}
\end{figure}

Shown in Fig.~\ref{GaAsAlGaAsT1R1EL} are the calculated electron spin coherence times $T_{1}$'s and the corresponding spin relaxation rates $T^{-1}_{1}$'s as a function of electron confinement energy $E_{\text{1e}}$ for $(30-100)$-\AA\ GaAs/Al$_{0.4}$Ga$_{0.6}$As quantum wells at $300$ K assuming optical phonon (solid line) and neutral impurity (dashed line) scattering. The results of experimental measurements~\cite{terauchi99} are represented by closed circles and the calculations of DK theory are represented by dot-dashed line. Our numerical results are in excellent agreement with the experimental measurements both in the absolute magnitude and trend. On the other hand, the calculations of the DK theory differ approximately by a factor of 4 and the predicted trend is stronger than the observed one~\cite{lau01}. The electron spin coherence times and electron spin relaxation rates as a function of quantum-well width are illustrated in Fig.~\ref{GaAsAlGaAsT1R1EL}.

\begin{figure}[htbp]
\setcaptionwidth{8.0 cm}
\scalebox{1}{\includegraphics[width = 8.0 cm]{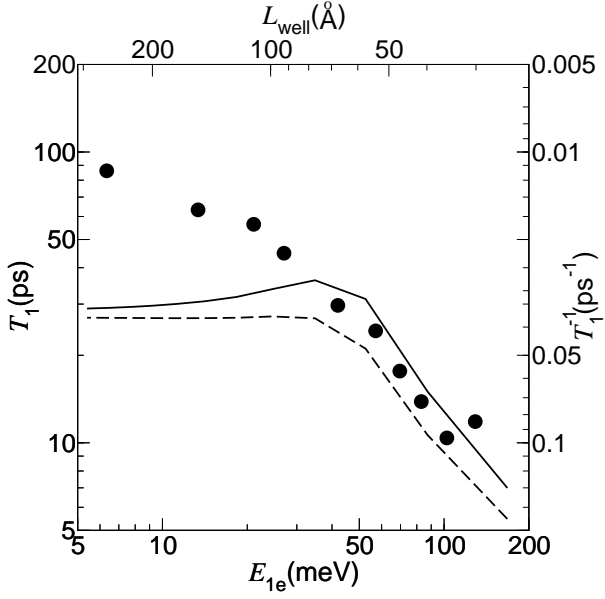}}
\caption[]{Electron spin coherence times and spin relaxation rates as a function of electron confiment energy and quantum-well width at room temperature for a set of $(25-300)$-\AA\ GaAs/$150$-\AA\ Al$_{0.35}$Ga$_{0.65}$As quantum wells~\cite{malinowski00} with electron density $n=1 \times 10^{17}$ cm$^{-3}$. Closed circles represent the results of experimental measurements, while solid and dashed lines represent the numerical calculations with $\mu = 4600$ cm$^{2}/$Vs for optical phonon and neutral impurity scattering, respectively.}
\label{GaAsAlGaAsT1EL}
\end{figure}

\begin{figure}[htbp]
\setcaptionwidth{8.0 cm}
\scalebox{1}{\includegraphics[width = 8.0 cm]{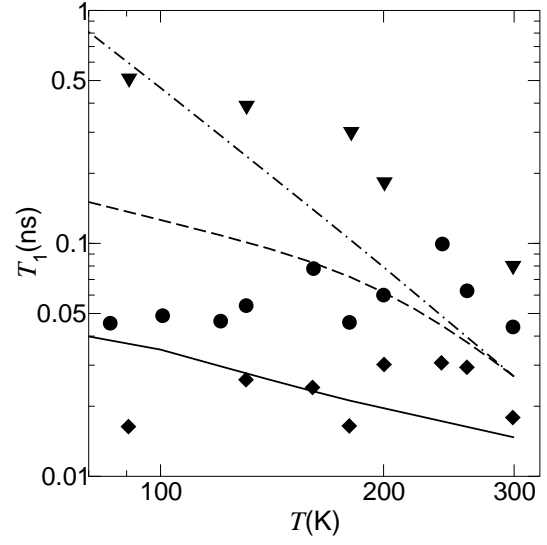}}
\caption[]{Electron spin coherence times as a function of temperature. Closed squares, circles, and triangles represent the experimental measurements for a $60$-\AA\ GaAs/$150$-\AA\ Al$_{0.35}$Ga$_{0.65}$As, a $100$-\AA\ GaAs/$150$-\AA\ Al$_{0.35}$Ga$_{0.65}$As, and a $200$-\AA\ GaAs/$150$-\AA\ Al$_{0.35}$Ga$_{0.65}$As quantum well, respectively, and the corrresponding calculated $T_1$'s are respectively represented by solid, dashed, and dot-dashed lines.}
\label{GaAsAlGaAsT1T}
\end{figure}

Another set of comprehensive measurements of the electron spin coherence time $T_1$ for $(25-300)$-\AA\ GaAs/$150$-\AA\ Al$_{0.35}$Ga$_{0.65}$As quantum wells was recently carried out by Malinowski {\it et al.}~\cite{malinowski00}, and the results of the experiments are shown in Fig.~\ref{GaAsAlGaAsT1EL}. The calculated $T_1$'s as a function of electron confinement energy for electron density $n$ $=$ $1 \times 10^{17}$ and an estimated constant electron mobility $\mu$ $=$ $4600$ cm$^{2}$/Vs are in reasonable agreement with the experimental measurements, in particular for quantum wells with relatively large electron confinement energy $E_{\text{1e}}$. The slower measured electron spin relaxation rates $T^{-1}_1$'s are most likely associated with the non-uniform electron mobility in different thickness quantum wells. As the quantum-well width increases, the calculated electron spin coherence times approach the values of bulk GaAs. The calculated spin coherence times of bulk GaAs at room temperature with $n$ $=$ $1 \times 10^{17}$ and $\mu$ $=$ $4600$ cm$^{2}$/Vs for optical phonon and neutral impurity scatterings are approximately $29$ ps and $27$ ps, respectively. Figure~\ref{GaAsAlGaAsT1T} shows the temperature dependence of $T_1$ for three different width quantum wells ($60$-\AA, $100$-\AA, and $200$-\AA) with the same electron density and estimated mobility as in Figs.~\ref{GaAsAlGaAsT1EL}. Given the uncertainty in the electron mobility, the calculated $T_1$'s compared with the results of experiments are quite reasonable.

\begin{figure}[htbp]
\setcaptionwidth{8.0 cm}
\scalebox{1}{\includegraphics[width = 8.0 cm]{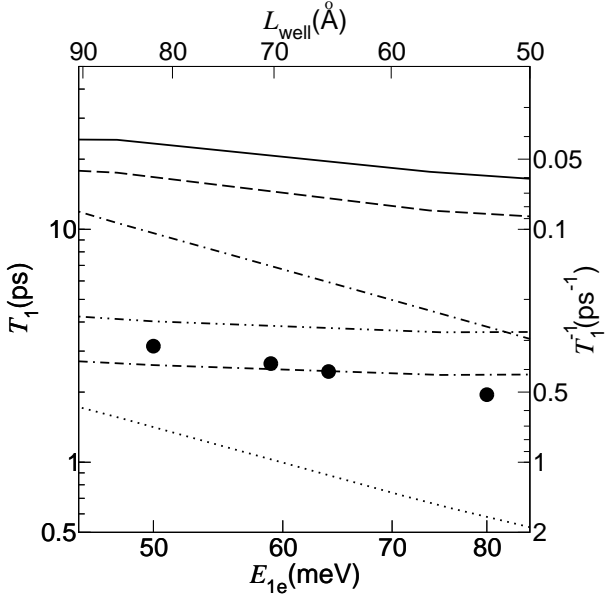}}
\caption[]{Electron spin coherence times and spin relaxation rates of $(25-100)$-\AA\ In$_{0.53}$Ga$_{0.47}$As/97-\AA\ InP QWs \cite{tackeuchi97} as a function of electron confinement energy and quantum-well width at room temperature with $\mu = 6700$ cm$^{2}/$Vs for non-degenerate ($n=6 \times 10^{15}$ cm$^{-3}$) and degenerate ($n=5 \times 10^{18}$ cm$^{-3}$) electron densities. Closed circles represent the results of experiments. Solid (dot-dot-dashed) and dashed (dot-dashed-dashed) lines respectively represent the results of the non-perturbative theory for OP and NI scattering with non-degenerate (degenerate) electron density. Dot-dashed (dotted) lines represent results of the DK theory for non-degenerate (degenerate) electron density.}
\label{InGaAsInPT1R1EL}
\end{figure}

The electron spin coherence times $T_1$'s and the corresponding spin relaxation rates $T^{-1}_1$'s of $(25-100)$-\AA\ In$_{0.53}$Ga$_{0.47}$As/$97$-\AA\ InP quantum wells~\cite{tackeuchi99} as a function confinement energy $E_{\text{1e}}$ and quantum-well width $L_{\text{well}}$ at $300$ K for electron densities $n$ $=$ $6 \times 10^{15}$ cm$^{-3}$ and $n$ $=$ $5 \times 10^{18}$ cm$^{-3}$ are shown in Fig.~\ref{InGaAsInPT1R1EL}. Since the experimental electron mobility is not available, a typical value of $6700$ cm$^{2}$/Vs is used in our calculations. The confinement-energy dependence of $T_{1}$'s from our theoretical calculations agrees with those of the experimental measurements. On other hand, as in the case of GaAs/AlGaAs quantum wells, the DK theory gives a stronger confinement-energy dependence of the electron spin coherence time than that of the experimental results. Better agreement between our theorectical calculations and experimental results is obtained if the electron density is degenerate.

\begin{figure}[htbp]
\setcaptionwidth{8.0 cm}
\scalebox{1}{\includegraphics[width= 8.0 cm]{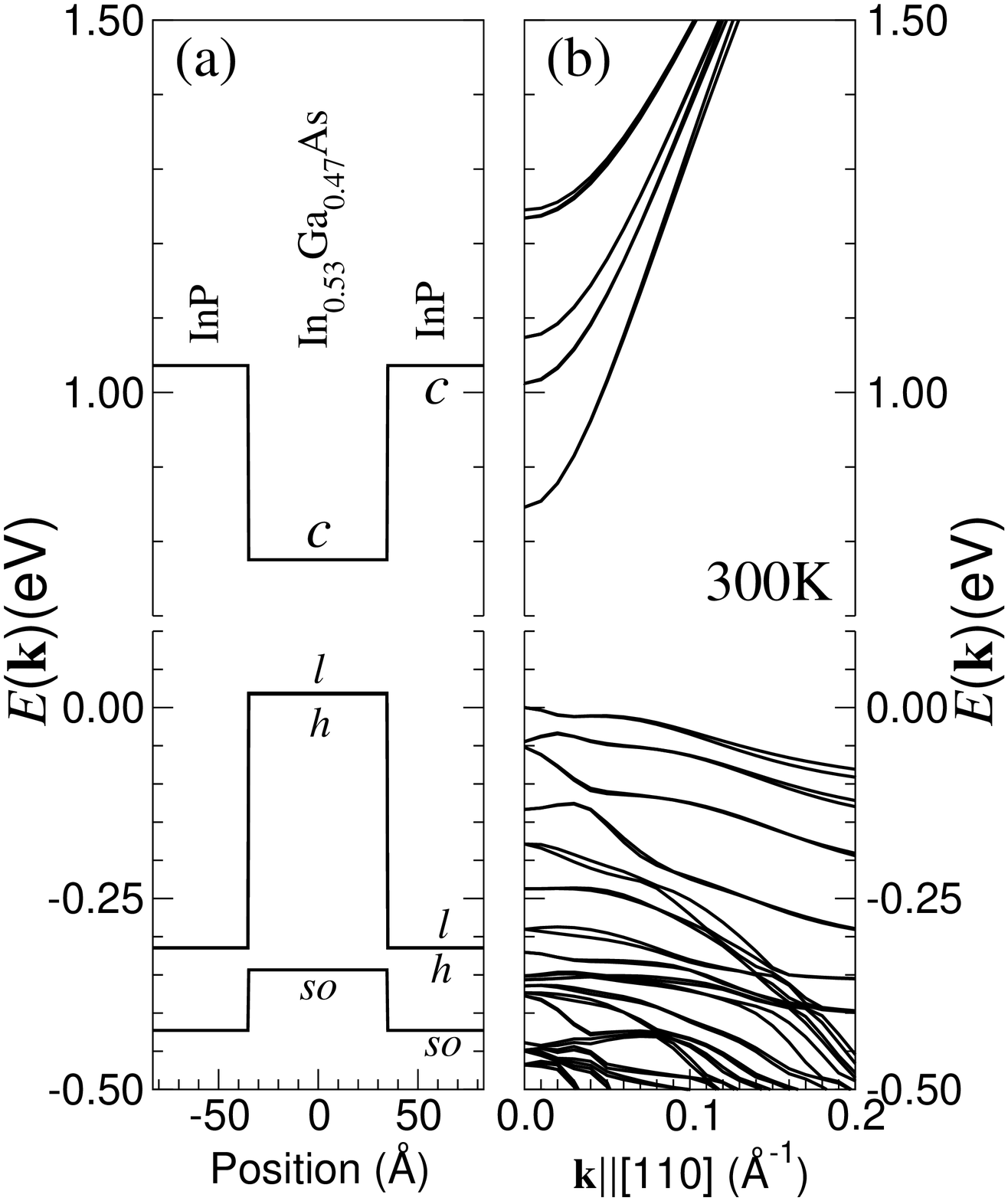}}
\caption[]{Band-edge diagram (a) and band structure (b) for a $70$-\AA\ In$_{0.53}$Ga$_{0.47}$As/$97$-\AA\ InP (001)-oriented quantum well at 300 K. The labels {\it c}, {\it l}, {\it h}, and {\it so} identify the conduction, light-hole, heavy-hole, and spin-orbit split-off hole band edges, respectively, at the zone center of the bulk constituent semiconductors.}
\label{InGaAsInPBands}
\end{figure}

The band-edge diagram and electronic band structure for a $70$-\AA\ In$_{0.53}$Ga$_{0.47}$As/$97$-\AA\ InP quantum well are illustrated in Figs.~\ref{InGaAsInPBands}(a)--(b) with band-gap energy of $976$ meV ($1.27$ ${\mu}$m). The conduction-band spin splitting of the lowest states as a function of crystal momentum ${\bf k}||[110]$ and excess energy is shown in Figs.~\ref{qwDeltaE}(a)--(b). The measured $T_{1}$ for this quantum well by Tackeuchi {\it et al.}~\cite{tackeuchi97} is $2.6$ ps, and calculated $T_{1}$ is $4.0$ ps ($6.5$ ps) using an estimated electron mobility $\mu$ $=$ $6700$ cm$^{2}$/Vs and a value of degenerate electron density $n$ $=$ $3.0 \times 10^{18}$ cm$^{-3}$ assuming neutral impurity (optical phonon) scattering~\cite{lau01}. InGaAs/InP quantum wells are a so called non-common-atom system~\cite{krebs96}, which gives rise to interface bonding asymmetry or native interface asymmetry (NIA). Native interface asymmetry causes additional electron spin splittings, and consequently it leads to faster electron spin decoherence in the system. We do not include NIA contributions in our calculations for these wells.

\begin{figure}[htbp]
\setcaptionwidth{8.0 cm}
\scalebox{1}{\includegraphics[width= 8.0 cm]{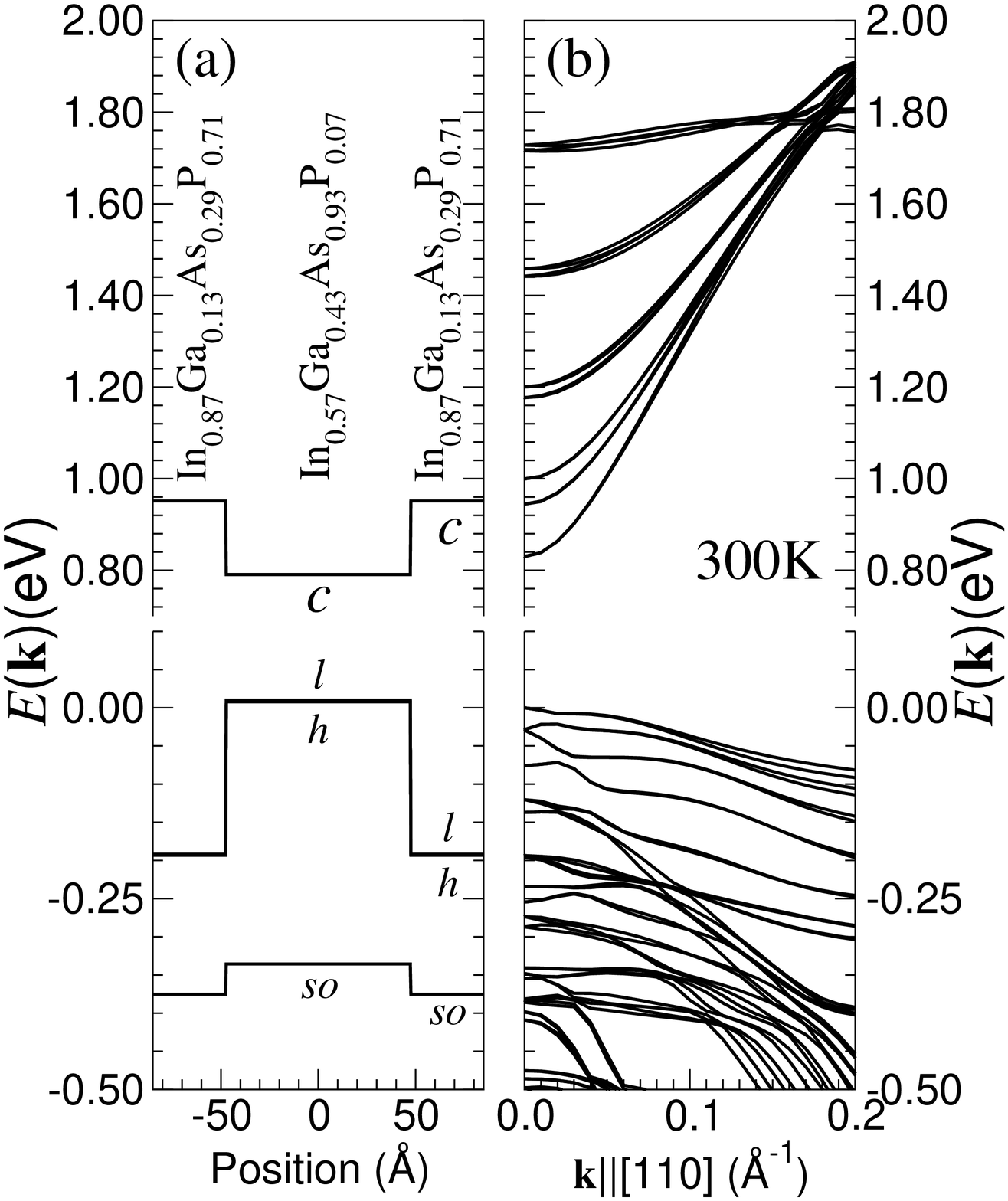}}
\caption[]{Band-edge diagram (a) and band structure (b) for a $95$-\AA\ In$_{0.57}$Ga$_{0.43}$As$_{0.93}$P$_{0.07}$/$75$-\AA\ In$_{0.87}$Ga$_{0.13}$As$_{0.29}$P$_{0.71}$ (001)-oriented quantum well~\cite{hyland99} at 300 K. The labels {\it c}, {\it l}, {\it h}, and {\it so} identify the conduction, light-hole, heavy-hole, and spin-orbit split-off hole band edges, respectively, at the zone center of the bulk constituent semiconductors.}
\label{MR850Bands}
\end{figure}

We also present  the electron spin coherence times for InGaAsP quaternary quantum wells with different compositions. The calculated $T_{1}$ for neutral impurity scattering at $300$ K with electron density $n$ $=$ $1.3 \times 10^{16}$ cm$^{-3}$ and electron mobility $\mu$ $=$ $800$ cm$^{2}$/Vs for a $55$-\AA\ In$_{0.53}$Ga$_{0.47}$As/$100$-\AA\ In$_{0.87}$Ga$_{0.13}$As$_{0.29}$P$_{0.71}$, $65$-\AA\ In$_{0.53}$Ga$_{0.47}$As/$100$-\AA\ In$_{0.87}$Ga$_{0.13}$As$_{0.29}$P$_{0.71}$, $90$-\AA\ In$_{0.53}$Ga$_{0.47}$As$_{0.93}$P$_{0.07}$/$75$-\AA\ In$_{0.87}$Ga$_{0.13}$As$_{0.29}$P$_{0.71}$, and $95$-\AA\ In$_{0.53}$Ga$_{0.47}$As$_{0.93}$P$_{0.07}$/$75$-\AA\ In$_{0.87}$Ga$_{0.13}$As$_{0.29}$P$_{0.71}$ is listed in Table~\ref{coherencetimes}. The calculated electron spin coherence times of these quantum wells are comparible to those of InGaAs/InP quantum wells [c.f., Fig.~\ref{InGaAsInPT1R1EL}]. These results are consistent with the experimental measurements by Hyland {\it et al.} in Ref.~\onlinecite{hyland99} (see Table~\ref{coherencetimes}). The band-edge diagram and electronic band structure for a $95$-\AA\ In$_{0.53}$Ga$_{0.47}$As$_{0.93}$P$_{0.07}$/$75$-\AA\ In$_{0.87}$Ga$_{0.13}$As$_{0.29}$P$_{0.71}$ quantum well, and the band-gap energy of $830$ meV ($1.49$ ${\mu}$m) are displayed in Figs.~\ref{MR850Bands}(a)-(b). The lowest conduction-band spin splitting as a function of crystal momentum ${\bf k}||[110]$ and excess energy is depicted in Figs.~\ref{qwDeltaE}(c)--(d).

\begin{figure}[htbp]
\setcaptionwidth{8.0 cm}
\scalebox{1}{\includegraphics[width = 8.0 cm]{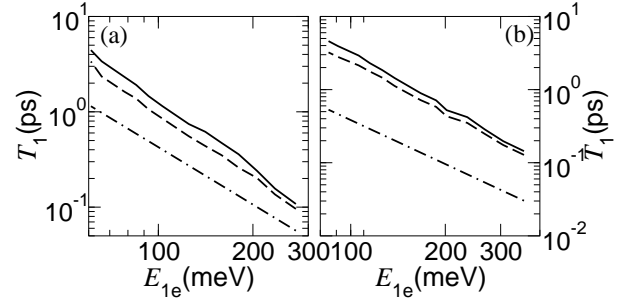}}
\caption[]{$T_1$ as a function of confinement energy $E_{\text{1e}}$ at $300$ K
for (a) $(30-100)$-\AA\ GaSb/AlSb QWs \cite{terauchi99} and
(b) $(30-100)$-\AA\ GaAsSb/AlSb QWs \cite{tackeuchi97} with $n=2 \times 10^{16}$ cm$^{-3}$ and $\mu = 2000$ cm$^{2}/$Vs. Solid and dashed lines respectively represent the results of the non-perturbative theory for OP and NI scattering. Dot-dashed lines represent results of the DK theory.}
\label{GaAsSbAlSbT1E}
\end{figure}

The confinement-energy dependence of the electron spin coherence times at room temperature for GaSb/AlSb and GaAsSb/AlSb quantum wells with electron density $n$ $=$ $2 \times 10^{16}$ cm$^{-3}$ and estimated mobility $\mu$ $=$ $2000$ cm$^{2}$/Vs are respectively shown in Figs.~\ref{GaAsSbAlSbT1E}(a) and \ref{GaAsSbAlSbT1E}(b). The $T_{1}$'s for both structures have very similar confinement-energy dependence, and the coherence times for the GaAsSb/AlSb quantum wells are slightly longer than those of the GaSb/AlSb quantum wells. The longer electron spin coherence times in the GaSb/AlSb quantum wells are attributed to the smaller conduction-band spin splittings due to the $19\%$ indium substitution in the well-acting layers. 

\begin{figure}[htbp]
\setcaptionwidth{8.0 cm}
\scalebox{1}{\includegraphics[width= 8.0 cm]{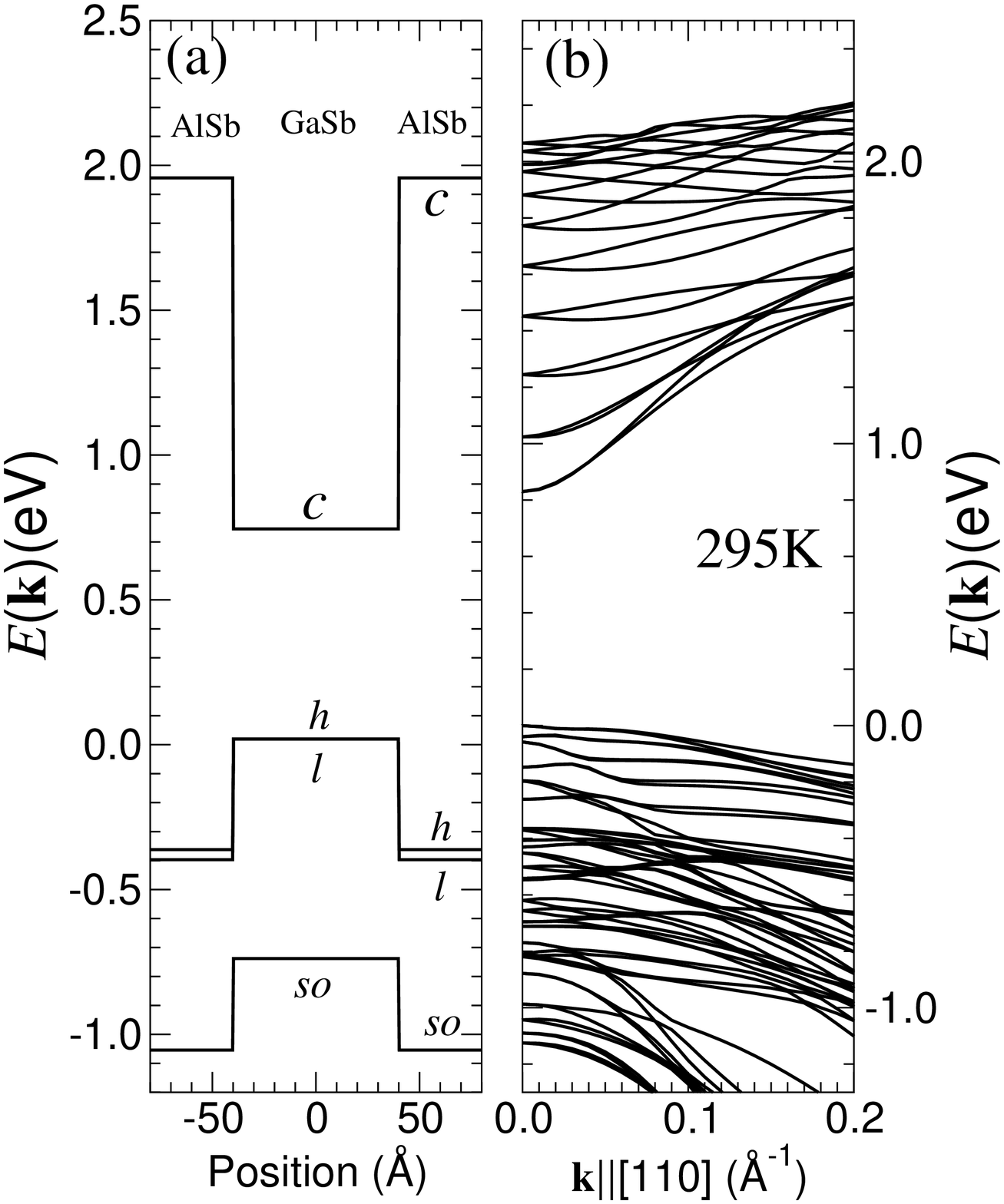}}
\caption[]{Band-edge diagram (a) and band structure (b) for a $80$-\AA\ GaSb /$80$-\AA\ AlSb (001)-oriented quantum well at 295 K. The labels {\it c}, {\it l}, {\it h}, and {\it so} identify the conduction, light-hole, heavy-hole, and spin-orbit split-off hole band edges, respectively, at the zone center of the bulk constituent semiconductors.}
\label{GaSbAlSbBands}
\end{figure}

\begin{figure}[htbp]
\setcaptionwidth{8.0 cm}
\scalebox{1}{\includegraphics[width= 8.0 cm]{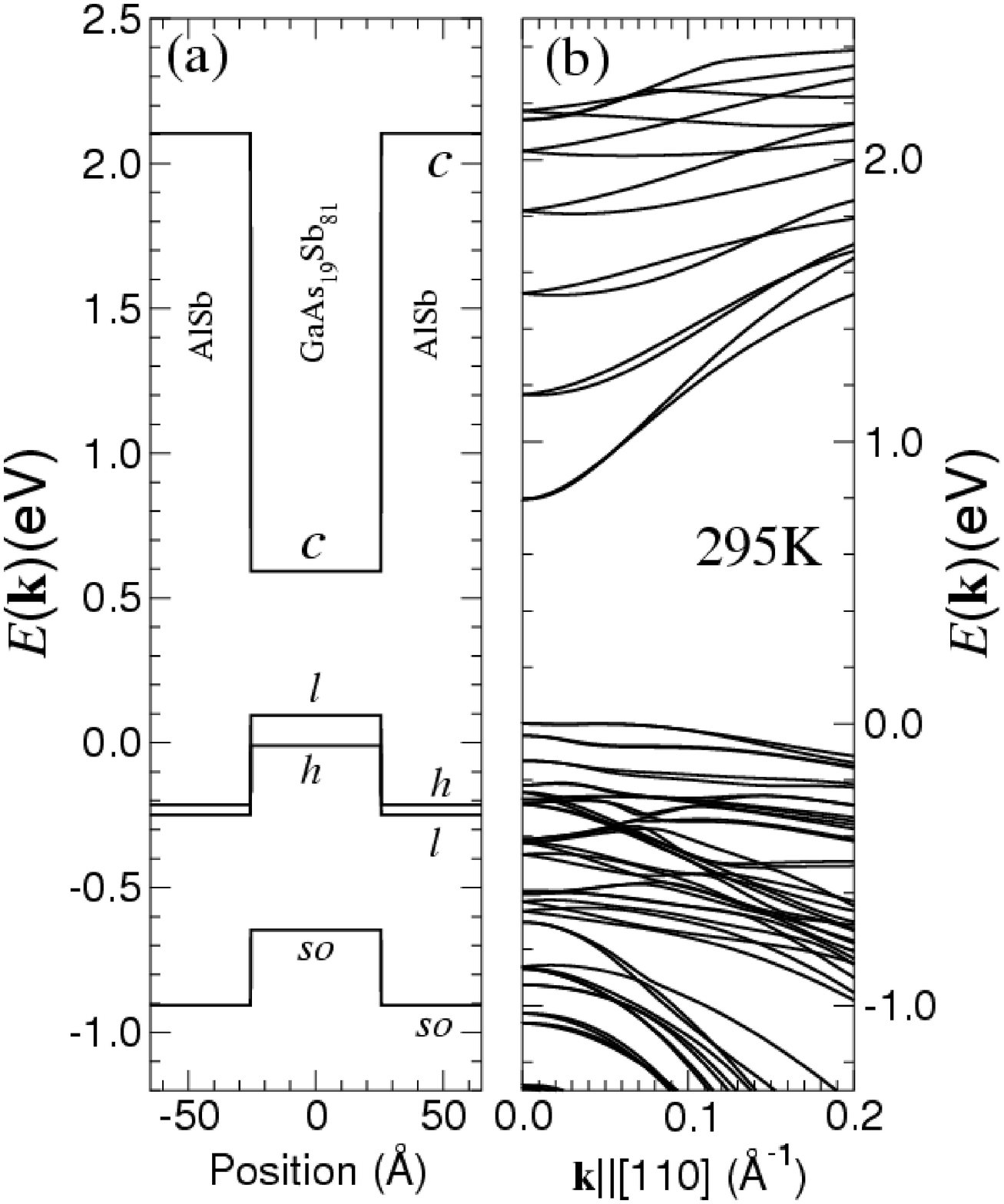}}
\caption[]{Band-edge diagram (a) and band structure (b) for a $51$-\AA\ Ga$_{0.19}$As$_{0.81}$Sb /$80$-\AA\ AlSb (001)-oriented quantum well at 295 K. The labels {\it c}, {\it l}, {\it h}, and {\it so} identify the conduction, light-hole, heavy-hole, and spin-orbit split-off hole band edges, respectively, at the zone center of the bulk constituent semiconductors.}
\label{GaAsSbAlSbBands}
\end{figure}

The band-edge diagram and electronic band structure for a $80$-\AA\ GaSb /$80$-\AA\ AlSb QW and a $51$-\AA\ Ga$_{0.19}$As$_{0.81}$Sb /$80$-\AA\ AlSb QW are shown in Figs.~\ref{GaSbAlSbBands}(a)--(b) and \ref{GaAsSbAlSbBands}(a)--(b), respectively. For the former, the band-gap energy is $829$ meV ($1.50$ ${\mu}$m), while for the latter the band-gap energy is $795$ meV ($1.56$ ${\mu}$m). The spin splitting of the lowest conduction band as a function of crystal momentum ${\bf k}||[110]$ and excess energy is plotted in Figs.~\ref{qwDeltaE}(a)--(b). Hall {\it et al.}~\cite{hall99} have experimentally investigated the electron spin dynamics for both quantum wells, and measured $T_{1}$ at $295$ K is $0.52$ ps for the GaSb/AlSb QW with optical excited electron density $n$ $=$ $2.8 \times 10^{18}$ cm$^{-3}$, and for the GaAsSb/AlSb QW, the measured $T_{1}$ at $295$ K is $0.42$ ps with optical excited electron density $n$ $=$ $3.4 \times 10^{18}$ cm$^{-3}$. The calculated electron spin coherence time $T_{1}$ for the GaSb/AlSb QW is $0.53$ ps ($1.4$ ps) assuming neutral impurity (optical phonon) scattering; and the calculated $T_{1}$ for the GaAsSb/AlSb QW is $0.87$ ps ($1.4$ ps) assuming neutral impurity (optical phonon) scattering~\cite{lau01}. The estimated electron mobility is $\mu$ $=$ $2000$ cm$^{2}$/Vs for both quantum wells. In this case, the agreement between theory and experiment is quite good. It should be noted that the optical excited electron densities in these experiments are highly degenerate.

\begin{figure}[htbp]
\setcaptionwidth{8.0 cm}
\scalebox{1}{\includegraphics[width = 8.0 cm]{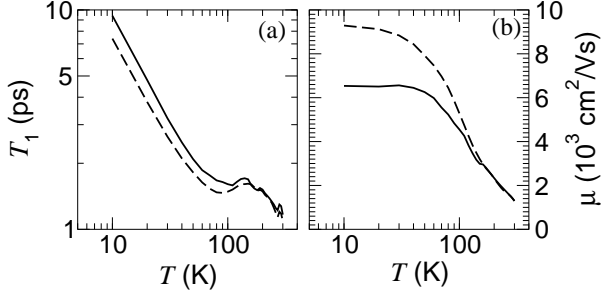}}
\caption[]{Electron spin coherence times. (a) Calculated $T_1$ as a function of temperature $T$ for a $80$-\AA\ GaSb/$300$-\AA\ Al$_{0.4}$Ga$_{0.6}$Sb QW~\cite{ghezzi01} with NI scattering. (b) Measured mobility as a function of temperature~\cite{ghezzi01}. Solid and dashed lines are for carrier density $n=2.11 \times 10^{17}$ cm$^{-3}$ and $n=2.76 \times 10^{17}$ cm$^{-3}$, respectively.}
\label{GaSbAlGaSbT1MNP}
\end{figure}

Due to the relatively large electron spin precession vector of GaSb, the electron spin coherence times in GaSb quantum wells are expected to be relative short, for example, compared to the electron spin coherence times of GaAs quantum wells [c.f., Figs.~\ref{GaAsAlGaAsT1R1EL}(a) and \ref{GaAsSbAlSbT1E}(a)]. Figure~\ref{GaSbAlGaSbT1MNP}(a) displays the dependence of the electron spin coherence times on temperature for a $80$-\AA\ GaSb/$300$-\AA\ Al$_{0.4}$Ga$_{0.6}$Sb QW. The experimental measurements of electron mobilities as a function of temperature are shown in Fig.~\ref{GaSbAlGaSbT1MNP}(b). At room temperature, the expected electron spin coherence time $T_{1}$ $=$ $1.2$ ps for $\mu$ $=$ $1300$ cm$^{2}$/Vs. The electron spin coherence times are proportional to the electron mobility but inversely proportional to the temperature. This example illustrates that in order to determine the dominant electron spin relaxation mechanism in a system, quantitative calculations of the electron spin coherence times are necessary for the electron spin coherence times depend on the electron mobility, density, and temperature.

\begin{figure}[htbp]
\setcaptionwidth{8.0 cm}
\scalebox{1}{\includegraphics[width= 8.0 cm]{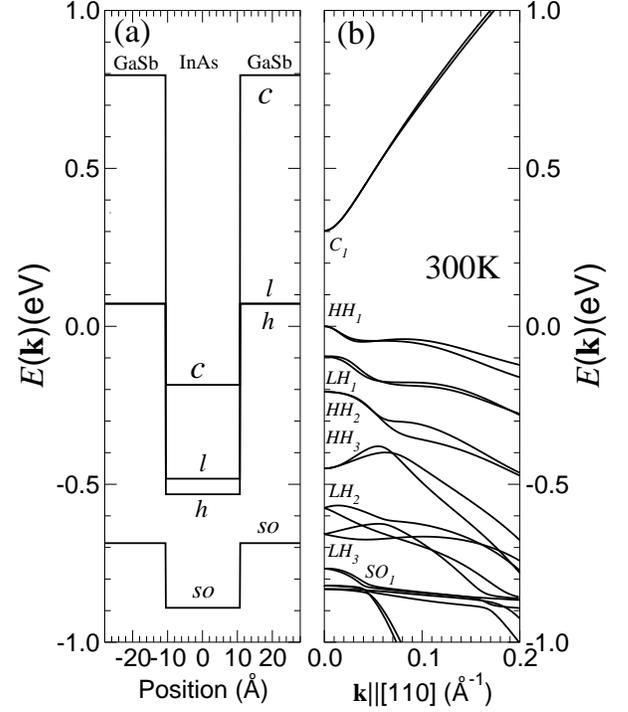}}
\caption[]{Band-edge diagram (a) and band structure (b) for a $21.2$-\AA\ InAs/$36.6$-\AA\ GaSb (001)-oriented strained-layer superlattice at 300 K. The labels {\it c}, {\it l}, {\it h}, and {\it so} identify the conduction, light-hole, heavy-hole, and spin-orbit split-off hole band edges, respectively, at the zone center of the bulk constituent semiconductors. The conduction, light-hole, heavy-hole, and spin-orbit split-off hole superlattice states are labelled by {\it C}$_{i}$, {\it LH}$_{i}$, {\it HH}$_{i}$, and {\it SO}$_{i}$, respectively.}
\label{SLInAsGaSbBands}
\end{figure}

Next we calculate the electron spin coherence times at room temperature for a thin-layer type-II superlattice, which is a non-common-atom structure. The calculated $T_1$'s for a $21.2$\AA\ InAs/$36.6$\AA\ GaSb superlattice are respectively $0.77$ ps and $1.6$ ps for optical phonon and neutral scattering with electron density $n=2 \times 10^{16}$ cm$^{-3}$ and mobility $\mu = 3000$ cm$^{2}/$Vs. The band-edge diagram and electronic band structure are displayed in Figs.~\ref{SLInAsGaSbBands}(a)--(b) with band-gap energy of $303$ meV ($4.09$ ${\mu}$m). The lowest conduction-band spin splitting as a function of crystal momentum ${\bf k}||[110]$ and excess energy for the superlattice is displayed in Figs.~\ref{qwDeltaE}(a)--(b). The effects due to native interface asymmetry are not included in these calculations, and it has been shown by Olesberg {\it et al.}~\cite{olesberg01} that native interface asymmetry plays an important role in electron spin decoherence in this particular structure (for details see Ref.~\onlinecite{olesberg01}).

\begin{figure}[htbp]
\setcaptionwidth{8.0 cm}
\scalebox{1}{\includegraphics[width= 8.0 cm]{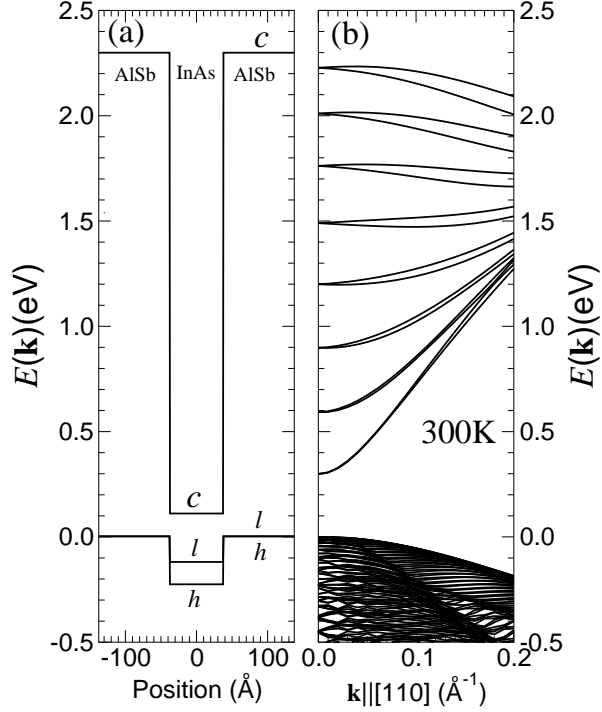}}
\caption[]{Band-edge diagram (a) and band structure (b) for a $75$-\AA\ InAs/$200$-\AA\ AlSb (001)-oriented quantum well at 300 K. The labels {\it c}, {\it l}, and  {\it h} identify the conduction, light-hole, and heavy-hole band edges, respectively, at the zone center of the bulk constituent semiconductors.}
\label{InAsAlSbBands}
\end{figure}

\begin{figure}[htbp]
\setcaptionwidth{8.0 cm}
\scalebox{1}{\includegraphics[width = 8.0 cm]{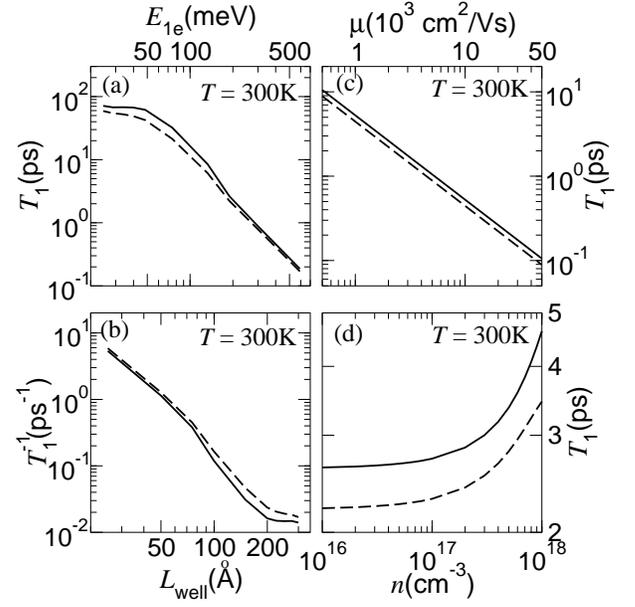}}
\caption[]{$T_1$ as a function of (a) confinement energy $E_{\text{1e}}$ and (b) quantum-well width at $300$ K for $(25-300)$-\AA\ InAs/AlSb QWs with $n=2.00 \times 10^{16}$ cm$^{-3}$ and $\mu = 2000$ cm$^{2}/$Vs. $T_1$ as a function of (c) mobility with $n=2 \times 10^{16}$ cm$^{-3}$ and (d) electron density with $\mu = 2000$ cm$^{2}/$Vs at $300$ K for a $75$-\AA\ InAs/AlSb QW. Solid and dashed lines represent the results for OP and NI scattering, respectively.}
\label{InAsAlSbT1EL}
\end{figure}

Another non-common-atom type-II system of interest in spintronic devices is InAs/AlSb quantum wells in which the conduction-band offset $\Delta E_{\text{c}}$ is approximately one order in magnitude larger than the valence-band offset $\Delta E_{\text{v}}$ [Fig.~\ref{InAsAlSbBands}(a)]. The band-edge diagram and electronic band structure for a $75$-\AA\ InAs/$200$-\AA\ AlSb quantum well are illustrated in Figs.~\ref{InAsAlSbBands}(a)--(b), and the band-gap energy is approximately $300$ meV ($4.13$ ${\mu}$m). The confinement-energy dependence of the electron spin coherence times at room temperature for $(25-100)$-\AA\ InAs/$200$-\AA\ AlSb quantum wells with electron density $n$ $=$ $2 \times 10^{16}$ cm$^{-3}$ and an estimated mobility $\mu$ $=$ $2000$ cm$^{2}$/Vs is shown in Fig.~\ref{InAsAlSbT1EL}(a), and the corresponding electron spin relaxation rate $T^{-1}_{1}$ as a function of quantum-well width $L_{\text{well}}$ is displayed in Fig.~\ref{InAsAlSbT1EL}(b). The range of tunability of the electron spin coherence time in this system is quite large, over two orders of magnitude. For a wide quantum well, e.g., $L_{\text{well}}$ $=$ $300$\AA\, $T_{1}$ is approximately $70$ ps; in contrast, $T_{1}$ is approximately $0.2$ ps for a narrow quantum well of width $L_{\text{well}}$ $=$ $30$ \AA. 

The electron spin coherence times as a function of electron mobility and density for a $75$-\AA\ InAs/$200$-\AA\ AlSb quantum well are shown in Figs.~\ref{InAsAlSbT1EL}(c)--(d). As expected, for non-degenerate electron density, $T_{1}$ $\propto$ $\mu^{-1}$. It should be noted that, for systems with non-degenerate carrier density, the carriers in the conduction band occupy relatively low energy states ($E_\text{occupied}$ $<$ $k_\text{B} T$). Within this energy range, the electron spin coherence time decreases with increasing carrier density because the effective magnetic field increases with increasing energy as the higher energy states are occupied [see Fig.~\ref{spinledT1MN}(d)]. For degenerate systems, the electron spin coherence time may decrease or increase with increasing carrier density depending on where the Fermi level lies because the effective magnetic field oscillates as the energy increases [see Figs.~\ref{qwOmegaE}(a)--(d)].
\subsubsection{SPIN-LED STRUCTURES}
\label{sec:spinled}
Another technique used to detect the spin polarization of carriers in a semiconductor spintronic device is to measure the degree of spin polarization of the electroluminescence (EL) in a light-emiting diode (LED) fabricated within the device~\cite{ohno99, fiederling99, zhu01}. This technique provides a quantitative and model-independent measurement of electrical spin injection~\cite{meier84}. A crucial factor in determining the spin polarization is the carrier spin relaxation after spin injection; therefore, it is important to know the spin coherence times (both $T_{1}$ and $T_{2}$) of the carriers in order to optimize the design of the device. In the following, we will calculate the spin coherence time for some common spin-LED devices used to investigate the efficiency of electrical spin injection.

\begin{figure}[htbp]
\setcaptionwidth{8.0 cm}
\scalebox{1}{\includegraphics[width= 8.0 cm]{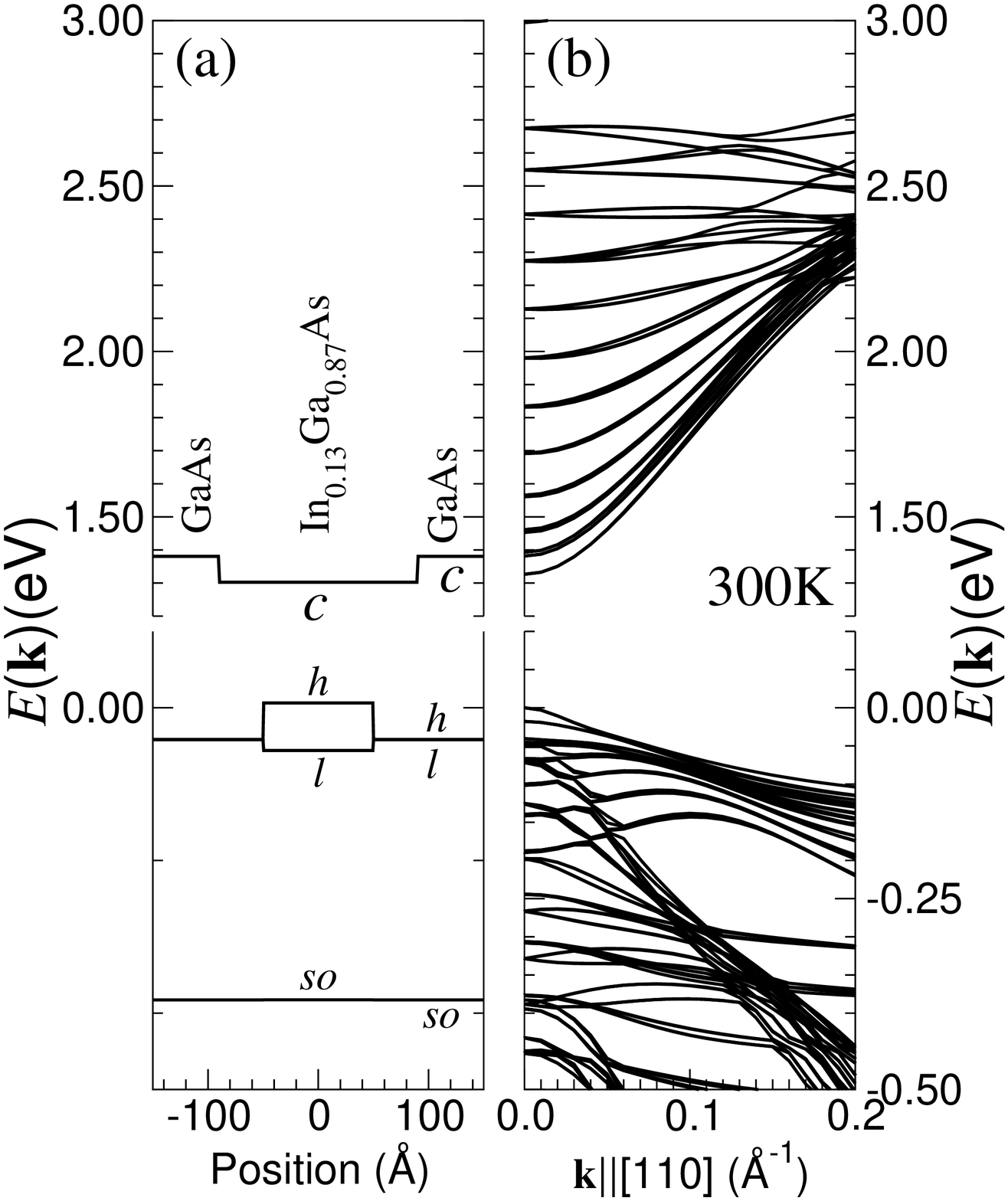}}
\caption[]{Band-edge diagram (a) and band structure (b) for a $100$-\AA\ In$_{0.13}$Ga$_{0.87}$As/$200$-\AA\ GaAs (001)-oriented quantum well~\cite{ohno99} at 300 K. The labels {\it c}, {\it l}, {\it h}, and {\it so} identify the conduction, light-hole, heavy-hole, and spin-orbit split-off hole band edges, respectively, at the zone center of the bulk constituent semiconductors.}
\label{InGaAsGaAsBands}
\end{figure}

\begin{figure}[htbp]
\setcaptionwidth{8.0 cm}
\scalebox{1}{\includegraphics[width = 8.0 cm]{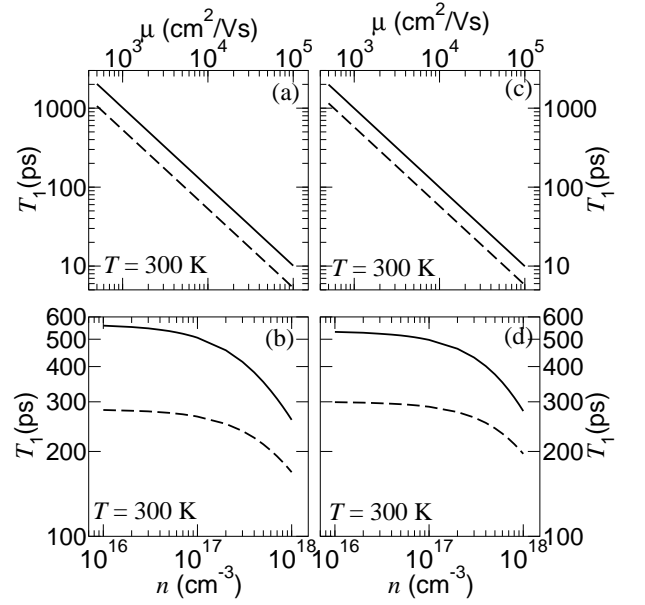}}
\caption[]{Mobility and carrier-density dependence of spin coherence time in spin-LED structures at 300K. (a) $T_1$ vs $\mu$ with $n=1 \times 10^{17}$ cm$^{-3}$ and (b) $T_1$ vs $n$ with $\mu = 2000$ cm$^{2}/$Vs for a $100$-\AA\ In$_{0.13}$Ga$_{0.87}$As/$200$-\AA\ GaAs (001)-oriented quantum well~\cite{ohno99}. (c) $T_1$ vs $\mu$ with $n=1 \times 10^{17}$ cm$^{-3}$ and (d) $T_1$ vs $n$ with $\mu = 2000$ cm$^{2}/$Vs for a $40$-\AA\ In$_{0.20}$Ga$_{0.80}$As/$100$-\AA\ GaAs/$40$-\AA\ In$_{0.20}$Ga$_{0.80}$As/$200$-\AA\ GaAs (001)-oriented quantum well~\cite{zhu01}. Solid and dashed lines respectively represent the results for OP and NI scattering.}
\label{spinledT1MN}
\end{figure}

The electronic band structure and band-edge diagram for a $100$-\AA\ In$_{0.13}$Ga$_{0.87}$As/$200$-\AA\ GaAs QW~\cite{ohno99} are shown in Figs.~\ref{InGaAsGaAsBands}(a)--(b), and the band-gap energy is $1326$ meV ($0.94$ ${\mu}$m). The lowest electron spin splitting of the lowest states as a function of crystal momentum ${\bf k}||[110]$ and excess energy for the quantum well is displayed in Figs.~\ref{qwDeltaE}(c)--(d). Figure~\ref{spinledT1MN}(a) shows the mobility dependence of the spin coherence times at room temperature. The $T_{1}$'s for optical phonon scattering differ from those for neutral impurity scattering by a factor of $2$, and the inverse relationship between the electron spin coherence time and mobility is the signature of DP spin relaxation mechanism for systems with a non-degenerate carrier density [see Eq.~(\ref{t1})]. The electron spin coherence times as a function of carrier density are illustrated in Fig.~\ref{spinledT1MN}(b), and the non-linear behavior of the electron spin coherence times are due to $T_{1}^{-1}$ $\propto$ $\Omega_{i l}^{2}({\cal L'}, E)$ [see Figs.~\ref{qwOmegaE}(a)--(d)]. 

\begin{figure}[htbp]
\setcaptionwidth{8.0 cm}
\scalebox{1}{\includegraphics[width= 8.0 cm]{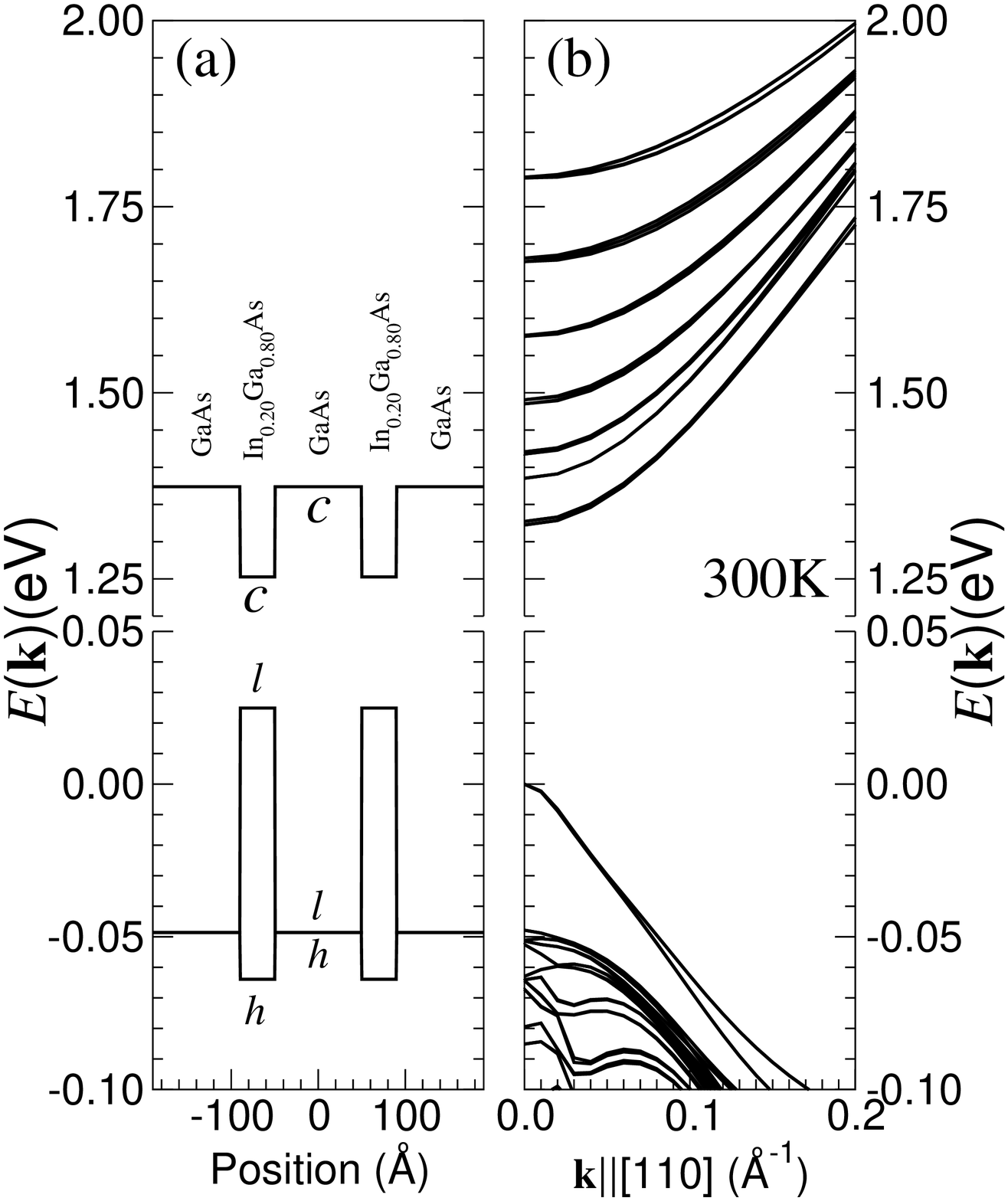}}
\caption[]{Band-edge diagram (a) and band structure (b) for a $40$-\AA\ In$_{0.20}$Ga$_{0.80}$As/$100$-\AA\ GaAs/$40$-\AA\ In$_{0.20}$Ga$_{0.80}$As/$200$-\AA\ GaAs (001)-oriented quantum well~\cite{zhu01} at 300 K. The labels {\it c}, {\it l}, and {\it h} identify the conduction, light-hole, and heavy-hole band edges, respectively, at the zone center of the bulk constituent semiconductors.}
\label{spinledBands}
\end{figure}

The band-edge diagram and electronic band structure for a $40$-\AA\ In$_{0.20}$Ga$_{0.80}$As/$100$-\AA\ GaAs/$40$-\AA\ In$_{0.20}$Ga$_{0.80}$As/$200$-\AA\ GaAs double quantum well~\cite{zhu01} at room temperature are depicted in Figs.~\ref{spinledBands}(a)--(b) with band-gap energy of $1322$ meV ($0.94$ ${\mu}$m). The lowest conduction-band spin splitting as a function of ${\bf k}||[110]$ and excess energy for the quantum well is plotted in Figs.~\ref{qwDeltaE}(c)--(d). The mobility and carrier-density dependence of the spin coherence times at room temperature are respectively shown in Figs.~\ref{spinledT1MN}(c) and \ref{spinledT1MN}(d). Similar dependences of the electron spin coherence times on mobility and carrier density are found.

\begin{figure}[htbp]
\setcaptionwidth{8.0 cm}
\scalebox{1}{\includegraphics[width= 8.0 cm]{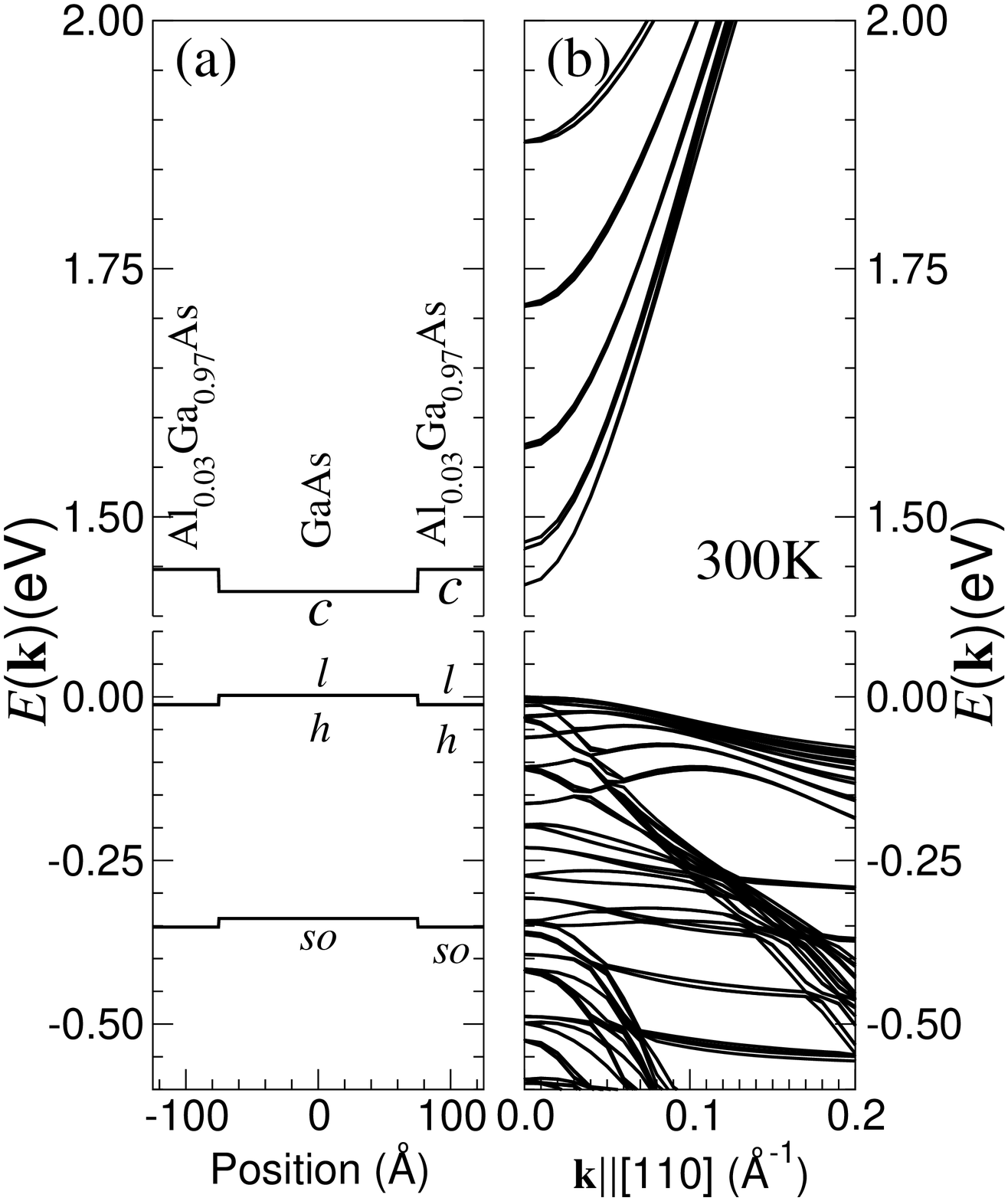}}
\caption[]{Band-edge diagram (a) and band structure (b) for a $150$-\AA\ GaAs/$100$-\AA\ Al$_{0.03}$Ga$_{0.97}$As (001)-oriented quantum well~\cite{fiederling99} at 300 K. The labels {\it c}, {\it l}, {\it h}, and {\it so} identify the conduction, light-hole, heavy-hole, and spin-orbit split-off hole band edges, respectively, at the zone center of the bulk constituent semiconductors.}
\label{GaAsAlGaAsBands}
\end{figure}

\begin{figure}[htbp]
\setcaptionwidth{8.0 cm}
\scalebox{1}{\includegraphics[width= 8.0 cm]{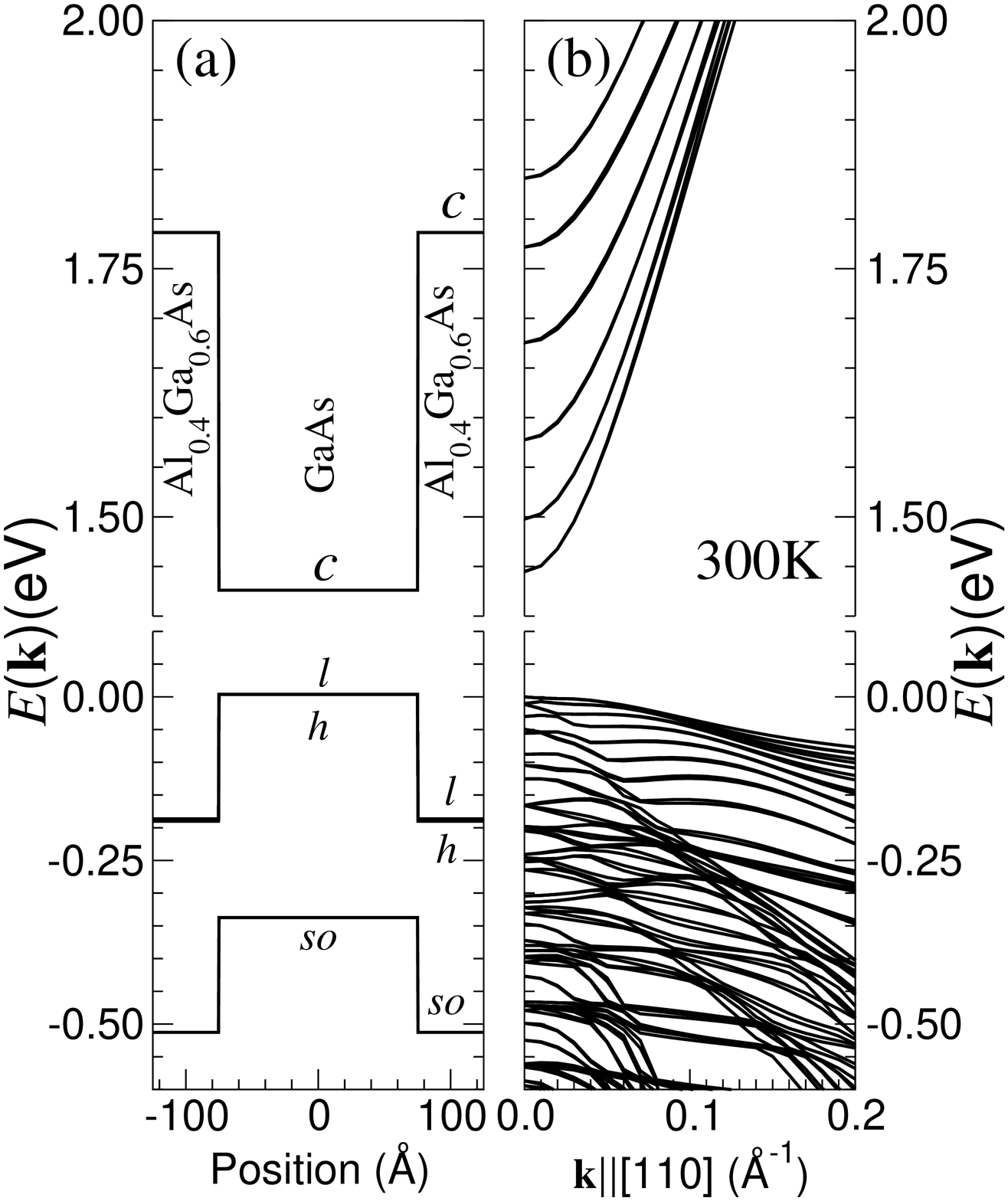}}
\caption[]{Band-edge diagram (a) and band structure (b) for a $150$-\AA\ GaAs/$100$-\AA\ Al$_{0.4}$Ga$_{0.6}$As (001)-oriented quantum well at 300 K. The labels {\it c}, {\it l}, {\it h}, and {\it so} identify the conduction, light-hole, heavy-hole, and spin-orbit split-off hole band edges, respectively, at the zone center of the bulk constituent semiconductors.}
\label{GaAsAlGaAsBands2}
\end{figure}

\begin{figure}[htbp]
\setcaptionwidth{8.0 cm}
\scalebox{1}{\includegraphics[width = 8.0 cm]{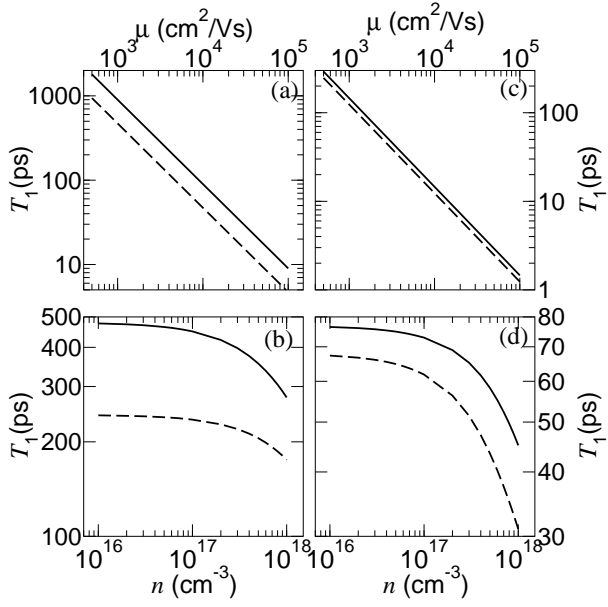}}
\caption[]{Mobility and carrier-density dependence of spin coherence time in spin-LED structures at room temperature. $T_1$ vs $\mu$ with $n=1 \times 10^{17}$ cm$^{-3}$ (a) and $T_1$ vs $n$ with $\mu = 2000$ cm$^{2}/$Vs (b) for a $150$-\AA\ GaAs/$100$-\AA\ Al$_{0.03}$Ga$_{0.97}$As (001)-oriented quantum well~\cite{fiederling99}. $T_1$ vs $\mu$ with $n=1 \times 10^{17}$ cm$^{-3}$ (c) and $T_1$ vs $n$ with $\mu = 2000$ cm$^{2}/$Vs (d) for a $150$-\AA\ GaAs/$100$-\AA\ Al$_{0.4}$Ga$_{0.6}$As (001)-oriented quantum well. Solid and dashed lines respectively represent the results for OP and NI scattering.}
\label{spinledT1MN2}
\end{figure}

Next we would like to compare the electron spin coherence times of two spin-LED structures with different aluminum concentration in the barriers for otherwise identical quantum wells. The band-edge diagrams and electronic band structures for these two structures are shown in Figs.~\ref{GaAsAlGaAsBands}(a)--(b) and \ref{GaAsAlGaAsBands2}(a)--(b). The band-gap energy for the spin-LED structure shown in Figs.~\ref{GaAsAlGaAsBands}(a)--(b) is $1432$ meV ($0.87$ ${\mu}$m), and the band-gap energy for the spin-LED structure displayed in Figs.~\ref{GaAsAlGaAsBands2}(a)--(b) is $1445$ meV ($0.86$ ${\mu}$m). The conduction-band spin splitting of the lowest states as a function of crystal momentum ${\bf k}||[110]$ and excess energy for these two spin-LED structures is illustrated in Figs.~\ref{qwDeltaE}(c)--(d). The mobility and carrier-density dependence of the electron spin coherence times for a $150$-\AA\ GaAs/$100$-\AA\ Al$_{0.03}$Ga$_{0.97}$As QW~\cite{fiederling99} at room temperature is plotted respectively in Figs.~\ref{spinledT1MN2}(a) and \ref{spinledT1MN2}(b), while the mobility and carrier-density dependence of the electron spin coherence times for a $150$-\AA\ GaAs/$100$-\AA\ Al$_{0.4}$Ga$_{0.6}$As QW at room temperature is respectively shown in Figs.~\ref{spinledT1MN2}(c) and \ref{spinledT1MN2}(d). The electron spin coherence times for the quantum well with $3\%$ aluminum concentration in the barriers are approximately a factor of $5$ longer than those of the quantum well with $40\%$ aluminum concentration in the barriers. As the aluminum concentration increases, the penetration of the electron wavefunctions into the barriers decreases due to the increase of potential energy in the barriers. As a result, the conduction spin splittings increase as the confinement energy increases [see Figs.~\ref{qwOmegaE}(c) and \ref{qwOmegaE}(d)]. Consequently the electron spin coherence times decrease as the aluminum concentration increases. 

\begin{figure}[htbp]
\setcaptionwidth{8.0 cm}
\scalebox{1}{\includegraphics[width = 8.0 cm]{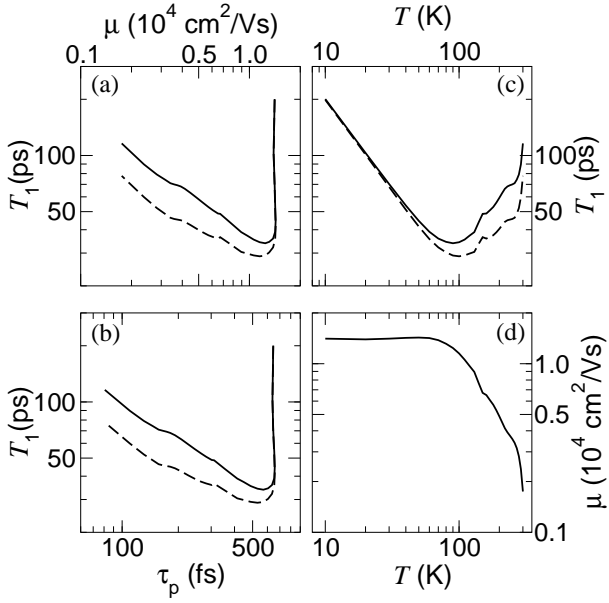}}
\caption[]{Calculated $T_1$ as a function of (a) mobility $\mu$, (b) average transport time $\tau_\text{p}$, and (c) Temperature $T$, for a $50$-\AA\ GaAs/$100$-\AA\ Ga$_{0.51}$In$_{0.49}$P QW~\cite{elhamri96} with carrier density $n=2.67 \times 10^{17}$ cm$^{-3}$. (d) Measured mobility as a function of temperature~\cite{ghezzi01}. Solid and dashed lines respectively represent the results for OP and NI scattering.}
\label{GaAsGaInPT1MNP}
\end{figure}

Next we consider a thin $50$-\AA\ GaAs quantum well with $100$-\AA\ Ga$_{0.51}$In$_{0.49}$P as barriers. Shown in Figs.~\ref{GaAsGaInPT1MNP}(a)--(c) are the dependence of the electron spin coherence times on the electron mobility, average electron momentum scattering time, and temperature. Although the temperature dependence of the electron mobility is similar to that of the GaSb quantum well considered above [see Figs.~\ref{GaSbAlGaSbT1MNP}(d) and \ref{GaAsGaInPT1MNP}(d)], the dependence of the electron spin coherence times on temperature is quite different [c.f., Figs.~\ref{GaSbAlGaSbT1MNP}(c) and \ref{GaAsGaInPT1MNP}(c)]. In this case the electron spin coherence times are proportional to the electron mobility at high temperature and inversely proportional to the electron mobility at low temperature. 
\subsubsection{II-VI LAYERED SEMICONDUCTORS}
\label{sec:qw26}
\begin{figure}[htbp]
\setcaptionwidth{8.0 cm}
\scalebox{1}{\includegraphics[width = 8.0 cm]{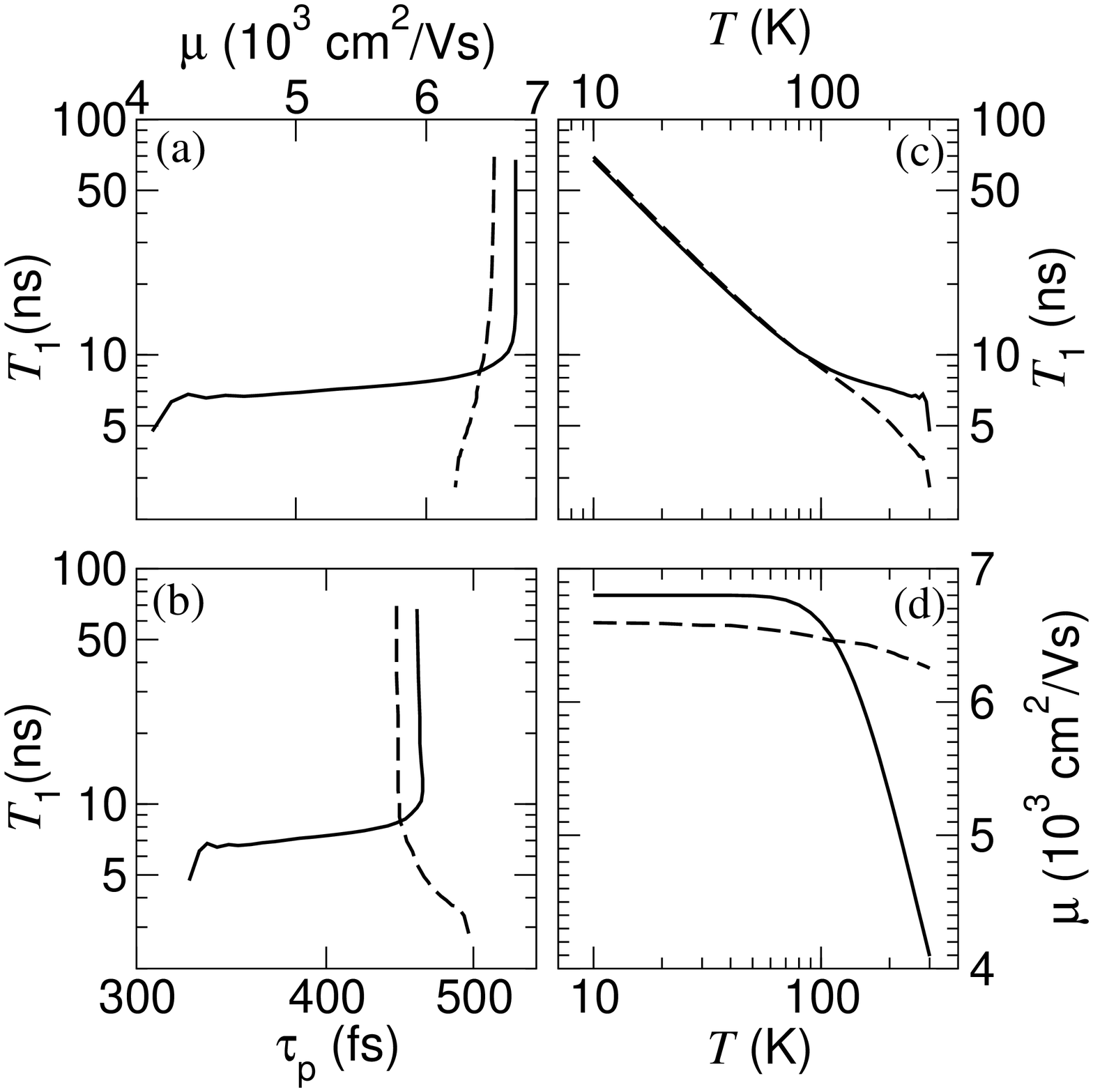}}
\caption[]{Calculated $T_1$ as a function of (a) mobility $\mu$, (b) average transport time $\tau_\text{p}$, and (c) Temperature $T$, for a $105$-\AA\ CdSe/$500$-\AA\ ZnSe QW~\cite{ng99} with carrier density $n=7.93 \times 10^{16}$ cm$^{-3}$. (d) Extrapolated mobility as a function of temperature~\cite{ghezzi01}. Solid and dashed lines respectively represent the results for OP and NI scattering.}
\label{CdSeZnSeT1MNP}
\end{figure}

Recently it has been demonstrated experimentally by Kikkawa {\it et al.}~\cite{kikkawa97} that II-VI semiconductor quantum wells have remarkably long electron spin coherence times, of the order of nanoseconds, which persist to room temperature. Here we look at the electron spin decoherence of two II-VI semiconductor quantum wells: CdSe/ZnSe and CdZnSe/ZnSe quantum wells. The electron spin coherence times as a function of mobility, average momentum scattering time, and temperature for a $105$-\AA\ CdSe/$500$-\AA\ ZnSe quantum well~\cite{ng99} are plotted in Figs.~\ref{CdSeZnSeT1MNP}(a)--(c). The electron mobilities are extrapolated from a measured value of $6800$ cm$^2$/Vs at $4.2$ K assuming neutral impurity or optical phonon scattering, and the results are displayed in Fig.~\ref{CdSeZnSeT1MNP}(d). The calculated electron spin coherence times are proportional to the temperature, but they are inversely proportional to the electron mobility because the extrapolated electron mobilities are inversely proportional to the temperature. The calculated electron spin coherence times at room temperature are approximately $4.7$ ns and $2.7$ ns for optical phonon and neutral impurity scattering, respectively. The calculated electron spin coherence times are comparable to those of the experimental measurements reported in Ref.~\onlinecite{kikkawa97}.

\begin{figure}[htbp]
\setcaptionwidth{8.0 cm}
\scalebox{1}{\includegraphics[width= 8.0 cm]{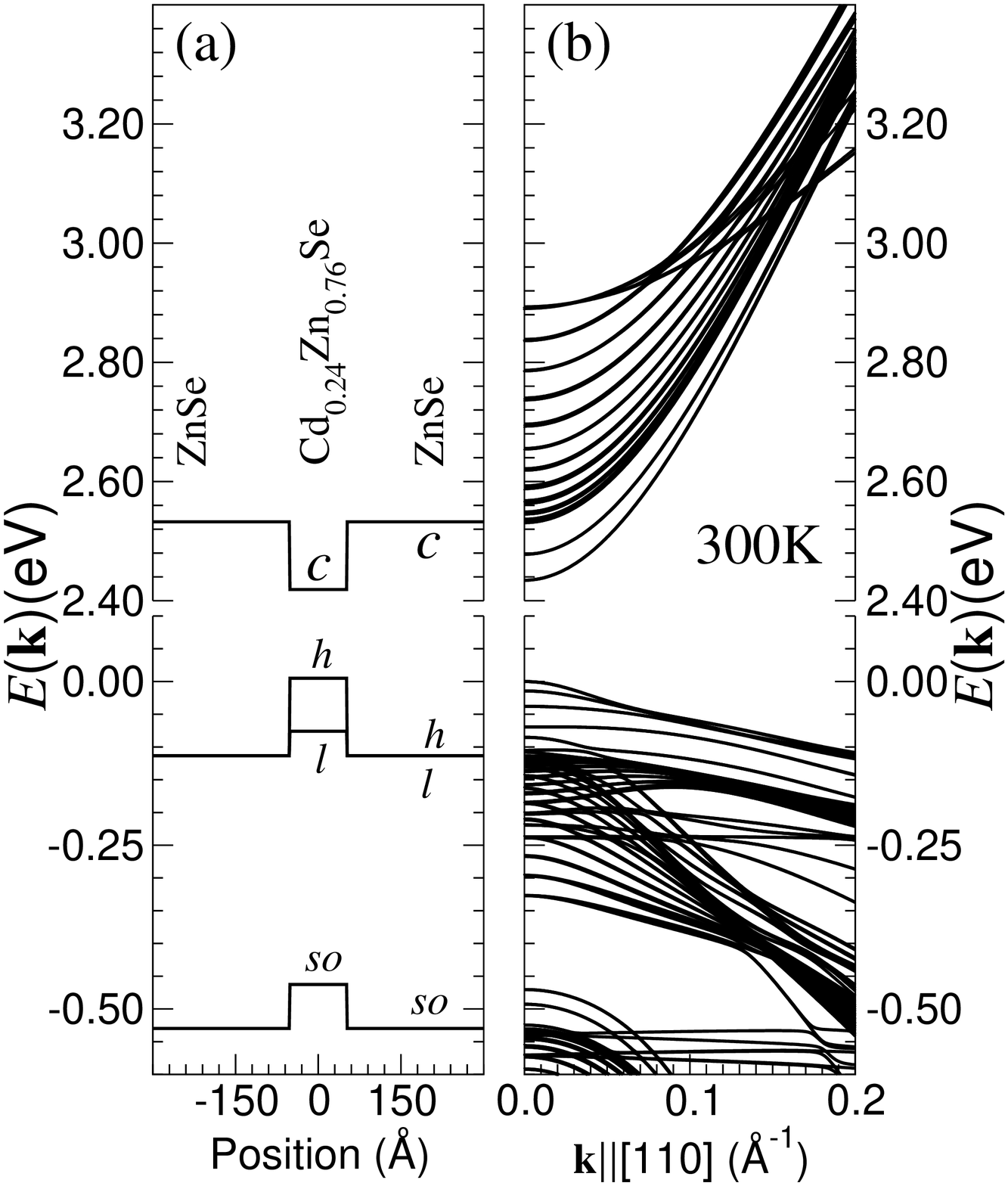}}
\caption[]{Band-edge diagram (a) and band structure (b) for a $105$-\AA\ Cd$_{0.24}$Zn$_{0.76}$Se/$500$-\AA\ ZnSe (001)-oriented quantum well~\cite{ng99} at 300 K. The labels {\it c}, {\it l}, {\it h}, and {\it so} identify the conduction, light-hole, heavy-hole, and spin-orbit split-off hole band edges, respectively, at the zone center of the bulk constituent semiconductors.}
\label{CdZnSeZnSeBands}
\end{figure}

\begin{figure}[htbp]
\setcaptionwidth{8.0 cm}
\scalebox{1}{\includegraphics[width = 8.0 cm]{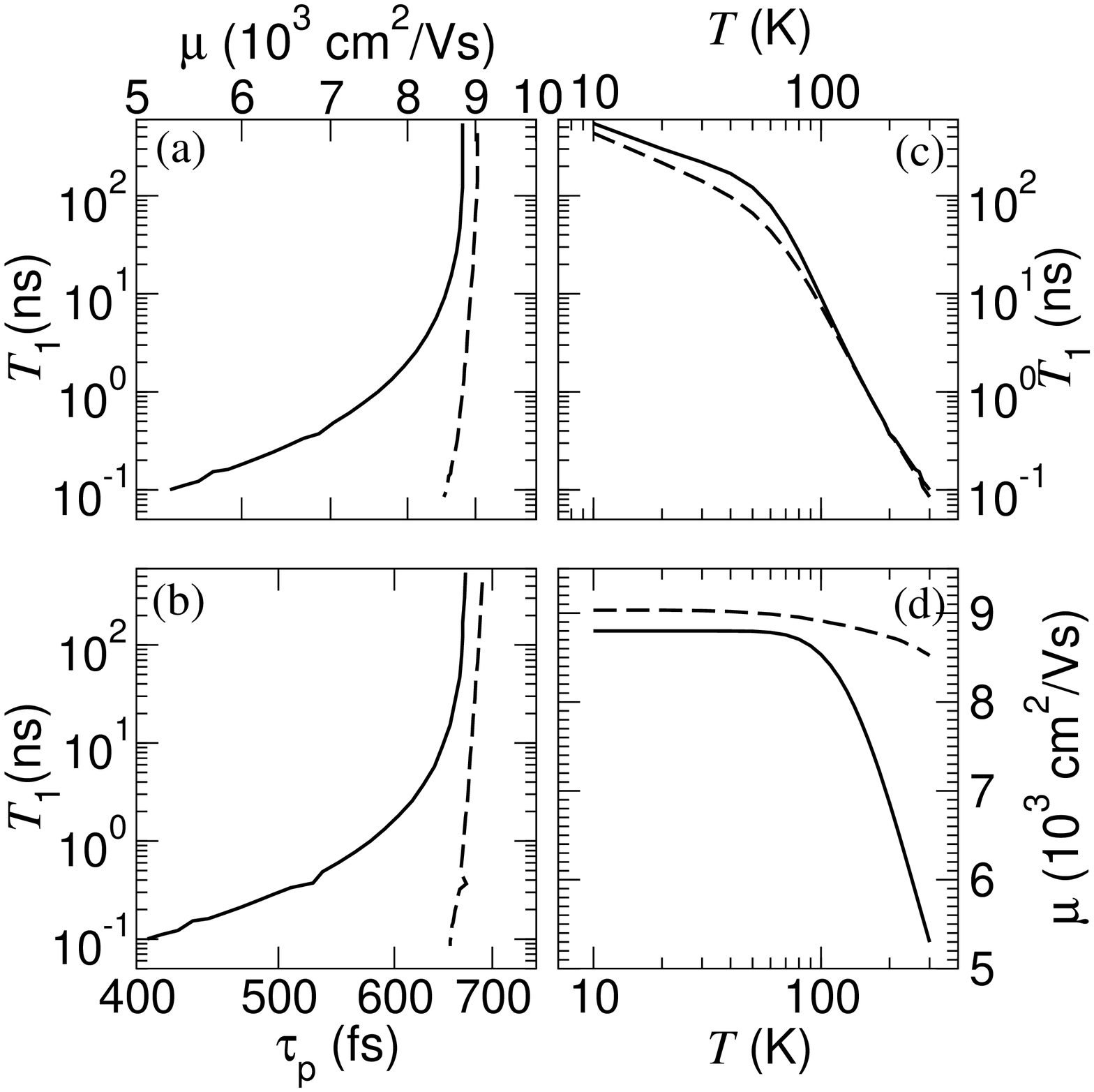}}
\caption[]{Calculated $T_1$ as a function of (a) mobility $\mu$, (b) average transport time $\tau_\text{p}$, and (c) Temperature $T$, for a $105$-\AA\ Cd$_{0.24}$Zn$_{0.76}$Se/$500$-\AA\ ZnSe QW~\cite{ng99} with carrier density $n=7.44 \times 10^{16}$ cm$^{-3}$. (d) Extrapolated mobility as a function of temperature~\cite{ghezzi01}. Solid and dashed lines respectively represent the results for OP and NI scattering.}
\label{CdZnSeZnSeT1MNP}
\end{figure}

Shown in Figs.~\ref{CdZnSeZnSeBands}(a)--(b) are the band-edge diagram and electronic band structure for a $105$-\AA\ Cd$_{0.24}$Zn$_{0.76}$Se/$500$-\AA\ ZnSe quantum well~\cite{ng99} with band-gap energy of $2425$ meV ($0.51$ ${\mu}$m). Figures~\ref{CdZnSeZnSeT1MNP}(a)--(c) illustrate the dependence of the electron spin coherence times on electron mobility, average momentum scattering time, and temperature. The temperature dependence of the electron mobilities are shown in Fig.~\ref{CdZnSeZnSeT1MNP}(d), which is obtained by extrapolating the electron mobility from a measured value of $8800$ cm$^2$/Vs at $4.2$ K assuming neutral impurity and optical phonon scattering. The dependence of the electron spin coherence times on mobility, average momentum scattering time, and temperature is very similar to that of the CdSe quantum well considered above. 

\section{VALIDITY OF DK APPROXIMATIONS}
\label{sec:vdkt}
The origin of the discrepancies between the DK theoretical and the experimental results can be traced back to the approximations made in the derivation of $T_{1}$ in the DK theory. To summarize, the expression for the electron spin coherence time $T_{1}$ in the DK theory for $(001)$-quantum wells is obtained as follows (see Sec.~\ref{sec:paqw}). First, it is assumed that the penetration of the electronic wavefunctions into the barriers in the quantum well is negligible, hence $E_{\text{1e}} \ll \Delta E_c$, where $E_{\text{1e}}$ is the electron confinement energy of the first quantum-well state and $\Delta E_c$ is the conduction band offset. Second, the perturbative expressions \cite{ivchenko97} $\widetilde{\Omega}_{x1}^2(1,E)\propto E(4E_{\text{1e}}-E)^2$ and $\widetilde{\Omega}_{x3}^2(1,E)\propto E^3$ are used. Third, the energies of relevant states are assumed to be $E \ll E_{\text{1e}}$, and thus (i) the contribution from $\widetilde{\Omega}_{x3}(1,E)$ is ignored, and (ii) it is assumed that $\widetilde{\Omega}_{x1}^2(1,E)\propto E$. It is not generally recognized that the conditions $k_BT\ll E_{\text{1e}}\ll \Delta E_c$ are quite restrictive and are difficult to satisfy at room temperature. It should be noted that in this approximation only the first quantum-well state is occupied, and hence it can be only applied to quantum wells with large energy separation between the first and the second quantized states.

Under the DK assumptions, the electron spin coherence time $T_{1}$ is inversely proportional to the mobility {\it independent of the dominant scattering mechanism}. In addition, $T_1$ is inversely proportional to the confinement energy $E_{\text{1e}}^2$. These trends are not supported by recent experimental measurements \cite{terauchi99} illustrated in Figs.~\ref{GaAsAlGaAsT1MNP}(a) and \ref{GaAsAlGaAsT1R1EL}(a). In both cases the experimental trends are weaker than the predicted theoretical ones. The weaker dependence of $T_1$ on $E_{\text{1e}}$ in our theory versus DK theory is due to wavefunction penetration into the barriers and non-perturbative effects (the spin splitting for damped waves in the barrier is of the opposite sign as that for propagating waves in the well).

\begin{figure}[htbp]
\setcaptionwidth{8.0 cm}
\scalebox{1}{\includegraphics[width = 8.0 cm]{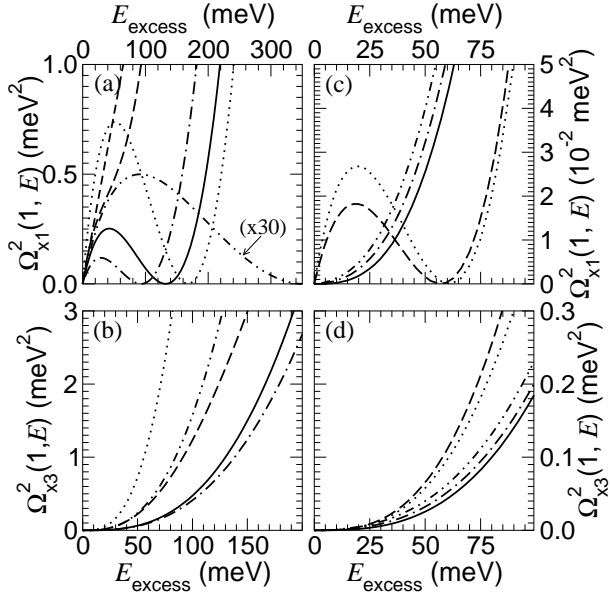}}
\caption[]{Energy dependence of electron spin precession vector for III-V quantum wells and superlattices at room temperature. (a) $\widetilde{\Omega}_{x1}^2(1, E)$ and (b) $\widetilde{\Omega}_{x3}^2(1, E)$ for a $75$-\AA\ GaAs/$100$-\AA\ Al$_{0.4}$Ga$_{0.6}$As QW (solid line), a thin-layer $21.2$-\AA\ InAs/$36.6$-\AA\ GaSb SL (dashed line), a thin-layer $75$-\AA\ InAs/$200$-\AA\ AlSb QW (dot-dot dashed line), an $70$-\AA\ In$_{0.53}$Ga$_{0.47}$As/97-\AA\ InP QW (dot-dashed line), and a $80$-\AA\ GaSb$_{0.81}$/$80$-\AA\ AlSb QW (dotted line). The short-dashed line is the DK approximation for the $75$-\AA\ GaAs/$100$-\AA\ Al$_{0.4}$Ga$_{0.6}$As quatum well. (c) $\widetilde{\Omega}_{x1}^2(1, E)$ and (d) $\widetilde{\Omega}_{x3}^2(1, E)$ for a $150$-\AA\ GaAs/$100$-\AA\ Al$_{0.03}$Ga$_{0.97}$As QW (solid line), a $150$-\AA\ GaAs/$100$-\AA\ Al$_{0.4}$Ga$_{0.6}$As QW (dashed line), a $100$-\AA\ In$_{0.13}$Ga$_{0.87}$As/$200$-\AA\ GaAs QW (dot-dashed), and  a $95$-\AA\ In$_{0.57}$Ga$_{0.43}$As$_{0.93}$P$_{0.07}$/$75$-\AA\ In$_{0.87}$Ga$_{0.13}$As$_{0.29}$P$_{0.71}$ QW (dotted line), and a $40$-\AA\ In$_{0.20}$Ga$_{0.80}$As/$100$-\AA\ GaAs/$40$-\AA\ In$_{0.20}$Ga$_{0.80}$As/$200$-\AA\ GaAs QW (dot-dot-dashed line).}
\label{qwOmegaE}
\end{figure}

The validity of DK approximations can be further analyzed quantitatively by comparing the energy dependence of $\widetilde{\Omega}_{x1}^2(1,E)$ and $\widetilde{\Omega}_{x3}^2(1,E)$ for these material systems. In DK theory, $\widetilde{\Omega}_{x1}^2(1,E)$ is a linear function of energy $E$ (see Sec.~\ref{sec:paqw} and Ref.~\onlinecite{dyakonov86}); however, Figs.~\ref{qwOmegaE}(a) and \ref{qwOmegaE}(c) show that for quantum wells $\widetilde{\Omega}_{x1}^2(1,E)$ is only linear for a small energy range ($E < k_\text{B} T$) above the band edge before it begins to deviate [c.f., the short dashed line based on the DK theory and the solid line in Fig.~\ref{qwOmegaE}(a)]. More energetic states than this certainly contribute to the electron spin decoherence at room temperature. The wider the well the lower the energy where $\widetilde{\Omega}_{x1}^2(1,E)$ deviates from linear behavior [c.f., the solid line in Figs.~\ref{qwOmegaE}(a) and \ref{qwOmegaE}(c)], as it approaches a bulk-like $E^3$ behavior [Fig.~\ref{BulkOmegaE1}]. $\widetilde{\Omega}_{x3}^2(1,E)$ for the material systems considered above is shown in Figs.~\ref{qwOmegaE}(b) and \ref{qwOmegaE}(d), and for a large width well is comparable in magnitude to $\widetilde{\Omega}_{x1}^2(1,E)$. 

The applicability of the perturbative approach for calculations of electron spin coherence times in two-dimensional systems can be explored by carefully examining the energy dependence of $\widetilde{\Omega}_{x1}(1,E)$ and $\widetilde{\Omega}_{x3}(1,E)$ for various structures. As the wells become narrower, even the perturbative expressions for $\widetilde{\Omega}_{x1}(1,E)$ and $\widetilde{\Omega}_{x3}(1,E)$ break down. Figures~\ref{qwOmegaE}(a) and \ref{qwOmegaE}(b) show $\widetilde{\Omega}_{x1}^2(1,E)$ and $\widetilde{\Omega}_{x3}^2(1,E)$ (long dashed lines) for a thin-layer type-II InAs/GaSb superlattice, indicating very different behavior from the other structures, poorly reproduced by even the general forms of the perturbative expression. 

\section{CONCLUDING REMARKS}
\label{sec:cr}
We have presented a {\it quantitatively accurate} non-perturbative nanostructure theory for electron spin decoherence in bulk and quantum-well zincblende semiconductors based on a fourteen-band model. The calculated electron spin coherence times in bulk and layered zincblende semiconductors are in agreement with experimental measurements, indicating the importance of accurate band-structure calculations for zincblende type nanostructures. We have explicitly identified the important assumptions of the DK theory regarding both quantum-well electronic structure and carrier dynamics, and the origins of the discrepancies between the results of the DK theory and a full fourteen-band theory. The validity and limitation of the perturbative approach are investigated numerically through specific examples, and the results show that the perturbative approach breaks down for systems with narrow wells. We have shown the important role of momentum scattering mechanisms in determining the electron spin coherence times by studying the mobility dependence of $T_1$ for various momentum scattering mechanisms. 
\begin{acknowledgments}
This work was supported by DARPA/ARO DAAD19-01-1-0490.
\end{acknowledgments}

\begin{table*}
\caption[]{The ratio $\eta^{(3D)}$ of the transport time $\widetilde{\tau}_1(n,k)$ 
and the momentum relaxation time $\widetilde{\tau}_l(n,k)$, 
and the power law of the energy-dependent transport time 
[$\widetilde{\tau}_1(n,k)$ $\propto$ $E^{\nu}_{n \sigma}({\bf k})$] for ionized impurity, 
optical phonon, and neutral impurity scattering in bulk semiconductors 
\cite{meier84}.}
\begin{ruledtabular}
\begin{tabular}{cccc}
&II  
&OP  
&NI\\ 
\colrule
$\nu$
&$\frac{3}{2}$
&$\frac{1}{2}$ 
&$0$\\
  
$\frac{\widetilde{\tau}_1(n,k)}{\widetilde{\tau}_l(n,k)}$           
&$\sum_{j=0}^{[l/2]}\frac{(-1)^j(2l-2j)!}{2^l j! (l-j)! (l-2j-1)!} $       
&$\sum_{j=0}^{[l/2]}\sum_{m=0}^{l-2j}\frac{(-1)^j(2l-2j)! [1-(-1)^m]}{2^{l+1} j! (l-j)! (l-2j)! m}$      
&$1$\\
\label{bulkMS}
\end{tabular}
\end{ruledtabular}
\footnotetext{$[l/2] = l/2$ for $l$ even, and $(l-1)/2$ for odd $l$.}
\end{table*} 
\begin{table*}
\caption[]{The ratio $\eta^{(2D)}$ of the transport time $\widetilde{\tau}_1({\cal L}, K_{\|}, K_z)$ 
and the momentum relaxation time $\widetilde{\tau}_l ({\cal L}, K_{\|}, K_z)$ 
for different angular index $l$, and the power law
of the energy-dependent transport time 
[$\widetilde{\tau}_1({\cal L}, K_{\|}, K_z)$ $\propto$ $E^{\nu}_{L S}({\bf K})$] for various
scattering mechanisms in quantum wells or superlattices \cite{shik97}.}
\begin{ruledtabular}
\begin{tabular}{cccc}
&II  
&OP  
&NI\\ 
\colrule
$\nu$
&$2$
&$1$ 
&$0$\\
  
$\frac{\widetilde{\tau}_1 ({\cal L}, K_{\|}, K_z)}{\widetilde{\tau}_l ({\cal L}, K_{\|}, K_z)}$           
&$l^2$        
&$l$       
&$1$\\
\label{mqwMS}
\end{tabular}
\end{ruledtabular}
\end{table*} 
\begin{table*}
\caption[]{Bulk basis states $\arrowvert n, \sigma, {\bf 0}\rangle$ 
used in the fourteen-band model for the $\Gamma_{6}$ 
bulk band edge and the bonding and antibonding $\Gamma_{7}$ 
and $\Gamma_{8}$ bulk band edges.}
\begin{ruledtabular}
\begin{tabular}{cccccc}
n 
&$\sigma$ 
&Representation   
&Designation  
&$\arrowvert J,J_{z}\rangle$   
&$\arrowvert n, \sigma, {\bf 0}\rangle$\\
\colrule
1
&$\uparrow$   
&$\Gamma_{6} \left( \overline{s} \right)$  
&$c_{\uparrow}$  
&$\arrowvert \frac{1}{2},+ \frac{1}{2}\rangle$     
&$\arrowvert S_{\uparrow} \rangle$\\
  
2  
&$\uparrow$ 
&$\Gamma_{8} \left( p \right)$  
&$h_{\downarrow}$  
&$\arrowvert \frac{3}{2},- \frac{3}{2}\rangle$     
&$\sqrt{\frac{1}{2}} \arrowvert X_{\downarrow} - iY_{\downarrow} \rangle$\\ 

3 
&$\uparrow$  
&$\Gamma_{8} \left( p \right)$  
&$l_{\uparrow}$  &$\arrowvert \frac{3}{2},+ \frac{1}{2}\rangle$       
&$\sqrt{\frac{2}{3}} \arrowvert Z_{\uparrow} \rangle - 
  \sqrt{\frac{1}{6}} \arrowvert X_{\downarrow} + iY_{\downarrow} \rangle$\\ 

4 
&$\uparrow$  
&$\Gamma_{7} \left( p \right)$  
&$so_{\uparrow}$  &$\arrowvert \frac{1}{2},+ \frac{1}{2}\rangle$   
&$\sqrt{\frac{1}{3}} \arrowvert Z_{\uparrow} \rangle + 
  \sqrt{\frac{1}{3}} \arrowvert X_{\downarrow} + iY_{\downarrow} \rangle$\\
  
5
&$\uparrow$
&$\Gamma_{8} \left( \overline{p} \right)$  
&$\overline{h}_{\downarrow}$  &$\arrowvert \frac{3}{2},- \frac{3}{2}\rangle$   
&$\sqrt{\frac{1}{2}} \arrowvert X_{\downarrow}^{c} - iY_{\downarrow}^{c} \rangle$\\ 

6 
&$\uparrow$ 
&$\Gamma_{8} \left( \overline{p} \right)$  
&$\overline{l}_{\uparrow}$  &$\arrowvert \frac{3}{2},+ \frac{1}{2}\rangle$     
&$\sqrt{\frac{2}{3}} \arrowvert Z_{\uparrow}^{c} \rangle - 
  \sqrt{\frac{1}{6}} \arrowvert X_{\downarrow}^{c} + iY_{\downarrow}^{c} \rangle$\\
 
7   
&$\uparrow$
&$\Gamma_{7} \left( \overline{p} \right)$  
&$\overline{so}_{\uparrow}$  &$\arrowvert \frac{1}{2},+ \frac{1}{2}\rangle$   
&$\sqrt{\frac{1}{3}} \arrowvert Z_{\uparrow}^{c} \rangle + 
  \sqrt{\frac{1}{3}} \arrowvert X_{\downarrow}^{c} + iY_{\downarrow}^{c} \rangle$\\

1  
&$\downarrow$ 
&$\Gamma_{6} \left( \overline{s} \right)$  
&$c_{\downarrow}$ 
&$\arrowvert \frac{1}{2},- \frac{1}{2}\rangle$   
&$\arrowvert S_{\downarrow} \rangle$\\

2 
&$\downarrow$   
&$\Gamma_{8} \left( p \right)$ 
&$h_{\uparrow}$  
&$\arrowvert \frac{3}{2},+ \frac{3}{2}\rangle$   
&$\sqrt{\frac{1}{2}} \arrowvert X_{\uparrow} + iY_{\uparrow} \rangle$\\

3  
&$\downarrow$ 
&$\Gamma_{8} \left( p \right)$  
&$l_{\downarrow}$  &$\arrowvert \frac{3}{2},- \frac{1}{2}\rangle$      
&$\sqrt{\frac{2}{3}} \arrowvert Z_{\downarrow} \rangle + 
  \sqrt{\frac{1}{6}} \arrowvert X_{\uparrow} - iY_{\uparrow} \rangle$\\ 

4  
&$\downarrow$ 
&$\Gamma_{7} \left( p \right)$  
&$so_{\downarrow}$  
&$\arrowvert \frac{1}{2},- \frac{1}{2}\rangle$   
&$\sqrt{\frac{1}{3}} \arrowvert Z_{\downarrow} \rangle - 
  \sqrt{\frac{1}{3}} \arrowvert X_{\uparrow} - iY_{\uparrow} \rangle$\\

5 
&$\downarrow$  
&$\Gamma_{8} \left( \overline{p} \right)$  
&$\overline{h}_{\uparrow}$  
&$\arrowvert \frac{3}{2},+ \frac{3}{2}\rangle$     
&$\sqrt{\frac{1}{2}} \arrowvert X_{\uparrow}^{c} + iY_{\uparrow}^{c} \rangle$\\

6  
&$\downarrow$ 
&$\Gamma_{8} \left( \overline{p} \right)$  
&$\overline{l}_{\downarrow}$  &$\arrowvert \frac{3}{2},- \frac{1}{2}\rangle$     
&$\sqrt{\frac{2}{3}} \arrowvert Z_{\downarrow}^{c} \rangle + 
  \sqrt{\frac{1}{6}} \arrowvert X_{\uparrow}^{c} - iY_{\uparrow}^{c} \rangle$\\
 
7
&$\downarrow$   
&$\Gamma_{7} \left( \overline{p} \right)$  
&$\overline{so}_{\downarrow}$  
&$\arrowvert \frac{1}{2},- \frac{1}{2}\rangle$   
&$\sqrt{\frac{1}{3}} \arrowvert Z_{\downarrow}^{c} \rangle - 
  \sqrt{\frac{1}{3}} \arrowvert X_{\uparrow}^{c} - iY_{\uparrow}^{c} \rangle$\\
\label{basis}
\end{tabular}
\end{ruledtabular}
\end{table*} 
\begin{table}
\caption[]{The in-plane vector ${\bf \Omega}_{\|} ({\cal L}, {\bf K})$
under symmetry operations of $(001)$-oriented structure. 
Here $\Phi = [0, \pi / 4]$ and 
$\overline{\Phi} \equiv \pi /4 - \Phi$.}
\begin{ruledtabular}
\begin{tabular}{ccc}
Quadrants
&$\Omega_x({\cal L}, {\bf K})$  
&$\Omega_y({\cal L}, {\bf K})$\\ 
\colrule
I&
$\Omega_x({\cal L}, K, \Theta, \Phi)$
&$\Omega_y({\cal L}, K, \Theta, \Phi)$\\
II&
$- \Omega_y({\cal L}, K, \Theta, \overline{\Phi})$
&$- \Omega_x({\cal L}, K, \Theta, \overline{\Phi})$\\
III&
$\Omega_y({\cal L}, K, \Theta, \Phi)$
&$- \Omega_x({\cal L}, K, \Theta, \Phi)$\\
IV&
$- \Omega_x({\cal L}, K, \Theta, \overline{\Phi})$
&$\Omega_y({\cal L}, K, \Theta, \overline{\Phi})$\\
V&
$- \Omega_x({\cal L}, K, \Theta, \Phi)$
&$- \Omega_y({\cal L}, K, \Theta, \Phi)$\\
VI&
$\Omega_y({\cal L}, K, \Theta, \overline{\Phi})$
&$\Omega_y({\cal L}, K, \Theta, \overline{\Phi})$\\
VII&
$- \Omega_y({\cal L}, K, \Theta, \Phi)$
&$\Omega_x({\cal L}, K, \Theta, \Phi)$\\
VII&
$\Omega_x({\cal L}, K, \Theta, \overline{\Phi})$
&$- \Omega_y({\cal L}, K, \Theta, \overline{\Phi})$\\
\label{toxy}
\end{tabular}
\end{ruledtabular}
\end{table} 
\begin{sidewaystable}
\caption[]{Parameters for fourteen-band ${\bf k} \cdot {\bf p}$ band structures 
for Al$_{x}$Ga$_{1-x}$Sb, Al$_{x}$Ga$_{1-x}$As, In$_{x}$Ga$_{1-x}$As, GaAs$_{x}$Sb$_{1-x}$, and Ga$_{x}$In$_{1-x}$P at room temperature. $E_g$ is the fundamental gap (between $\overline{s}$ and $p \left(\Gamma_{8} \right)$ states), $E_c$ is the zone-center energy difference between $p \left(\Gamma_{8} \right)$ and $\overline{p} \left(\Gamma_{7} \right)$, and $\Delta$ and $\Delta_{c}$ are the spin-orbit splitting of the $p$ and $\overline{p}$ states, respectively, $a_{\text{o}}$ is the lattice constant, $c_{11}$ and $c_{12}$ are the elastic constants, and $V_{a}$ and $V_{b}$ are the deformation potentials.}
\begin{ruledtabular}
\begin{tabular}{cccccc}
&Al$_{x}$Ga$_{1-x}$Sb  
&Al$_{x}$Ga$_{1-x}$As  
&In$_{x}$Ga$_{1-x}$As  
&GaAs$_{x}$Sb$_{1-x}$
&Ga$_{x}$In$_{1-x}$P\\ 
\colrule
$E_g$ (eV)                 
&$0.724+1.105x+0.471x^2$   
&$1.422+1.198x+0.433x^2$  
&$1.422-1.484x+0.418x^2$  
&$0.724-0.506x+1.204x^2$
&$1.351+0.647x+0.7859x^2$ \\
  
$\Delta$ (eV)            
&$0.673x+0.758(1-x)$       
&$0.30x+0.341(1-x)$       
&$0.38x+0.341(1-x)$       
&$0.34x+0.758(1-x)$
&$0.080x+0.108(1-x)$\\
 
$E_c$ (eV)               
&$3.50x+3.19(1-x)$         
&$4.48x+4.61(1-x)$        
&$4.63x+4.61(1-x)$        
&$4.77x+3.19(1-x)$
&$5.24x+4.80(1-x)$ \\
  
$\Delta_c$ (eV)          
&$0.0x+0.21(1-x)$          
&$0.0x+0.16(1-x)$         
&$0.24x+0.16(1-x)$        
&$0.16x+0.21(1-x)$
&$0.09x+0.07(1-x)$\\
  
$E_{P_0}$ (eV) 
&$20.9x+27.5(1-x)$         
&$23.5x+27.0(1-x)$        
&$21.6x+27.0(1-x)$        
&$27.0x+27.5(1-x)$
&$22.0x+20.2(1-x)$\\ 

$E_{P_1}$ (eV)
&$2.40x+11.5(1-x)$         
&$0.03x+5.9(1-x)$         
&$5.2x+5.9(1-x)$          
&$5.90x+11.5(1-x)$
&$3.0x+3.7(1-x)$\\ 

$E_Q$ (eV)       
&$12.0x+15.1(1-x)$         
&$12.9x+12.8(1-x)$        
&$15.8x+12.8(1-x)$        
&$12.8x+15.1(1-x)$
&$22.8x+15.9(1-x)$\\

$a_{\text{o}}$ (\AA)       
&$6.136x+6.096(1-x)$         
&$5.652x+5.653(1-x)$        
&$6.058x+5.653(1-x)$        
&$5.653x+6.096(1-x)$
&$5.447x+5.860(1-x)$\\

$c_{11}$ ($10^{11}$ dyn/cm$^2$)       
&$8.769x+8.834(1-x)$         
&$12.02x+12.3(1-x)$        
&$8.329x+12.3(1-x)$        
&$12.3x+8.834(1-x)$
&$14.388x+10.11(1-x)$\\

$c_{12}$ ($10^{11}$ dyn/cm$^2$)       
&$4.341x+4.023(1-x)$         
&$5.70x+5.71(1-x)$        
&$4.526x+5.71(1-x)$        
&$5.71x+4.023(1-x)$
&$7.144x+5.61(1-x)$\\

$V_{a}$ (eV)       
&$-5.9x-7.2(1-x)$         
&$-2.6x-9.8(1-x)$        
&$-5.8x-9.8(1-x)$        
&$-9.8x-7.2(1-x)$
&$-9.3x-6.4(1-x)$\\

$V_{b}$ (eV)       
&$-1.4x-1.8(1-x)$         
&$-1.5x-2.0(1-x)$        
&$-1.8x-2.0(1-x)$        
&$-2.0x-1.8(1-x)$
&$-1.8x-2.0(1-x)$\\
\label{materials1}
\end{tabular}
\end{ruledtabular}
\footnotetext{$E_{P_0} = 2 m_e^{-1} P_0^2$, $E_{P_1} = 2 m_e^{-1} P_1^2$, and $E_Q = 2 m_e^{-1} Q^2$}
\end{sidewaystable}
\begin{sidewaystable}
\caption[]{Parameters for fourteen-band ${\bf k} \cdot {\bf p}$ band structures 
for GaP$_{x}$As$_{1-x}$, InP$_{x}$As$_{1-x}$, In$_{x}$Ga$_{1-x}$Sb, and Cd$_{x}$Zn$_{1-x}$Se at room temperature. $E_g$ is the fundamental gap (between $\overline{s}$ and $p \left(\Gamma_{8} \right)$ states), $E_c$ is the zone-center energy difference between $p \left(\Gamma_{8} \right)$ and $\overline{p} \left(\Gamma_{7} \right)$, and $\Delta$ and $\Delta_{c}$ are the spin-orbit splitting of the $p$ and $\overline{p}$ states, respectively, $a_{\text{o}}$ is the lattice constant, $c_{11}$ and $c_{12}$ are the elastic constants, and $V_{a}$ and $V_{b}$ are the deformation potentials.}
\begin{ruledtabular}
\begin{tabular}{ccccc}    
&GaP$_{x}$As$_{1-x}$  
&InP$_{x}$As$_{1-x}$  
&In$_{x}$Ga$_{1-x}$Sb
&Cd$_{x}$Zn$_{1-x}$Se\\ 
\colrule
$E_g$ (eV)                 
&$1.422+1.172x+0.190x^2$  
&$0.356+0.675x+0.320x^2$  
&$0.723-0.926x+0.371x^2$
&$1.670-0.673x+0.302x^2$\\
  
$\Delta$ (eV)                 
&$0.080x+0.341(1-x)$      
&$0.108x+0.380(1-x)$       
&$0.850x+0.758(1-x)$
&$0.390x+0.417(1-x)$\\
 
$E_c$ (eV)                        
&$5.24x+4.61(1-x)$         
&$4.63x+4.39(1-x)$        
&$3.14x+3.19(1-x)$
&$6.50x+7.17(1-x)$\\
  
$\Delta_c$ (eV)                 
&$0.09x+0.16(1-x)$         
&$0.07x+0.24(1-x)$        
&$0.39x+0.21(1-x)$
&$0.24x+0.08(1-x)$\\
  
$E_{P_0}$ (eV)       
&$22.0x+27.0(1-x)$         
&$20.2x+21.6(1-x)$        
&$24.0x+27.5(1-x)$
&$23.1x+19.9(1-x)$\\ 

$E_{P_1}$ (eV)      
&$3.0x+5.9(1-x)$        
&$3.7x+5.2(1-x)$          
&$7.5x+11.5(1-x)$
&$0.05x+0.2(1-x)$\\ 

$E_Q$ (eV)               
&$22.8x+12.8(1-x)$        
&$15.9x+15.8(1-x)$        
&$15.5x+15.1(1-x)$
&$13.0x+12.9(1-x)$\\

$a_{\text{o}}$ (\AA)       
&$5.447x+5.653(1-x)$         
&$5.860x+6.058(1-x)$            
&$6.469x+6.096(1-x)$
&$6.077x+5.668(1-x)$\\

$c_{11}$ ($10^{11}$ dyn/cm$^2$)       
&$14.388x+12.3(1-x)$         
&$10.11x+8.329(1-x)$               
&$6.918x+8.834(1-x)$
&$7.49x+9.407(1-x)$\\

$c_{12}$ ($10^{11}$ dyn/cm$^2$)       
&$7.144x+5.71(1-x)$         
&$5.61x+4.526(1-x)$             
&$3.788x+4.023(1-x)$
&$4.609x+5.625(1-x)$\\

$V_{a}$ (eV)       
&$-9.3x-9.8(1-x)$         
&$-6.4x-5.8(1-x)$              
&$-7.7x-7.2(1-x)$
&$-2.1x-6.2(1-x)$\\

$V_{b}$ (eV)       
&$-1.8x-2.0(1-x)$         
&$-2.0x-1.8(1-x)$              
&$-2.0x-1.8(1-x)$
&$-1.1x-1.4(1-x)$\\
\label{materials2}
\end{tabular}
\end{ruledtabular}
\footnotetext{$E_{P_0} = 2 m_e^{-1} P_0^2$, $E_{P_1} = 2 m_e^{-1} P_1^2$, and $E_Q = 2 m_e^{-1} Q^2$}
\end{sidewaystable}
\clearpage
\begin{sidewaystable}
\caption[]{The conduction-band offset at $300$ K for various structures}
\begin{ruledtabular}
\begin{tabular}{ccc}  
Structure &$\Delta E_{\text{c}}$ (meV) &$\Delta E_{\text{s}}$ (meV)\\ 
\colrule
$75$-\AA\ GaAs/$100$-\AA\ Al$_{0.4}$Ga$_{0.6}$As QW &$370$ &$4$\\
$150$-\AA\ GaAs/$100$-\AA\ Al$_{0.03}$Ga$_{0.97}$As QW &$22$ &$0$\\
$150$-\AA\ GaAs/$100$-\AA\ Al$_{0.4}$Ga$_{0.6}$As QW &$360$ &$4$\\
$21.2$-\AA\ InAs/$36.6$-\AA\ GaSb SL &$981$ &$49$\\
$75$-\AA\ InAs/$200$-\AA\ AlSb QW &$2188$ &$106$\\
$70$-\AA\ In$_{0.53}$Ga$_{0.47}$As/$97$-\AA\ InP QW &$261$ &$0$\\
$100$-\AA\ In$_{0.13}$Ga$_{0.87}$As/$200$-\AA\ GaAs QW &$78$ &$62$\\
$95$-\AA\ In$_{0.57}$Ga$_{0.43}$As$_{0.93}$P$_{0.07}$/$75$-\AA\ In$_{0.87}$Ga$_{0.13}$As$_{0.29}$P$_{0.71}$ QW &$161$ &$2$\\
$40$-\AA\ In$_{0.20}$Ga$_{0.80}$As/$100$-\AA\ GaAs/$40$-\AA\ In$_{0.20}$Ga$_{0.80}$As/$200$-\AA\ GaAs QW &$121$ &$99$\\
$80$-\AA\ GaSb /$80$-\AA\ AlSb QW &$1212$ &$35$\\
$51$-\AA\ Ga$_{0.019}$As$_{0.81}$Sb /$80$-\AA\ AlSb QW &$1511$ &$105$\\
$105$-\AA\ Cd$_{0.24}$Zn$_{0.76}$Se/$75$-\AA\ ZnSe QW &$114$ &$82$\\
\label{vboffset}
\end{tabular}
\end{ruledtabular}
\footnotetext{$\Delta E_{\text{c}}$ is the conduction-band offset, and $\Delta E_{\text{s}}$ is the spin splitting energy due to lattice-mismatched-induced strain.}
\end{sidewaystable}
\begin{table}
\caption[]{Electron spin coherence times for In$_{x}$Ga$_{1-x}$As$_{y}$P$_{1-y}$ quantum wells}
\begin{ruledtabular}
\begin{tabular}{ccc}  
&Experiments~\cite{hyland99} &Theory\\
Well width (\AA) &$T_{1}$ (ps) &$T_{1}$ (ps)\\ 
\colrule
$55$ &$2.85$  &$3.2$ \\
$65$ &$3.95$  &$3.1$ \\
$90$ &$6.0$  &$3.9$ \\
$95$ &$10.0$  &$4.0$ \\
\label{coherencetimes}
\end{tabular}
\end{ruledtabular}
\footnotetext{Experimental data are taken directly from Table~2 of Ref.~\onlinecite{hyland99}}.
\end{table}
\clearpage
\clearpage
\clearpage
\end{document}